\documentclass[a4paper,11pt]{article}
\pdfoutput=1 

\usepackage{jcappub} 

\usepackage[T1]{fontenc} 
\usepackage{placeins}
\usepackage{float}

\title{\boldmath Investigating binary-neutron-star mergers as production sites of high-energy neutrinos}

\author[a]{S. Rossoni,}
\author[b,c]{D. Boncioli,}
\author[a]{G. Sigl}

\affiliation[a]{ II. Institute for Theoretical Physics, Hamburg University, \\ Luruper Chaussee 149, 22761, Hamburg, Germany}
\affiliation[b]{ Università degli Studi dell’Aquila, Dipartimento di Scienze Fisiche e Chimiche, \\ Via Vetoio, 67100, L’Aquila, Italy}
\affiliation[c]{INFN Laboratori Nazionali del Gran Sasso, \\ Assergi (L’Aquila), Italy}

\emailAdd{simone.rossoni@desy.de}
\emailAdd{denise.boncioli@univaq.it}
\emailAdd{guenter.sigl@desy.de}

\abstract{The end state of binary-neutron-star (BNS) mergers can manifest conditions to produce high-energy neutrinos. Inspired by the event GW170817, detected in gravitational waves and in optical/infrared emission, we investigate a scenario in which cosmic-ray (CR) particles are accelerated, in a population of BNS mergers, in the energy range that might contribute from the \textit{knee} to the \textit{ankle} of the CR measured spectrum. By taking into account the measured thermal and non-thermal energy density of the photon fields in the source environment as a function of the time after the merger, we model the CR interactions and the consequent neutrino production. We propagate the escaped CR and neutrino fluxes through the extragalactic space and compare the expected diffuse fluxes to the experimental data and current limits. Depending on the CR spectral and composition parameters at acceleration, and on the possible contribution to the \textit{sub-ankle} CR flux, we discuss the predicted diffuse neutrino flux associated to this class of astrophysical objects, as a function of the details of the photon field characterizing the merger stage, including its evolution in time. We constrain the fraction of accelerated baryons in the source site given the BNS merger rate per volume, taking into account at the same time the constraints from the measured CR and neutrino fluxes.

\vspace*{\fill}
\begin{flushleft}
\textbf{Key words: }binary-neutron-star merger, ultra-high-energy cosmic rays, astrophysical neutrinos, cosmogenic neutrinos, source interactions, extragalactic propagation 
\end{flushleft}}

\begin{document}
\maketitle
\flushbottom

\section{Introduction}
\label{sec_intro}
\par In the last ten years, the IceCube Neutrino Observatory \cite{IceCube:2016zyt} has reported the observation of a diffuse neutrino flux in the TeV - PeV energy range \cite{IceCube:2013low}. In particular, the muon-neutrino flux from the IceCube muon-track data \cite{IceCube:2021uhz} is consistent with a single power-law spectrum of the form $E_\nu^{-\gamma}$, with normalization at $100\,\text{TeV}$ of $\phi_{\nu_\mu+\bar{\nu}_\mu}\simeq 1.5\cdot 10^{-18}\,\text{GeV}^{-1}\,\text{cm}^{-2}\,\text{s}^{-1}\,\text{sr}^{-1}$ and spectral index $\gamma\simeq2.4$. Recently, the flux of neutrinos from cascade events \cite{IceCube:2020acn} has also been shown to be consistent with a single power law with normalization of $\phi_{\nu+\bar{\nu}}\simeq 1.7\cdot 10^{-18}\,\text{GeV}^{-1}\,\text{cm}^{-2}\,\text{s}^{-1}\,\text{sr}^{-1}$ at $100\,\text{TeV}$, and spectral index $\gamma\simeq2.5$. 
Upper limits have been set for contributions from different classes of sources, such as for instance the one from blazars \cite{IceCube:2016qvd}. Observations of astrophysical sources through different messengers led to the emergence of multi-messenger astronomy \cite{Ackermann:2019ows}, such as the successful observation in neutrinos and gamma rays of the blazar TXS 0506+056 \cite{IceCube:2018dnn}, and the recent association of 79 neutrino events to the Seyfert galaxy NGC 1068 \cite{IceCube:2022der}. It has also been shown that IceCube neutrino events can be associated with the optical counterparts of the emission from sites where the disruption of stars from a supermassive black hole is supposed to happen (tidal disruption events, TDEs) \cite{Stein:2020xhk,Reusch:2021ztx,vanVelzen:2021zsm}. 
\par The observation of high-energy neutrinos from astrophysical sites might reveal hadronic or photo-hadronic processes within the source region, involving hadronic particles accelerated in the environment as well as photons and/or matter of the source site. Phenomenological models are developed to explain at the same time the electromagnetic emission and the neutrino production, as for instance in \cite{Gao:2018mnu} for the blazar TXS 0506+056 or in \cite{Eichmann:2022lxh} for NGC 1068, without the need of invoking acceleration mechanisms of cosmic rays (CRs) in jets. In particular, in the latter case a two-zone model is investigated by modeling interactions of cosmic-ray particles in the corona as well as in the circumnuclear starburst region. Non-jetted sites are therefore nowadays increasingly interesting as possible high-energy neutrino factories, as recently investigated also in \cite{Padovani:2024tgx}. This is also supported by the interpretation of the neutrino emission from the TDE AT2019aalc as modeled in \cite{Winter:2022fpf}, where it is shown that the delay of the neutrino signal with respect to the optical-infrared emission can be due to the confinement of protons in regions not aligned with the jet. 
\par On the other hand, the first joint observation of a gravitational wave signal and the electromagnetic counterpart \cite{LIGOScientific:2017vwq,LIGOScientific:2017ync,LIGOScientific:2017zic} happened in 2017. Although, nowadays, no evidence of correlations between high-energy neutrinos and gravitational waves has been established \cite{IceCube:2020xks}, the source environments responsible for gravitational-wave signals are considered to be of great interest for the study of high-energy interactions, as already shown in \cite{Kotera:2011vs}. A particular class of gravitational-wave sources are binary systems of coalescing neutron stars (NSs). The probable end state of binary-neutron-star (BNS) mergers is a black hole (BH) with a relativistic jet, powered by the material in the accretion disk. The formation of the jet gives rise to a short gamma-ray burst (GRB), which represents a promising site for the production of high-energy neutrinos, as also reported in \cite{Farrar:2024zsm}. An alternative scenario for the production of astrophysical neutrinos is described in \cite{Decoene:2019eux}, where a small fraction of the ejected material is considered to fall back to the central compact object produced after the merger. This fallback outflow encounters the earlier ejected mass shell producing a shock wave where particles can be accelerated. It has been shown in previous studies \cite{Kotera:2015pya,Kimura:2017kan,Kimura:2018ggg,Kimura:2018vvz,Rodrigues:2018bjg} that these environments might be interesting acceleration sites of cosmic rays. In particular, in \cite{Rodrigues:2018bjg} it is discussed how the characteristic magnetic field might allow cosmic-ray particles to reach the energy of the \textit{ankle} in the CR energy spectrum. Therefore, these sources could be considered as candidates for contributing to the energy region of the cosmic-ray spectrum beyond the Galactic contribution.  
\par In the present work, we consider the modeling of the BNS merger remnant described in \cite{Decoene:2019eux} to study the interaction of accelerated ultra-high-energy cosmic rays (UHECRs, i.e. atomic nuclei with energy $\gtrsim 10^{17}\,\text{eV}$) with the local photon fields. In particular, we consider the source region to be populated by a thermal field produced by the nuclear decay of synthesized nuclei in the ejecta, and a non-thermal synchrotron component (see \cite{Rodrigues:2018bjg,Margutti:2018xqd}). Interactions of UHECRs with local photons give rise to the production of unstable mesons, and therefore to the production of high-energy neutrinos. In particular, after being produced, the latter ones leave the source undisturbed and travel through the outer space without undergoing any interaction or magnetic deflection. In contrast, the escape condition of UHECRs represents a non-trivial problem. As a first approximation, the typical UHECR escape time is given by the dimension of the source. In this work, we assume the radius of the ejecta material as the typical size of the interaction region, i.e. we adopt the ballistic approximation. 
\par In order to link the observed information in UHECRs and neutrinos, the re-processing of the accelerated UHECRs within the merger region needs to be combined with the effects of the extragalactic propagation from the production site to Earth, consisting of interactions with the cosmic photon fields, as the cosmic microwave background (CMB) and extragalactic background light (EBL) \cite{Greisen:1966jv,Zatsepin:1966jv}. Interactions with cosmic photons will give rise to a second population of neutrinos called cosmogenic. Hereafter we will refer to neutrinos produced in the source as source neutrinos, and  neutrinos produce during the propagation as cosmogenic neutrinos. 
\par A population of BNS mergers is considered in this work; therefore, the diffuse flux will depend on the event rate per volume of BNS mergers, $\dot{n}$, which also affects the amount of UHECRs injected in the extragalactic space. We use these quantities to constrain the baryonic loading $\eta$ (i.e. the ratio between the fallback luminosity and the UHECR luminosity at the acceleration) of BNS mergers.
\par This study is organized as follows: the modeling of the BNS merger remnant and its interaction efficiency, and numerical implementation are discussed in Sec.~\ref{sec_int_source_env}. Simulation results at the escape from the source and at Earth are shown in Sec.~\ref{sec_results} (in particular, in Sec.~\ref{subsec_parameter} the production of high energy neutrinos is analyzed in detail). In Sec.~\ref{subsec_time_int}, neutrino production is studied by integrating over the time evolution of the source environment. Discussion of the results and conclusions are given in Sec.~\ref{sec_discussion}.

\section{Modeling interactions in the source environment}
\label{sec_int_source_env}
The production of high-energy neutrinos is here investigated as the result of interactions between the fallback material and the local photon fields of the source environment. We assume, as done in \cite{Decoene:2019eux}, that acceleration mechanisms can happen in the fallback process, so that nuclei can reach energies up to $\sim10^{19}\,\text{eV}$ (see also Sec.~\ref{subsec_source-escape} for more details). A brief introduction to our modeling of the source environment and interactions can be found in \cite{Rossoni:2023put,Rossoni:2021dhn}. We define the time $t$ as the time after the coalescence, so that $t=0$ corresponds to the merger event. Due to the nuclear decay of the unstable species synthesized in the ejecta by the merger, a thermal photon field is produced in the source environment. Assuming that the heat from nuclear decays is homogeneously distributed, the photon emission can be modeled as a black-body (BB) photon field. The flux of energy (i.e. the energy per unit of time, area and frequency) emitted by a black body and observed at a distance $d$ is given by 
\begin{equation}
F_\nu^{\text{BB}} = \dfrac{2\pi h \nu}{c^2} \left(\dfrac{R}{d}\right)^2 \dfrac{\nu^2}{\exp (h\nu/k_\text{B} T)-1} \, ,
\end{equation}
where $\nu$ is the photon frequency, $T$ is the BB temperature, $k_B$ is the Boltzmann constant, $c$ is the speed of light, $h$ is the Planck constant and $R$ is the source radius. The temperature of the BB is given by the Stefan-Boltzmann law 
\begin{equation}
T = \left( \dfrac{3E_{\text{BB}}}{4\pi a R^3}\right)^{1/4} \, ,
\end{equation}
where $u=3E_{\text{BB}}/4\pi R^3$ is the energy density of the BB and $a=7.6\cdot 10^{-15} \,\text{erg}\,\text{cm}^{-3}\,\text{K}^{-4}$ \cite{Decoene:2019eux,Maoz:2007}. We modeled the temporal evolution of the BB photon field density as in Fig.~2 of \cite{Decoene:2019eux}. Therefore, the time after the merger $t$ and the BB temperature $T$ are such that
\begin{equation}
\label{temperature_time}
T= 10^6 \,\cdot \left(\dfrac{t}{10^3\,\text{s}}\right)^{-2} \,\text{K} \, .
\end{equation}
As expected, when time increases after the merger, the temperature of the BB decreases.
The spectral energy density (SED, i.e. the number of photons per unit of energy and volume) is given by
\begin{equation}
\label{sed_bb}
n_\text{BB}(\epsilon) = \dfrac{1}{\pi^2 \left(\hbar c\right)^3} \dfrac{\epsilon^2}{\exp (\epsilon/k_\text{B} T)-1} \, ,
\end{equation}
where $\epsilon$ is the photon energy. The number density of BB photons is given by the integral in the photon energy $\epsilon$ of Eq.~\eqref{sed_bb}, and it corresponds to $n_{\text{BB}}\simeq 20\cdot(T/1\,\text{K})^3\,\text{cm}^{-3}$.
\par
\begin{figure}[t]
\centering
\includegraphics[scale=0.5]{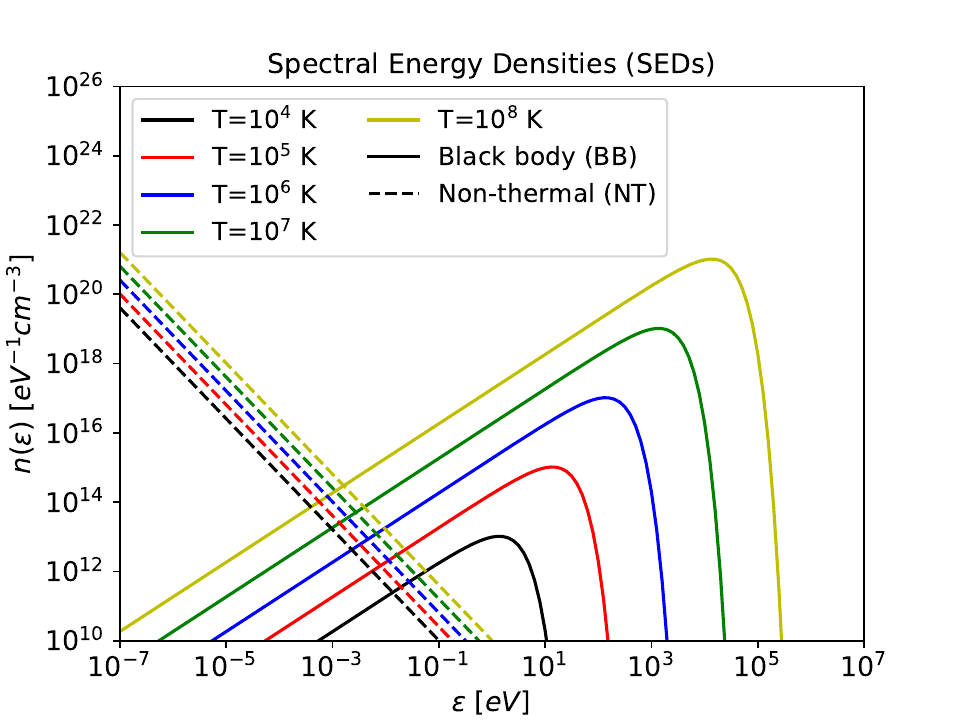}
\caption{Spectral energy densities of the photon fields used in this work: black body (solid lines) and non-thermal (dashed lines). Different times after the merger (corresponding to different temperatures of the black body) are shown in different colors.}
\label{img_sed}
\end{figure}

 Several days after the merger, the dominance of the thermal photon field is replaced by a non-thermal (NT) component, mainly due to synchrotron emission. In order to model this background field, we consider the radio-to-X-ray emission of the merger event GW170817 described in \cite{Margutti:2018xqd}, where the flux density function is modeled as $\phi_\text{NT}(\nu)\propto \nu^{-\beta}$, where $\nu$ is the photon frequency and the index $\beta$ is fixed to the value $0.6$, as discussed in \cite{Margutti:2018xqd}. From \cite{Margutti:2018xqd}, we obtain the flux density function\footnote{The units are given in $\mu\text{Jy}$, where $1\,\text{Jy}=10^{-23}\,\text{erg}\,\text{s}^{-1}\,\text{cm}^{-2}\,\text{Hz}^{-1}$.} as
\begin{equation}
\label{flux_den_func_nt}
\phi_\text{NT}\left( \nu\right) = 0.76\ \cdot \left( \dfrac{t}{1\,\text{s}}\right)^{1.2} \left( \dfrac{\nu}{1\,\text{Hz}}\right)^{-0.6}  \,\mu\text{Jy}\, ,
\end{equation}
where $t$ is the time after the merger. The flux density in Eq.~\eqref{flux_den_func_nt} can be converted to spectrum energy density, so that the non-thermal SED reads
\begin{equation}
\label{sed_nt}
n_\text{NT} \left( \epsilon\right)= 1.2\cdot 10^{27} \, \cdot \left(\dfrac{V}{1\,\text{km}^{3}}\right)^{-1}\left(\dfrac{t}{1\,\text{s}}\right)^{2.2} \left(\dfrac{\epsilon}{1\,\text{eV}}\right)^{-1.6} \, \text{eV}^{-1}\text{cm}^{-3}\, ,
\end{equation}
where $V$ is the volume of the source environment. Both the SEDs in Eqs.~\eqref{sed_bb} and~\eqref{sed_nt} are shown in Fig.~\ref{img_sed}, where solid lines correspond to the thermal (BB) field, while dashed lines correspond to the non-thermal (NT) component. Different colors are used for different temperatures (or times) after the merger. The BB component is the dominant background for $\epsilon\gtrsim0.01\,\text{eV}$. For $T\lesssim10^4 \, \text{K}$ (i.e. $t\gtrsim 3\,\text{h}$) the non-thermal field becomes dominant over the BB. However, we will show that after this time from the merger, the neutrino production by photohadronic interactions becomes inefficient.
\par The size of the production site for neutrinos can be approximated by taking into account the radius of the ejected material, in free expansion after the merger. In this work we make the simplistic assumption that the typical escape length for UHECRs is given by the radius of the ejected material, namely 
\begin{equation}
\label{source_radius}
\lambda_\text{esc}(t) = \beta_\text{ej} c\, t \, ,
\end{equation}
where $\beta_\text{ej}$ is the speed of the ejected material in units of speed of light. The typical escape rate is thus 
\begin{equation}
\label{escape_time}
\tau_\text{esc}^{-1}(t) = \dfrac{c}{\lambda_\text{esc}(t)} = \dfrac{1}{\beta_\text{ej}t}\, .
\end{equation}
We assume $\beta_\text{ej}=0.3$, as done in \cite{Decoene:2019eux}. This assumption may influence the neutrino production efficiency of the source. A higher value of $\beta_\text{ej}$ corresponds to a lower escape rate, and thus accelerated CRs may produce more neutrinos. We also note that the confinement of a nucleus is only due to the dimension of the source itself, and does not depend on its rigidity. Further details on the validity of the ballistic approximation can be found in Appendix~\ref{sec_ballistic}.
\par The typical source length defined in Eq.~\eqref{source_radius} can be used to compute the source volume in the non-thermal SED in Eq.~\eqref{sed_nt}. Therefore, the non-thermal SED $n_\text{NT}(\epsilon)$ can be written as 
\begin{equation}
\label{sed_nt_bis}
n_\text{NT} \left( \epsilon\right)= 4.2\cdot 10^{11} \, \cdot \left(\dfrac{\beta_\text{ej}}{0.3}\right)^{-3}\left(\dfrac{t}{1\,\text{s}}\right)^{-0.8} \left(\dfrac{\epsilon}{1\,\text{eV}}\right)^{-1.6} \, \text{eV}^{-1}\text{cm}^{-3}\, .
\end{equation}
We obtain that $n_\text{NT}(\epsilon)$ evolves in time  as $\propto t^{-0.8}$. 

\subsection{Interaction efficiency}
\label{subsec_int-efficiency}
\begin{figure}[t]
\centering
\begin{minipage}{6.5cm}
\centering
\includegraphics[scale=0.42]{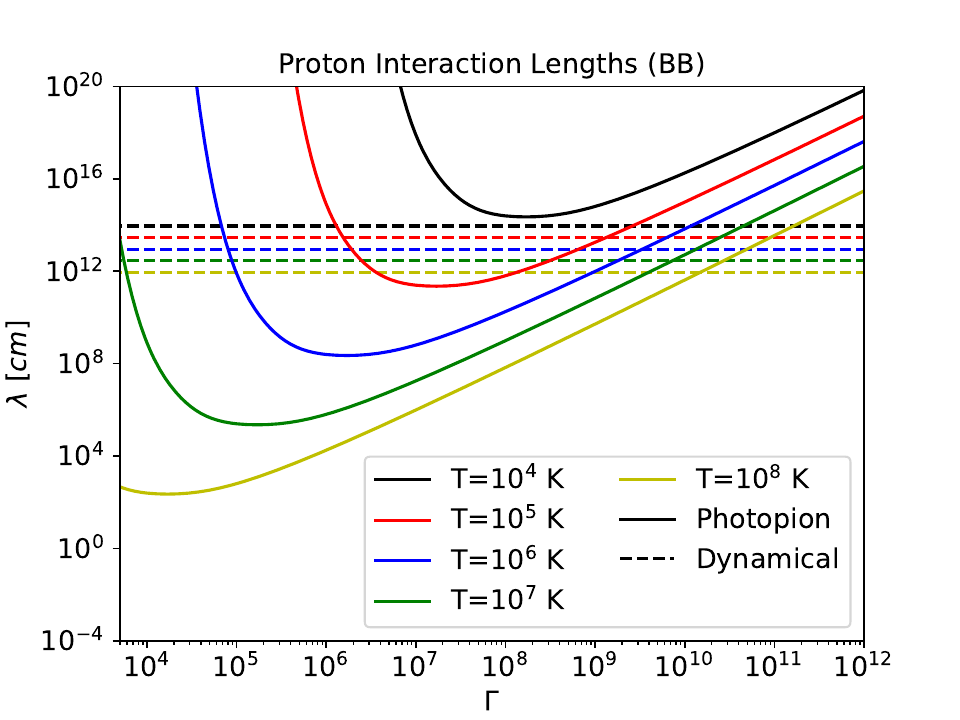}
\end{minipage}
\begin{minipage}{6.5cm}
\centering
\includegraphics[scale=0.42]{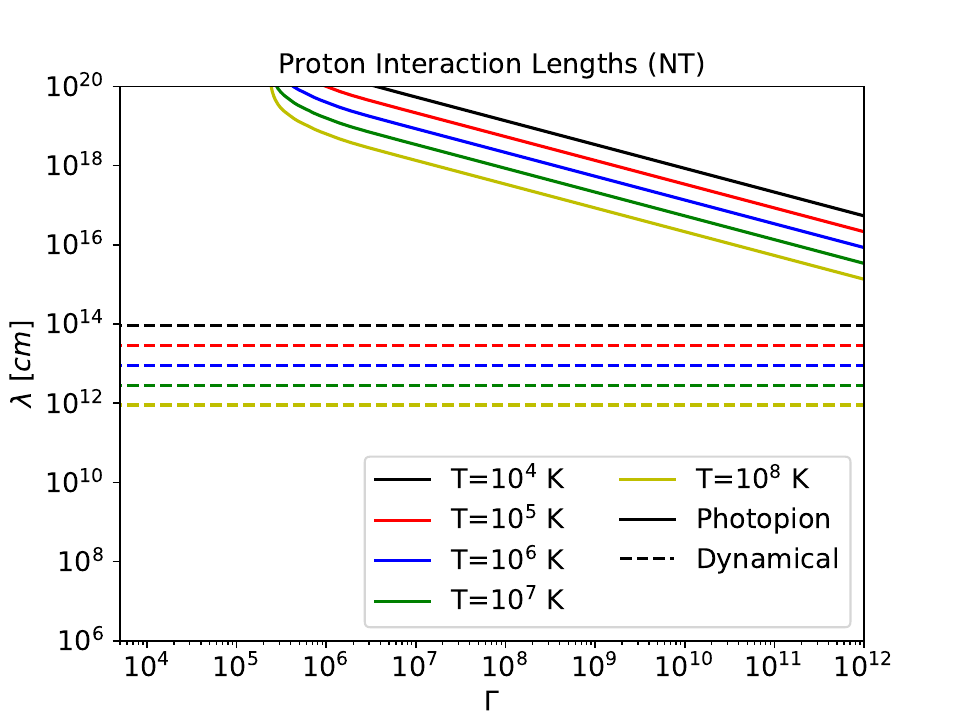}
\end{minipage}
\caption{Interaction lengths for protons for photopion production, corresponding to the BB photon field (solid line, left panel) and to the NT field (solid line, right panel), as a function of the Lorentz factor; the size of the source radii is indicated with dashed lines. Different colors refer to the BB temperatures as indicated in Fig.~\ref{img_sed}. Note the different ranges of the y-axes.}
\label{int_len_p}
\end{figure}

\begin{figure}[t]
\centering
\begin{minipage}{6.5cm}
\centering
\includegraphics[scale=0.42]{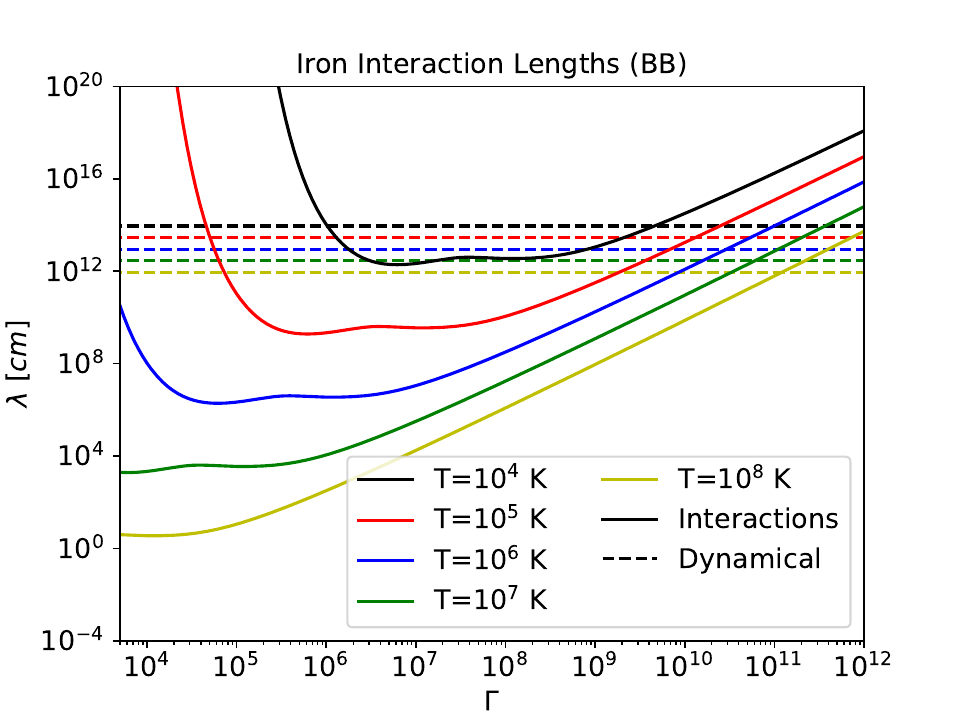}
\end{minipage}
\begin{minipage}{6.5cm}
\centering
\includegraphics[scale=0.42]{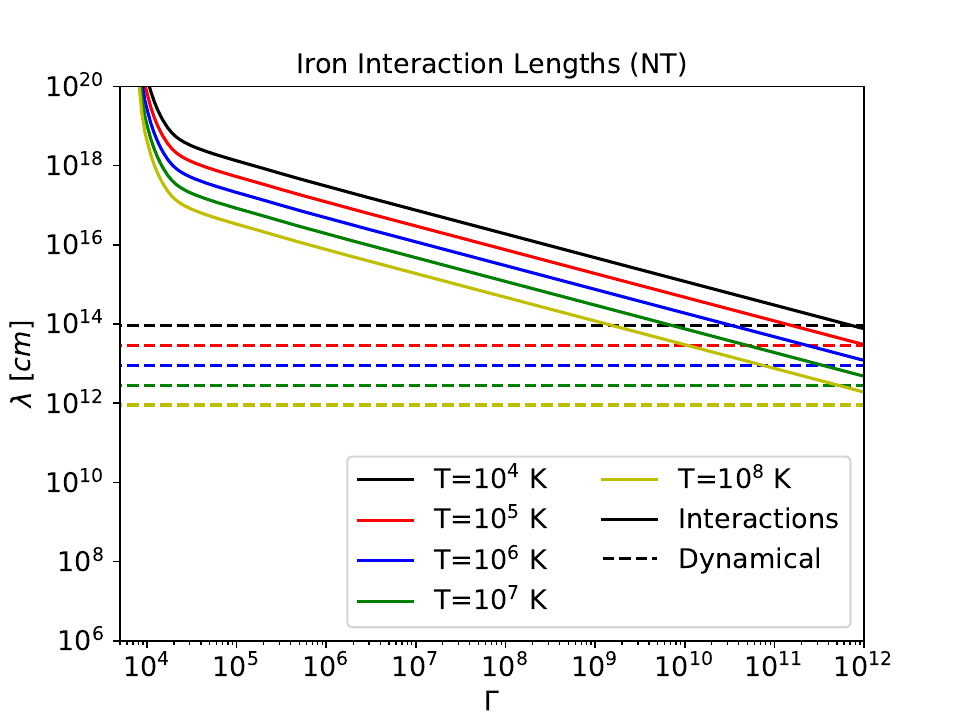}
\end{minipage}
\caption{Same as in Fig.~\ref{int_len_p}, for iron nuclei. Note the different ranges of the y-axes.}
\label{int_len_Fe}
\end{figure}

In this section we compute the photohadronic interaction lengths of the two nuclear species considered in this work, namely protons (p) and iron nuclei ($^{56}$Fe), at different times after the merger\footnote{As shown in \cite{Decoene:2019eux}, the photopion production is the dominant process for neutrino production, for both protons and iron nuclei, with respect to hadronic interactions. In particular, given the hadronic cross-section $\sigma\sim10^{-25}\,\text{cm}^2$ and inelasticity $\kappa\sim0.5$ at $1\,\text{EeV}$, we obtain, for post-merger time $t\simeq10^2\,\text{s}$ and baryon density from \cite{Decoene:2019eux}, an energy-loss length $\sim2\cdot10^{10}\,\text{cm}$. Comparing this estimate to the photohadronic interaction lengths shown in the left panels of Figures~\ref{int_len_p} and~\ref{int_len_Fe}, it can be seen that photohadronic interactions are the main neutrino production processes. Therefore, we consider only photohadronic interactions in this work.}. Bethe-Heitler pair production is also taken into account in this work. However, as shown in Figure 3 of \cite{Decoene:2019eux}, the energy-loss length of pair production is always several orders of magnitude greater than that associated with photopion production. Therefore, Bethe-Heitler pair production will not have a major effect on the total interaction efficiency and escape conditions of the accelerated nuclei. Further details on the calculation of photohadronic interaction lengths can be found in Appendix~\ref{sec_uhecr-int}. 
\par In the left panel of Fig.~\ref{int_len_p}, the photopion interaction lengths (solid lines) of protons interacting with BB photon fields are shown, as a function of the proton Lorentz factor $\Gamma$. Different colors correspond to different temperatures, as in the legend of Fig. \ref{img_sed}. The dashed lines correspond to the source typical lengths as defined in Eq.~\eqref{source_radius} at the corresponding temperatures. Two main effects can be observed: first, both the interaction lengths $\lambda_\text{p}$ and the source radii $\lambda_\text{esc}$ increase as the BB field cools; secondly, $\lambda_\text{p}>\lambda_\text{esc}$ for each value of $\Gamma$ only when $T\sim 10^4 \, \text{K}$. After the merger, the BB temperature decreases and the number of available high-energy photons decreases, making photopion production less efficient. At the same time, the expansion of the source environment is not fast enough to compensate for the field dilution. The result is that, for $T\gtrsim 10^4 \, \text{K}$, protons no longer interact with the local field and escape freely from the source. Conversely, in the early stages after the merger, protons interact frequently with the local field, and only very high energy protons escape undisturbed. The same result applies to protons of very low energy. However, in this case the escape energy threshold for low energy protons increases rapidly as the temperature of the BB decreases, to the point where protons escape for any energy (black line in left panel of Fig.~\ref{int_len_p}). In the right panel of Fig.~\ref{int_len_p} the same quantities of the left panel are shown, corresponding to the NT field as defined in Eq.~\eqref{sed_nt_bis}. In this case, the photopion production is never an efficient process since the number of high-energy photons is highly suppressed, even when the source temperature is $T=10^8 \,\text{K}$. 

\par The propagation of iron nuclei in the source environment is affected by the photodisintegration as well as by the photopion process. In the left panel of Fig.~\ref{int_len_Fe} the total interaction length of an iron nucleus is shown together with the typical source radius; the low-energy minimum can be attributed to the photodisintegration while the high energy minimum to the photopion production. The general evolution of the interaction-escape dynamics with the temperature is similar to the proton scenario, but for this case interactions are possible also for $T\sim 10^4\,\text{K}$. The total interaction length of iron nuclei with NT photons is slightly smaller than in the case of protons. However, interactions with this photon component of the source environment are still inefficient.
\par The evolution of the efficiency of photohadronic interactions as a function of the time after the merger can be summarized by considering the source opacity. The opacity of the source environment for a nucleus of mass $A$ can be defined as  
\begin{equation}
\label{opacity_def}
\zeta_A (\Gamma,t)= \dfrac{\lambda_\text{esc}(t)}{\lambda_A(\Gamma,t)} \, ,
\end{equation}
where $\lambda_A$ is the total interaction length of a nucleus with mass $A$ and $\lambda_\text{esc}$ is defined in Eq.~\eqref{source_radius}. 
\begin{figure}[t]
\centering
\begin{minipage}{6.5cm}
\centering
\includegraphics[scale=0.42]{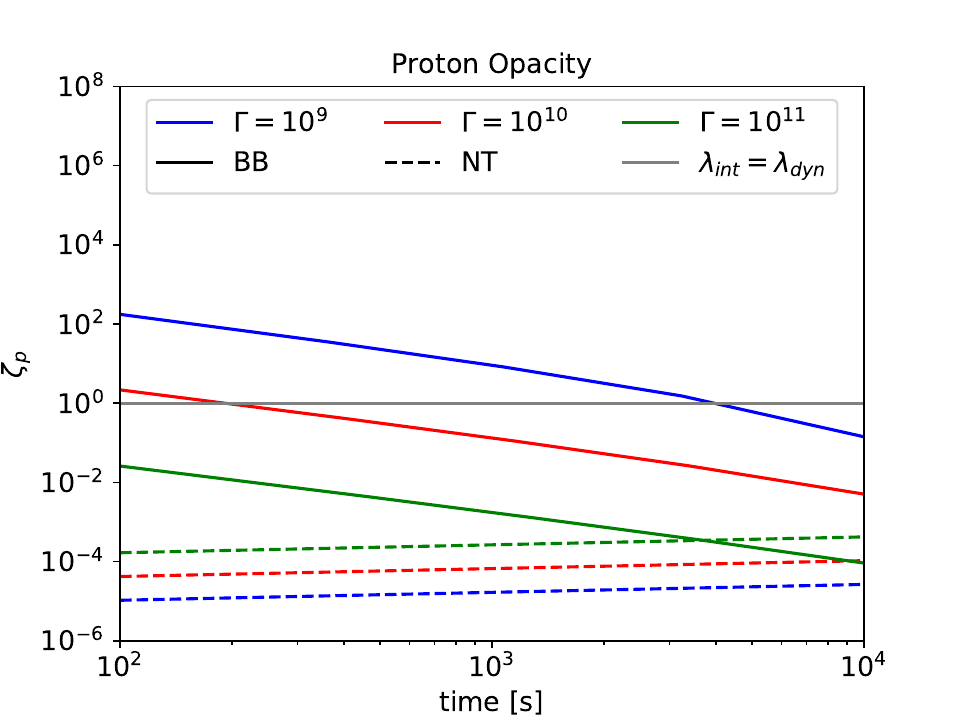}
\end{minipage}
\begin{minipage}{6.5cm}
\centering
\includegraphics[scale=0.42]{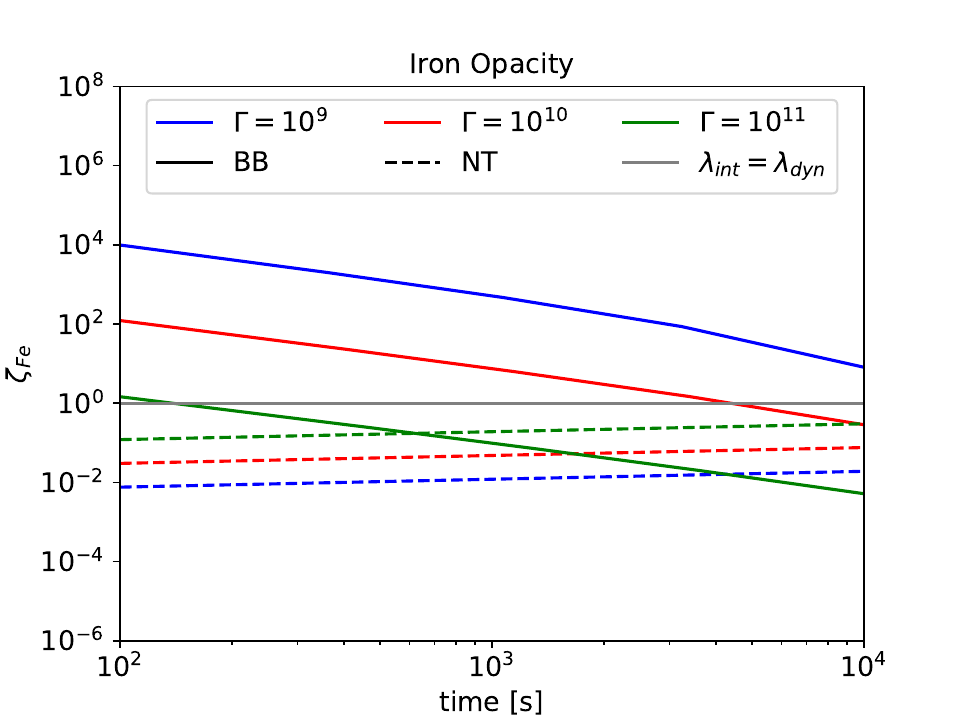}
\end{minipage}
\caption{Source opacity as defined in Eq.~\eqref{opacity_def} for Lorentz factors $\Gamma=10^9$ (blue), $\Gamma=10^{10}$ (red) and $\Gamma=10^{11}$ (green). Solid lines corresponds to the BB field, dashed lines to the NT field and the solid gray line represents the condition $\lambda_A=\lambda_\text{esc}$. The cases for protons (left) and for iron nuclei (right panel) are shown.}
\label{opacities}
\end{figure}
In particular, when $\zeta_A>1$ ($<1$), interactions dominate (are suppressed) over the escape condition. This quantity is shown in the left panel of Fig.~\ref{opacities} for protons and in the right panel for iron nuclei, as a function of time, for different fixed values of the Lorentz factor. Here, solid lines refer to the opacity of the BB, and dashed lines to the opacity of the NT field. The solid gray line represents the confinement-escape limit $\lambda_A=\lambda_\text{esc}$. The NT component is always transparent for protons and iron nuclei, while the BB opacity is greater than $1$ for $\Gamma \lesssim 10^8$ for protons, and $\Gamma\lesssim10^{10}$ for iron nuclei, for a considerable time after the merger. As also discussed in \cite{Rodrigues:2018bjg}, very high energy cosmic rays ($\Gamma \gtrsim 10^{12}$) will undergo interactions with local photon fields when $t\lesssim10^4\,\text{s}$. At later times, the source opacity becomes smaller than $1$, and the nuclei are free to escape from the source region without interacting.
\par The NT component of the source environment is, for both the nuclear species considered, not relevant for UHECR interactions, and therefore for the production of astrophysical neutrinos. For this reasons, we will consider only the interactions in the BB field in the following. 

\subsection{Numerical simulations}
\label{subsec_numerical_sim}
In this work, we modify the Monte Carlo code for the computation of the UHECR extragalactic propagation \texttt{SimProp-v2r4}\footnote{The code is available at \url{https://github.com/SimProp/SimProp-v2r4}} \cite{Aloisio:2017iyh}, in order to simulate the interactions within the source environment, until the escape condition is met. We name this modified version \texttt{SimProp-Mod}. All the interaction processes present in \texttt{SimProp-v2r4} involve CRs and the extragalactic background photon fields (CMB and EBL). The first modification concerns the target photon fields for photohadronic interactions. We replaced the cosmic fields with the local fields described in Sec.~\ref{sec_int_source_env}. As discussed earlier, the non-thermal component of the source environment is not considered in the simulations. In the original version of \texttt{SimProp-v2r4}, particles propagate from the source until they reach the redshift condition $z=0$. In the current version, we implemented an escape condition based on a leaky-box model, where the escape rate in Eq.~\eqref{escape_time} is compared to the interaction rates as defined in Eq.~\eqref{int_rate} (see Appendix~\ref{sec_uhecr-int} for more details). The escape condition is therefore determined by the random sampling of the escape rate among all the possible processes.
 
\section{Results}
\label{sec_results}
In this section, we make use of the simulation framework previously descibed considering several scenarios for the injection of UHECRs in the source region. When the spectra of the particles at the escape are obtained, these are propagated to the Earth and compared to available data, in order to constrain the source parameters.

\subsection{Source escape and interactions}
\label{subsec_source-escape}
We describe here the propagation of UHECRs in the source taking into account pair production, photopion production and photodisintegration with the BB component of the local photon fields (see Sec.~\ref{subsec_int-efficiency}). We consider post-merger times corresponding to BB temperatures between $10^8\,\text{K}$ and $10^4\,\text{K}$ (i.e. the temperatures shown in Fig.~\ref{img_sed}). Since the temperature of the BB is a fixed parameter in every simulation, the other source properties, which depend on the temperature (source radius, environment SEDs and escape rate) are also fixed. The photodisintegration of nuclei heavier than protons is simulated considering the default option of \texttt{SimProp-v2r4} with the parametrization of the cross-sections adapted from \cite{Puget:1976nz} and \cite{Stecker:1998ib}. 
\par In \cite{Decoene:2019eux}, it was shown that, assuming that the magnetic energy at the acceleration site is a fraction of the kinetic energy of the fallback, the only limiting process to the acceleration of nuclei is synchrotron cooling. Since hadronic and photoadronic interaction processes occur on a large time scale, we can separate the CR acceleration phase from the neutrino production phase. Furthermore, assuming an equipartition between magnetic and kinetic energy, a maximum CR acceleration energy of $\sim10^{19}\,\text{eV}$ can be reached for proton and iron nuclei. In \cite{Decoene:2019eux} we can also see that the maximum acceleration energy depends slightly on the post-merger time. Therefore, we assume a scenario in which  $E_\text{cut}$ does not change over time.
\par We study the production of high energy neutrinos by injecting CRs with energy between $10^{14}\,\text{eV}$ and $10^{20}\,\text{eV}$ into the source region. The injection rate at the acceleration (i.e. the number of cosmic rays per unit of energy, time and  volume) of the nuclear species with mass number $A$ is taken as
\begin{equation}
\label{spectrum_acc}
Q_\text{acc}^A (E) = Q_\text{0,acc}^A \left( \frac{E}{1\,\text{EeV}}\right)^{-\gamma} \exp \left( -\dfrac{E}{E_\text{cut}}\right) \, ,
\end{equation}
where $\gamma$ is the spectral index and $E_\text{cut}$ is the high-energy cutoff; the normalization $Q_\text{0,acc}^A$ will be adjusted thanks to the comparison of the propagated cosmic rays to the experimental data. For the spectral parameters we consider $\gamma=0.5,...,2.5$ with steps of $\Delta\gamma=0.25$, and $\log (E_\text{cut}/1\,\text{eV})=17.0,...,19.0$ with steps of $\Delta\log (E_\text{cut}/1\,\text{eV})=0.1$. 
\begin{figure}[t]
\centering
\begin{minipage}{6.5cm}
\centering
\includegraphics[scale=0.42]{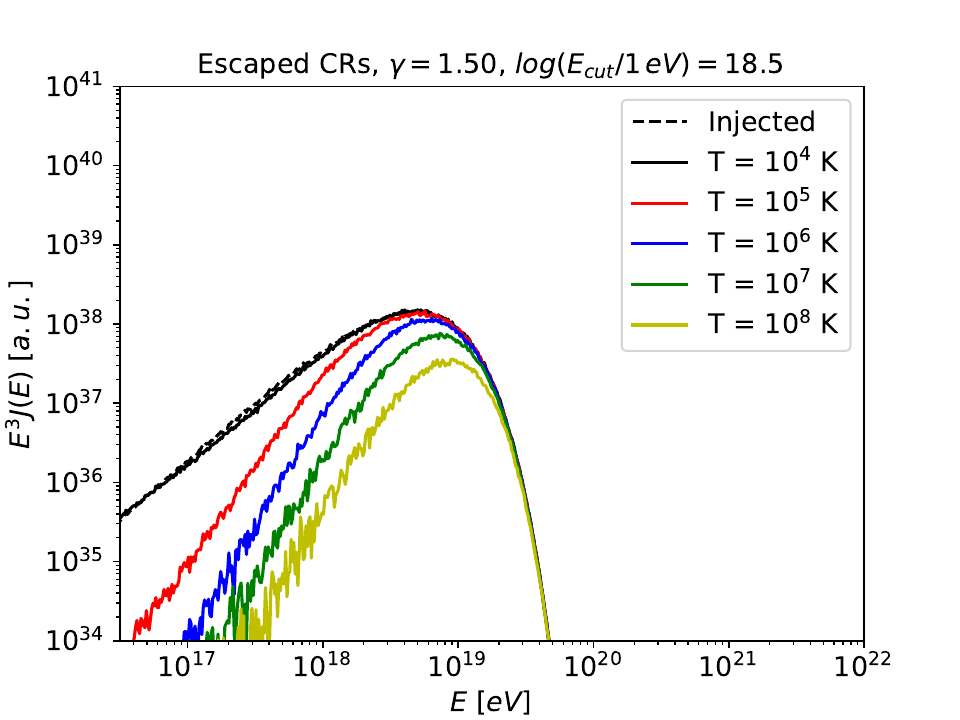}
\end{minipage}
\begin{minipage}{6.5cm}
\centering
\includegraphics[scale=0.42]{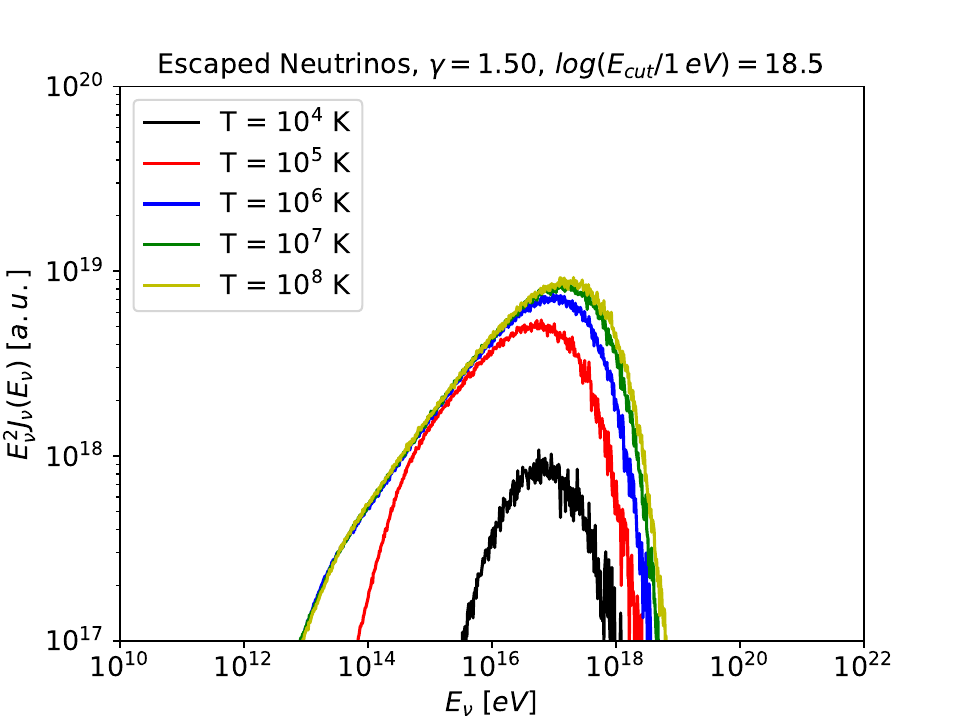}
\end{minipage}
\begin{minipage}{6.5cm}
\centering
\includegraphics[scale=0.42]{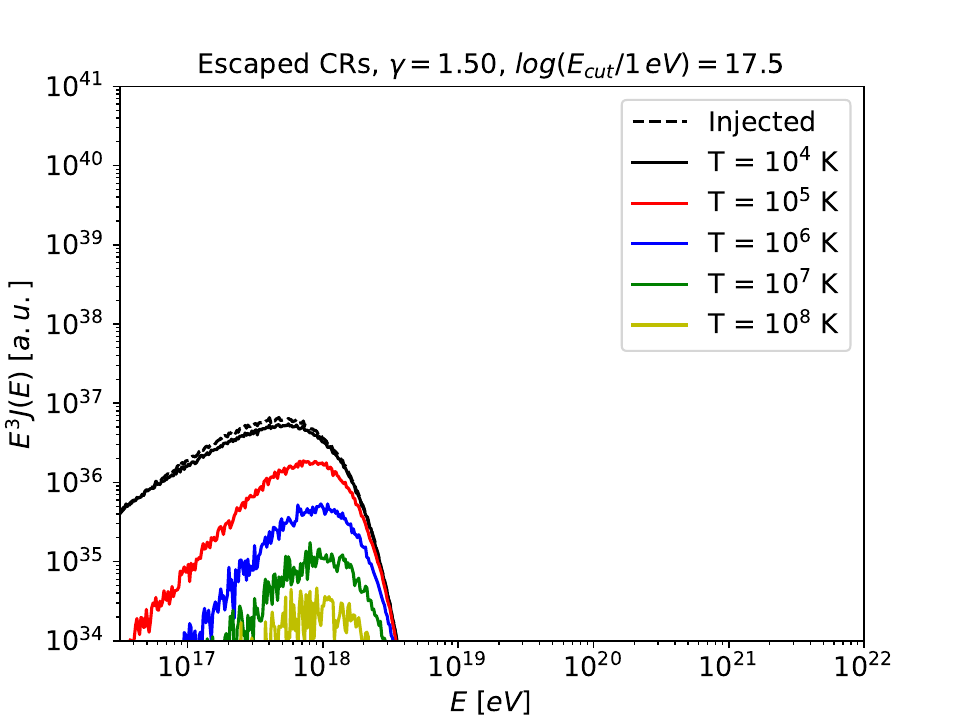}
\end{minipage}
\begin{minipage}{6.5cm}
\centering
\includegraphics[scale=0.42]{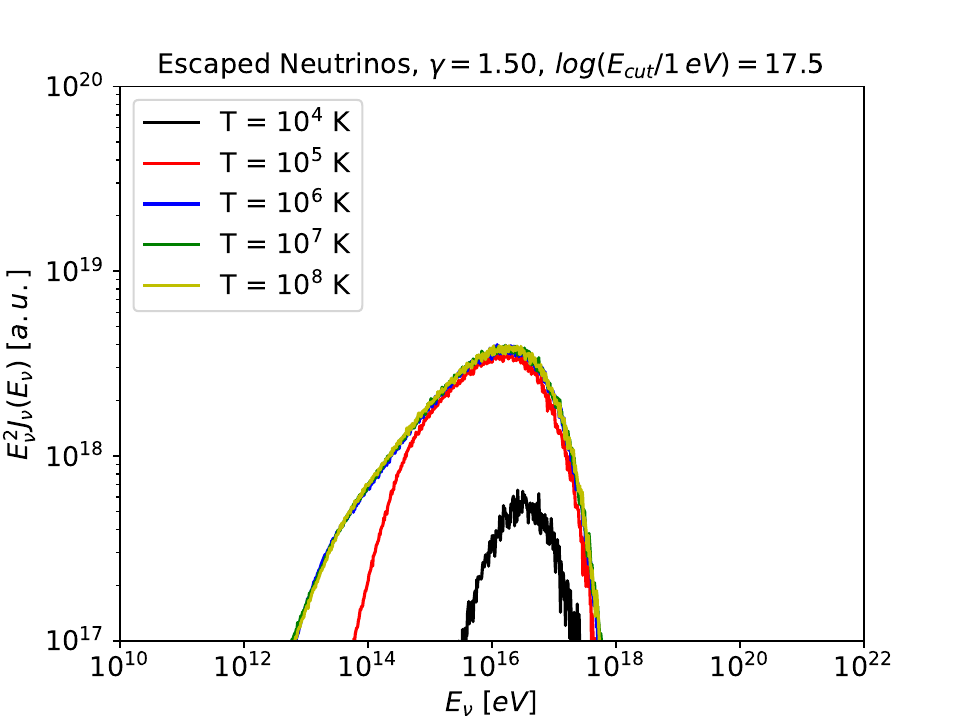}
\end{minipage}
\begin{minipage}{6.5cm}
\centering
\includegraphics[scale=0.42]{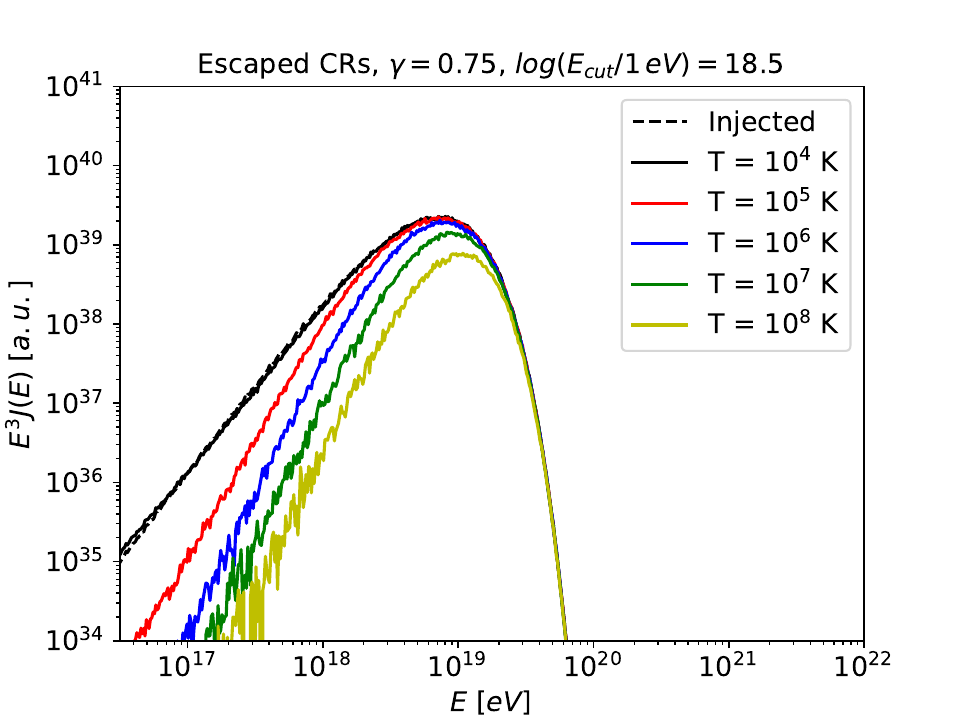}
\end{minipage}
\begin{minipage}{6.5cm}
\centering
\includegraphics[scale=0.42]{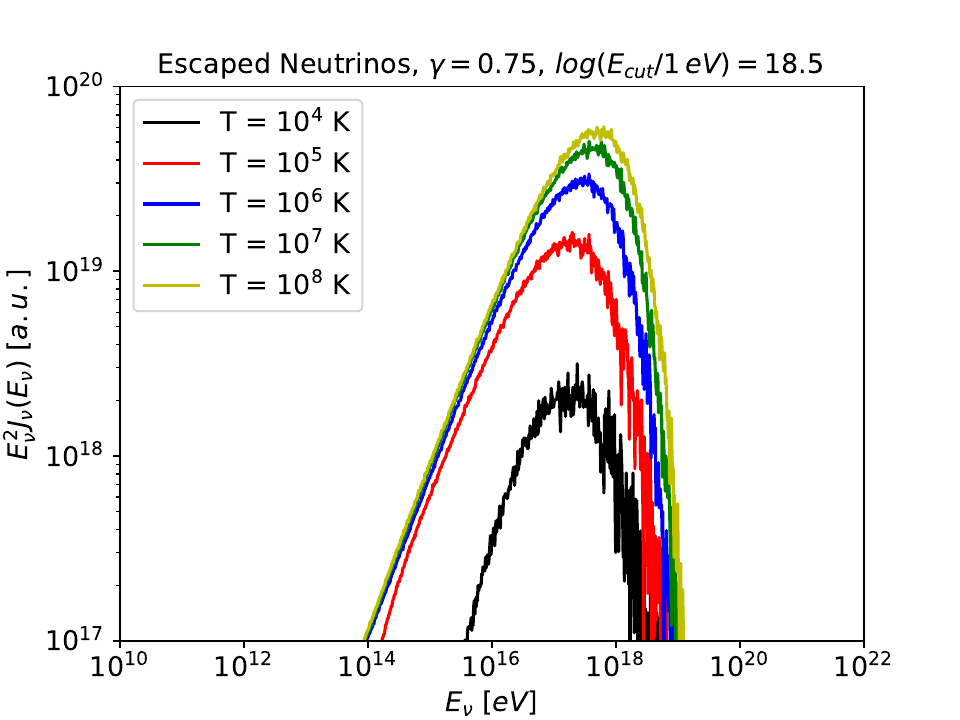}
\end{minipage}
\caption{Left panels: energy spectra at the escape from the source for pure-proton acceleration, in arbitrary units. Right panels: neutrino energy spectra at the escape from the source, in arbitrary units. The dashed black line in the CR spectra correspond to the injected energy spectrum. The injection parameters are $\gamma=1.5$ and $E_\text{cut}=10^{18.5}\,\text{eV}$ (upper panels), $\gamma=1.5$ and $E_\text{cut}=10^{17.5}\,\text{eV}$ (central panels) and $\gamma=0.75$ and $E_\text{cut}=10^{18.5}\,\text{eV}$ (bottom panels). Colors refer to different BB temperatures (see Fig.~\ref{img_sed}). }
\label{escape_p}
\end{figure}
\par The escaped UHECR and neutrino spectra for a pure-proton injection (for a simulation of $10^6$ protons) are shown in Fig.~\ref{escape_p} in arbitrary units. Several injection configurations are shown to appreciate the effect of different injection scenarios on the escaped spectra: $\gamma=1.5$ and $E_\text{cut}=10^{18.5}\,\text{eV}$ in the upper panels, $\gamma=1.5$ and $E_\text{cut}=10^{17.5}\,\text{eV}$ in the central panels and $\gamma=0.75$ and $E_\text{cut}=10^{18.5}\,\text{eV}$ in the bottom panels. Different colors correspond to different BB temperatures while the dashed black line in the UHECR spectra corresponds to the injected energy spectrum from Eq.~\eqref{spectrum_acc}. As expected, photohadronic interactions systematically suppress the injected proton spectrum, and, when the BB field cools down, the escaped spectrum tends to converge to the injected one. This is due to the fact that the total interaction length $\lambda_p(\Gamma,t)$ increases (decreases) faster than the escape length $\lambda_\text{esc}(t)$ with time (and temperature). The injected spectrum at $E\lesssim 10^{19}\,\text{eV}$ is in general more suppressed than the high energy tail. This is because the condition $\lambda_p\gtrsim\lambda_\text{esc}$ is satisfied for almost all the temperatures when $E\gtrsim 10^{19}\,\text{eV}$. The opposite behavior can be observed for the neutrino spectra, as shown in the right panels of Fig.~\ref{escape_p}. When the temperature of the BB is high, interactions are more efficient, and the production of high-energy neutrinos is favored, until the saturation level is reached. We can also see that the peak in the neutrino spectra is at $E_{\nu}\sim 5\% \cdot E_\text{p}$, where $E_\text{p}$ is the peak in the CR spectrum. 
\par
\begin{figure}[t]
\centering
\begin{minipage}{6.5cm}
\centering
\includegraphics[scale=0.42]{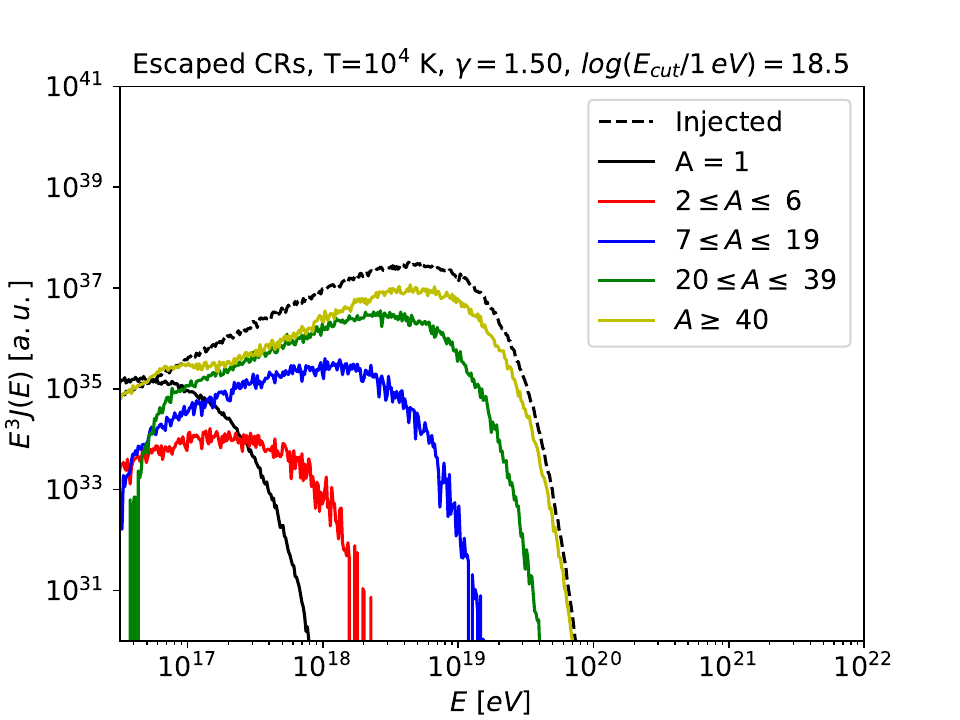}
\end{minipage}
\begin{minipage}{6.5cm}
\centering
\includegraphics[scale=0.42]{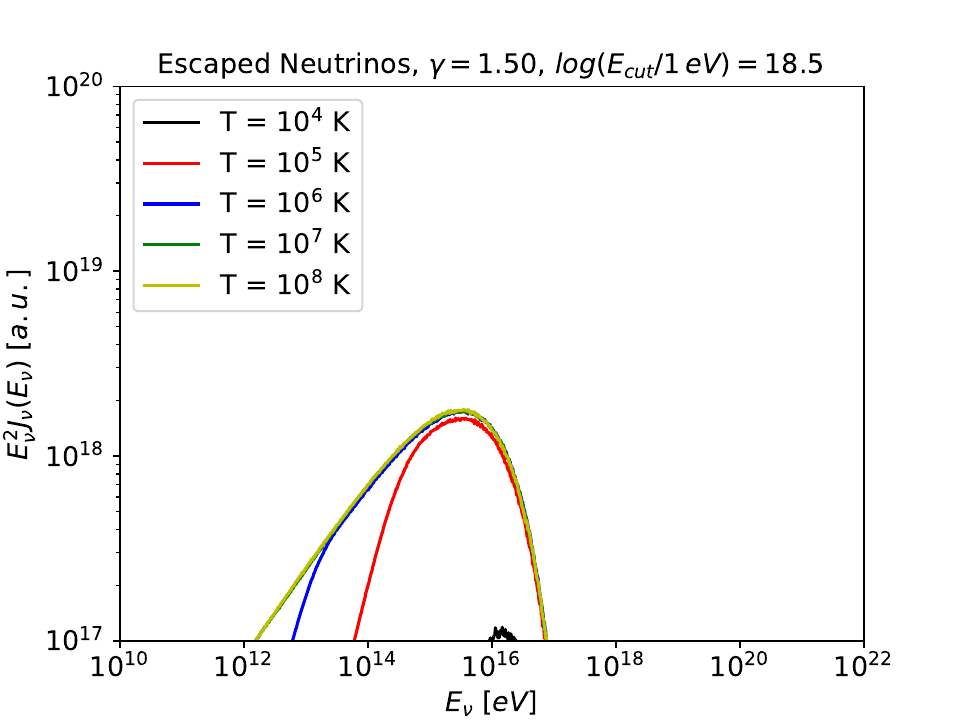}
\end{minipage}
\caption{Left panel: energy spectrum at the escape from the source for pure-iron acceleration, in arbitrary units. Only the BB temperature $T=10^4\,\text{K}$ is shown. Colors correspond to different mass groups at the escape, as indicated in the legend. Note the different range of the y-axis from the proton scenario. Right panel: neutrino energy spectrum at the escape from the source, in arbitrary units. Colors refer to the BB temperatures in Fig.~\ref{img_sed}. The dashed black line in the UHECR spectra correspond to the injected energy spectrum. The injection parameters are $\gamma=1.5$ and $E_\text{cut}=10^{18.5}\,\text{eV}$.}
\label{escape_Fe}
\end{figure}

 In Fig.~\ref{escape_Fe} the escaped spectra for a pure-iron injection (corresponding to a simulation of $10^5$ iron nuclei) are shown. The injection parameters are $\gamma=1.5$ and $E_\text{cut}=10^{18.5}\,\text{eV}$. Only the case with $T=10^4\,\text{K}$ is shown in the left panel, where different colors correspond to different mass groups. The effect of injecting nuclei can be appreciated in two different ways: the escaped UHECRs are a mix of nuclear species, due to photodisintegration, and the production of high-energy neutrinos is reduced with respect to the proton scenario, as due to the increased threshold for the photopion production. However, the effect of the propagation within the source environment for different temperatures is similar to the case of the proton injection: high temperatures correspond to an enhanced production of neutrinos, and a reduced flux of escaped cosmic rays. 
\par The inverse behavior of the CR escape and neutrino production while changing the target temperature characterizes the efficiency of the source in converting the accelerated baryonic material into neutrinos. We quantify this effect by computing the UHECR and neutrino emissivites. In particular, we compute the UHECR emissivity as 
\begin{equation}
\label{emissivity_acc}
\mathcal{E}_\text{acc} = \sum_A \int dE \, E \, Q^{A}_\text{acc}(E) \, ,
\end{equation}
\begin{equation}
\mathcal{E}_\text{esc} = \sum_A \int dE \, E \, Q_\text{esc}^{A}(E) \, ,
\end{equation}
where $Q^{A}_\text{acc}(E)$ and $Q_\text{esc}^A(E)$ are the rates of the nuclear species $A$ at the acceleration (i.e. the injection into the source environment) and at the escape, respectively. The integration is performed over the injection energy rage ($10^{14}-10^{20}\,\text{eV}$). The neutrino emissivity is calculated in the same way starting from the neutrino escape rate $Q_\text{esc}^\nu(E_\nu)$, but using an energy range shifted by a factor of $5\%$ relative to the energy range of cosmic rays, and it is denoted by $\mathcal{E}_\text{esc}^\nu$. Therefore, we define the efficiency parameters as
\begin{equation}
\label{eff_par_uhecr}
f_\text{CR} = \dfrac{\mathcal{E}_\text{acc}-\mathcal{E}_\text{esc}}{\mathcal{E}_\text{acc}} \, ,
\end{equation}
\begin{equation}
\label{eff_par_nu}
f_\nu = \dfrac{\mathcal{E}_\text{esc}^\nu}{\mathcal{E}_\text{acc}} \, ,
\end{equation}
where $f_\text{CR}$ represents the fraction of energy lost by accelerated UHECRs during the propagation within the source (source efficiency), and $f_\nu$ the fraction of energy transformed into neutrino energy by in-source interactions (neutrino production efficiency). Given the definitions of the efficiency parameters in Eqs.~\eqref{eff_par_uhecr} and~\eqref{eff_par_nu}, we can see that the maximum interaction efficiency corresponds to both $f_\text{CR}\, , \,f_\nu \simeq1$, i.e. total conversion of accelerated CR energy in source neutrinos.
\begin{figure}[t]
\centering
\begin{minipage}{6.5cm}
\centering
\includegraphics[scale=0.42]{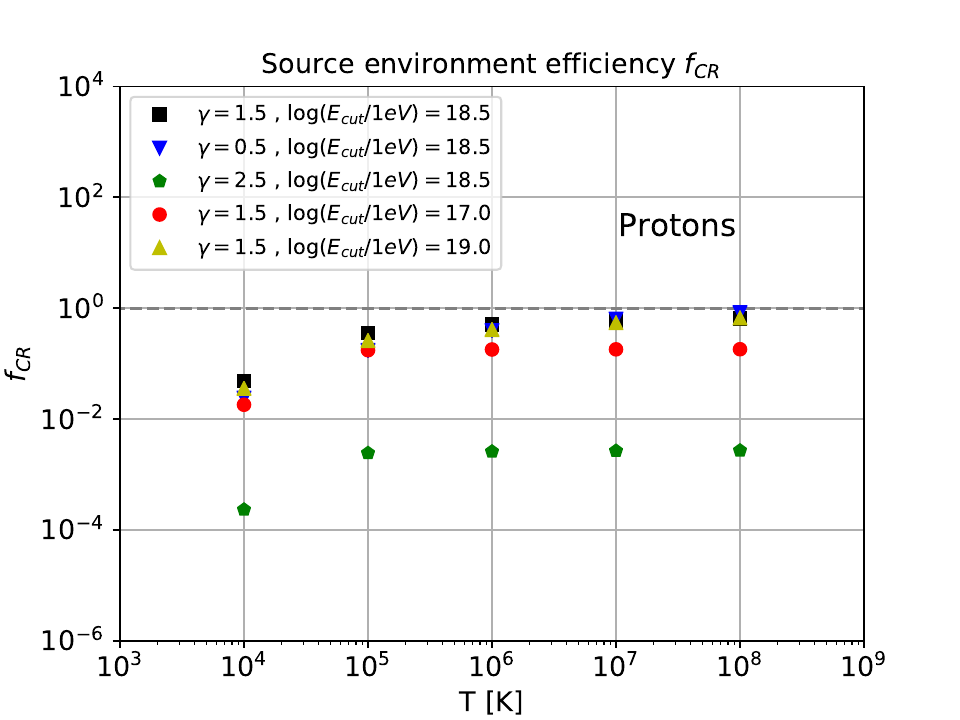}
\end{minipage}
\begin{minipage}{6.5cm}
\centering
\includegraphics[scale=0.42]{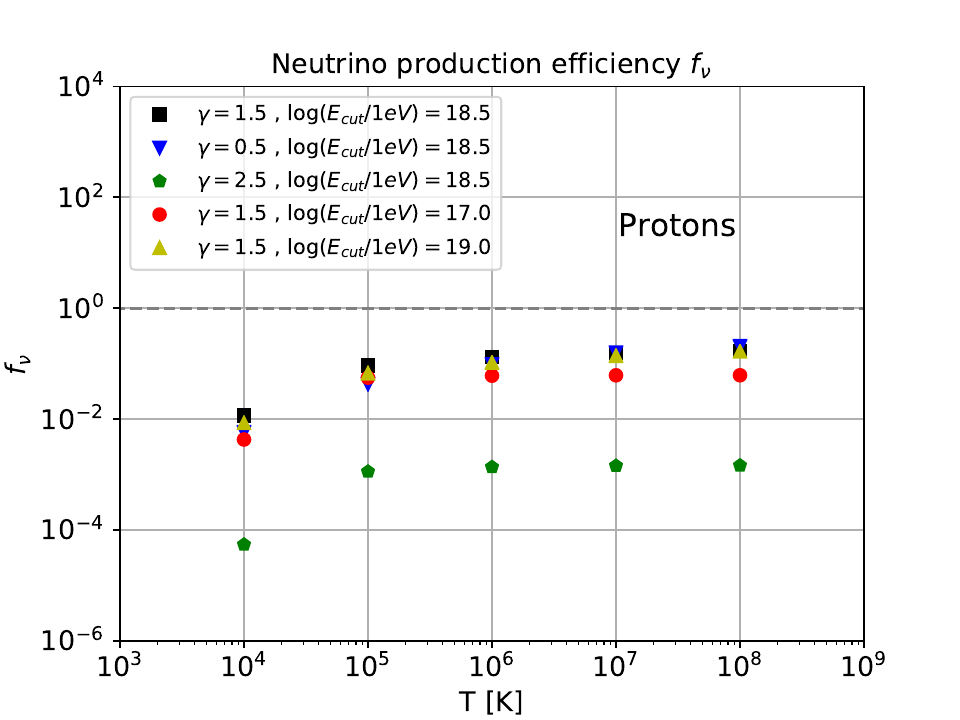}
\end{minipage}
\caption{Left panel: source environment efficiency, as defined in Eq.~\eqref{eff_par_uhecr}. Right panel: neutrino production efficiency, as defined in Eq.~\eqref{eff_par_nu}. Colors correspond to different injection scenarios: $\gamma=1.5$ and $\log (E_\text{cut}/1\,\text{eV})=18.5$ (black), $\gamma=0.5$ and $\log (E_\text{cut}/1\,\text{eV})=18.5$ (blue), $\gamma=2.5$ and $\log (E_\text{cut}/1\,\text{eV})=18.5$ (green), $\gamma=1.5$ and $\log (E_\text{cut}/1\,\text{eV})=17.0$ (red) and $\gamma=1.5$ and $\log (E_\text{cut}/1\,\text{eV})=19.0$ (yellow). Both panels refer to a pure-proton injection.}
\label{source_efficiency_p}
\end{figure}
\begin{figure}[t]
\centering
\begin{minipage}{6.5cm}
\centering
\includegraphics[scale=0.42]{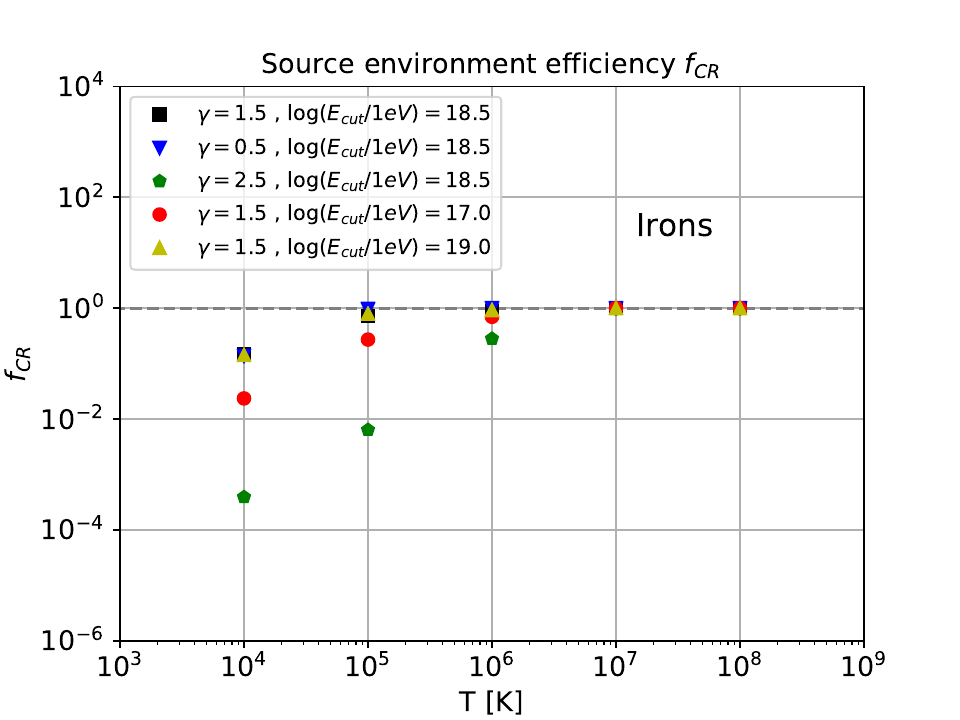}
\end{minipage}
\begin{minipage}{6.5cm}
\centering
\includegraphics[scale=0.42]{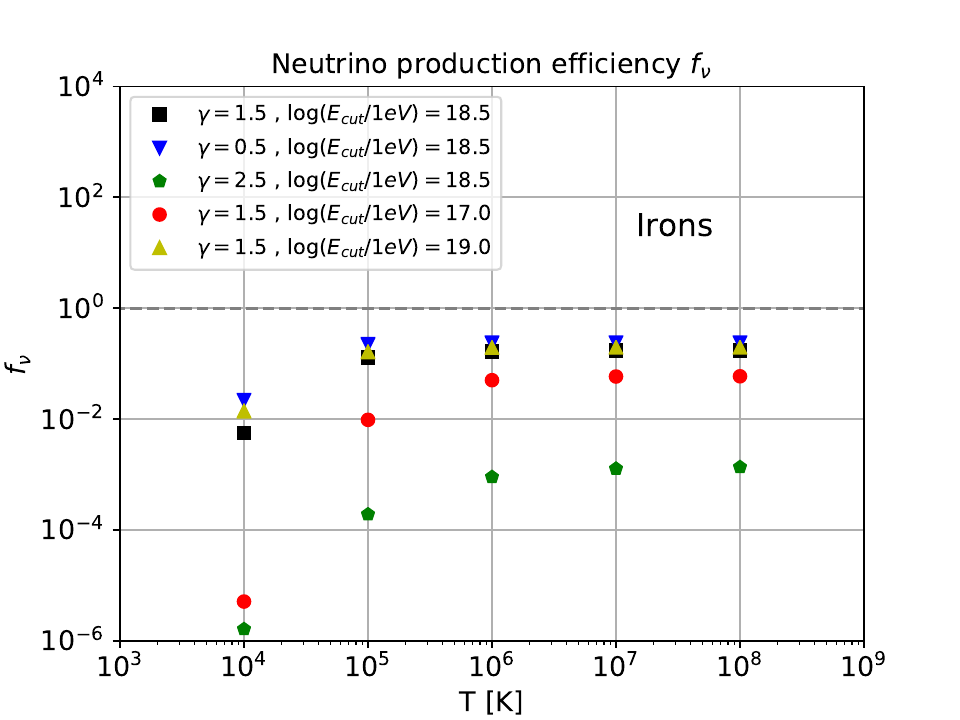}
\end{minipage}
\caption{Same as Fig.~\ref{source_efficiency_p}, corresponding to pure-iron injection.}
\label{source_efficiency_Fe}
\end{figure}
\par In the left panel of Fig.~\ref{source_efficiency_p}, the source efficiency $f_\text{CR}$ is shown for different BB temperatures in the scenarios of a pure-proton injection. Different colors correspond to different combinations of acceleration parameters $\gamma$ and $E_\text{cut}$. All configurations considered saturate at a specific value of $f_\text{CR}$, for $T\gtrsim10^{6}\,\text{K}$. This demonstrates that interactions within the source environment are very efficient during the early stages after the merger. However, scenarios with $\gamma \gtrsim 1.5$ are characterized  by a maximum fraction of energy lost of $f_\text{CR}\lesssim 10^{-2}$, due to the large number of low energy protons escaping the source environment almost undisturbed. In the right panel of Fig.~\ref{source_efficiency_p}, the neutrino production efficiency as defined in Eq.~\eqref{eff_par_nu} is shown, for the same combinations of parameters in the left panel. The neutrino production efficiency is clearly increasing as a function of the increasing temperature (decreasing time) of the source environment. Since protons can only produce neutrinos through photopion production, we observe that the neutrino production saturates at the same temperature as the one for protons. Most of the configurations considered saturate at $f_\nu\simeq0.1$, corresponding to a $\sim10\%$ conversion of cosmic ray energy into neutrino flux. The observed saturation of neutrino production is consistent with the known Waxman-Bahcall bound for source environments optically thin to photopion production \cite{Waxman:1998yy,Bahcall:1999yr}. As discussed earlier, scenarios characterized by $\gamma=2.5$ correspond to $f_\nu\simeq10^{-3}$, as a result of the fact that escape from the source is favored over interactions.
\par In Fig.~\ref{source_efficiency_Fe} the same quantities are shown, corresponding to a pure iron injection. In the left panel of Fig.~\ref{source_efficiency_Fe}, the evolution of $f_\text{CR}$ with the BB temperature is shown. In this case, all scenarios saturate to $f_\text{CR}\simeq1$, for $T> 10^6 \,\text{K}$, as can be expected from the total interaction length in Fig.~\ref{int_len_Fe}. We observe that the photodisintegration increases the source conversion efficiency in cosmic rays. The neutrino production is almost unchanged for scenarios characterized by a low $\gamma$ value or a large $E_\text{cut}$ value. In contrast, scenarios characterized by a large value of $\gamma$ or a low value of $E_\text{cut}$ (see red and green points in the right panel of Fig.~\ref{source_efficiency_Fe}) show reduced neutrino production when $T\lesssim10^{5}\,\text{K}$. In fact, as shown in Fig.~\ref{int_len_Fe}, the presence of the low-energy minimum in $\lambda_\text{Fe}$, associated to photodisintegration, corresponds to a low escape rate of iron nuclei, which continue to interact with BB photons by disintegrating, but without producing neutrinos. We note here that the large interaction efficiency in the case of iron nuclei will have an effect on the normalization of the propagated fluxes, compared to the proton scenario. In fact, in order to reproduce the observed UHECR flux, a higher normalization factor will be required with respect to the proton case, resulting in a higher injection rate at the source and an increased neutrino flux on Earth in the case of iron acceleration at the source. We will discuss these effects in the next section. We also note here that in BNS environments, nuclei heavier than the ones belonging to the iron group should be taken into account, as due to $r$-processes. We plan to include their treatment, as for instance done already in \cite{Zhang:2024sjp}, in future works.

\subsection{Extragalactic propagation}
\label{subsec_extragal-prop}
The escaped spectra described in the previous section are used as an input for the extragalactic propagation, in order to compute the expected diffuse spectra at Earth, from a population of BNS mergers. We use the original version of the simulation framework \texttt{SimProp-v2r4}. UHECRs are propagated in the extragalactic space taking into account stochastic interactions with the CMB and the EBL from \cite{Stecker:2005qs}, nuclear decay and redshift energy loss. The production and propagation of cosmogenic neutrinos (i.e. neutrinos produced by interactions between UHECRs and cosmic photon fields) are also simulated. The propagation of high-energy neutrinos produced within the source environment is taken into account by considering the energy evolution on cosmological distances, given by
\begin{equation}
\dfrac{dE_\nu}{dz} = \dfrac{E_\nu}{1+z}\, ,
\end{equation}
where $E_\nu$ is the neutrino energy, $z$ is the redshit and $E_{\nu,0} = E_\nu(z=0)$ is the energy expected at Earth. Therefore, the energy spectrum  of astrophysical neutrinos at Earth is given by
\begin{equation}
J_{\nu,\text{source}} (E_{\nu,0}) = \frac{c}{4\pi H_0} \int_0^{z_\text{max}}dz\, \dfrac{\xi (z)}{\sqrt{\Omega_\text{m}(1+z)^3+\Omega_\Lambda}} Q^{\nu}_{\text{esc}} (E_\nu (z)) \, ,
\end{equation}
where $z_\text{max}$ is the maximum redshift considered for the extragalactic propagation (in this work, $z_\text{max}=6$), $H_0\simeq 70 \,\text{km/s/Mpc}$ is the Hubble constant at present time, $\Omega_\text{m}\simeq 0.3$ is the density matter and $\Omega_\Lambda\simeq 0.7$ is the dark energy density, in the standard cosmological model ($\Lambda$CDM), and $Q^{\nu}_{\text{esc}} (E_\nu (z))$ is the neutrino production rate of the source (i.e. the number of produced neutrinos per unit energy, time and volume). The function $\xi(z)$ describes the redshift evolution of the source distribution, and it is generically of the form $\xi(z)\propto (1+z)^m$, where $m$ is the source evolution index. In this work, we consider the star formation rate evolution (SFR) parametrisation like in \cite{AlvesBatista:2019rhs}, 
\begin{equation}
\label{SFR_function}
\xi(z) = 
\begin{cases}
(1+z)^{3.4}, \,\,\,\,\,\,\,\,\,\,\,\,\,\,\,\,\,\,\,\,\,\,\,\,\,\,\,\,\,\, 0 \leq z \leq 1 \, ; \\
2^{3.7}(1+z)^{-0.3}, \,\,\,\,\,\,\,\,\,\,\,\,\,\,\,\,\,\, 1 \leq z \leq 4\, ;  \\
2^{3.7} 5^{3.2} (1+z)^{-3.5}, \,\,\,\,\,\,\,\,\, 4 \leq z \leq 6 \, ,
\end{cases}
\end{equation}
together with the case of no source cosmological evolution $m=0$.
\par In order to compare our simulations with measured UHECR and neutrino fluxes, we scale the propagated UHECR all-particle spectrum. In particular, we require that the propagated all-particle UHECR spectrum at $E=E_\text{cut}$ corresponds to the UHECR flux measured by the Pierre Auger Observatory \cite{PierreAuger:2015eyc} at the same energy\footnote{The cutoff energy $E_\text{cut}$ is defined at the acceleration in Eq.~\eqref{spectrum_acc} and it will be modified by interactions within the source environment and during extragalactic propagation. In the normalization condition in Eq.~\eqref{spectrum_norm} we mean that the normalization of the propagated UHECR spectrum to the observed spectrum is done at the numerical value of observed energy $E=E_\text{cut}$.} (see Appendix \ref{sec_uhecr-spectra}). In this way, we fix the accelerated injection rate $Q_{0,\text{acc}}^A$ in Eq.~\eqref{spectrum_acc}. For our reference scenario with protons we obtain $Q_{0,\text{acc}}^\text{p}=8.2\cdot 10^{39} \, \text{erg}^{-1}\,\text{Mpc}^{-3}\,\text{yr}^{-1}$, while for the same scenario with iron nuclei we obtain $Q_{0,\text{acc}}^\text{Fe}=3.4\cdot 10^{41}\,\text{erg}^{-1}\,\text{Mpc}^{-3}\,\text{yr}^{-1}$. We then introduce the scale factor $g$, such that
\begin{equation}
\label{spectrum_norm}
J_\text{prop} (E=E_\text{cut})=g\cdot J_\text{exp} (E=E_\text{cut}) \, ,
\end{equation}
where $J_\text{prop}(E)=\sum_A J_\text{prop}^A(E)$. In this way, the scale factor $g$ can be used to soften the assumption on the relative contribution of BNS mergers to the observed UHECR flux in the energy range below the ankle. The following values for the scale factor $g$ will be considered: $\log{g}=-5,...,0$ with $\Delta\log{g}=0.1$. We define a reference scenario characterized by the following choices: pure-proton injection, source temperature of $T=10^5 \,\text{K}$, and spectral parameters $\gamma=1.5$ and $E_\text{cut}=10^{18.5}\,\text{eV}$; we also fix the normalization factor $g=0.4$, as roughly what is obtained from the fit of the nuclear species at Earth at energy $10^{18.5}\,\text{eV}$ in \cite{PierreAuger:2023bfx}, for the proton component.
\par
\begin{figure}[t]
\centering
\begin{minipage}{6.5cm}
\centering
\includegraphics[scale=0.42]{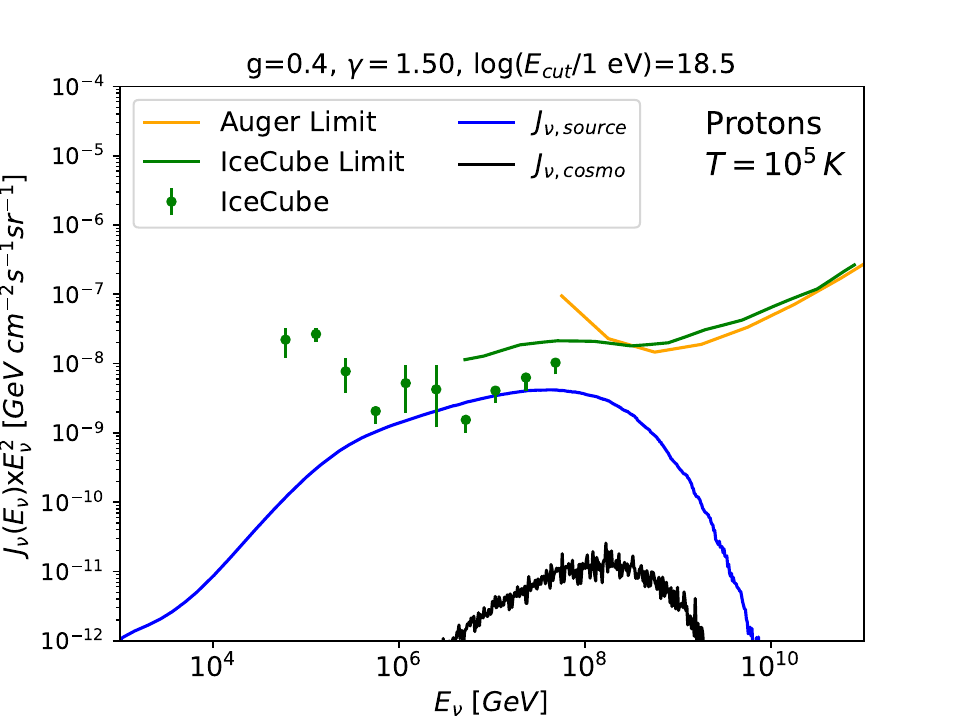}
\end{minipage}
\begin{minipage}{6.5cm}
\centering
\includegraphics[scale=0.42]{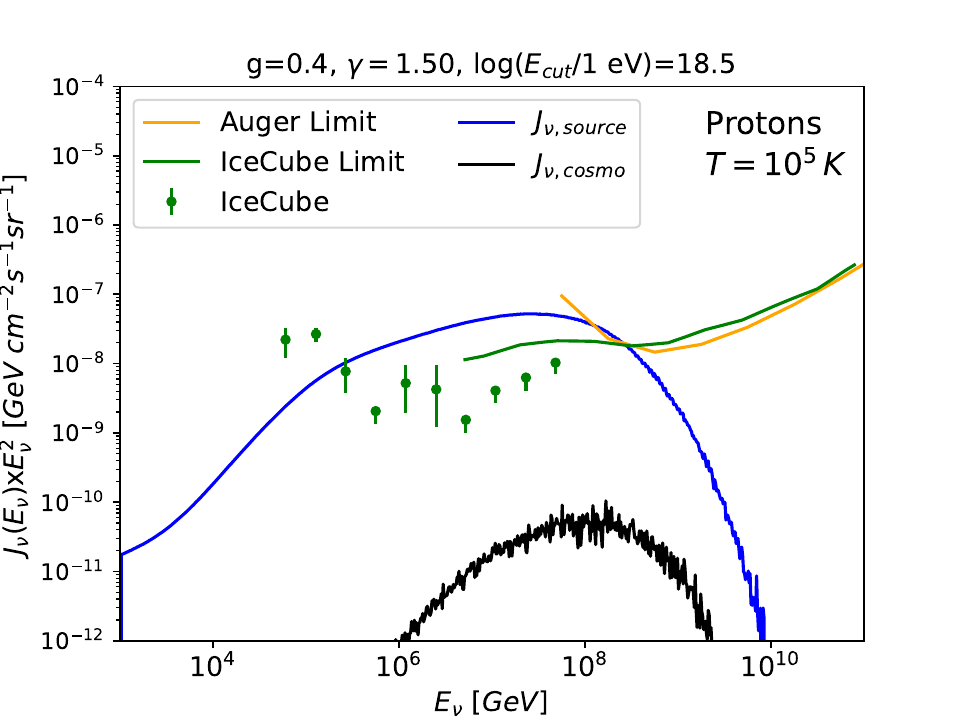}
\end{minipage}
\caption{Single-flavor neutrino energy spectra for a pure-proton injection, for a population of BNS mergers at a fixed time after the merger (no-evolution (left) and SFR evolution (right panel) cases): propagated source neutrinos (blue line) and cosmogenic neutrinos (black line). Observed neutrino flux and cosmogenic neutrino limit by IceCube \cite{IceCube:2017zho} and the cosmogenic neutrino limit by the Pierre Auger Observatory \cite{PierreAuger:2019ens} are also shown.}
\label{neu_flux_p}
\end{figure}

 In Fig.~\ref{neu_flux_p} the propagated source neutrino spectra for the reference scenario with source evolution $m=0$ (left) and SFR evolution in Eq.~\eqref{SFR_function} are shown. In both panels, the (single flavor) source neutrino spectrum and the cosmogenic neutrino spectrum (black line) are shown. The measured neutrino flux and cosmogenic neutrino limit by IceCube \cite{IceCube:2017zho} as well as the cosmogenic neutrino limit by the Pierre Auger Observatory \cite{PierreAuger:2019ens} are also shown. The source neutrino fluxes are compatible with the upper limits from IceCube, but only partially compatible with the observed neutrinos for $E_\nu\gtrsim 10^6 \,\text{GeV}$, in the case $m=0$. The produced cosmogenic neutrino flux is several orders of magnitude below the experimental limits. In the right panel the scenario corresponding to the SFR source evolution is shown. The production of cosmogenic neutrinos is slightly enhanced in this case, due to increased interactions with the cosmic fields corresponding to the SFR source evolution. However, the production of source neutrinos also increases, making the spectrum of propagated neutrino higher than the observed flux. 
\begin{figure}[t]
\centering
\begin{minipage}{6.5cm}
\centering
\includegraphics[scale=0.42]{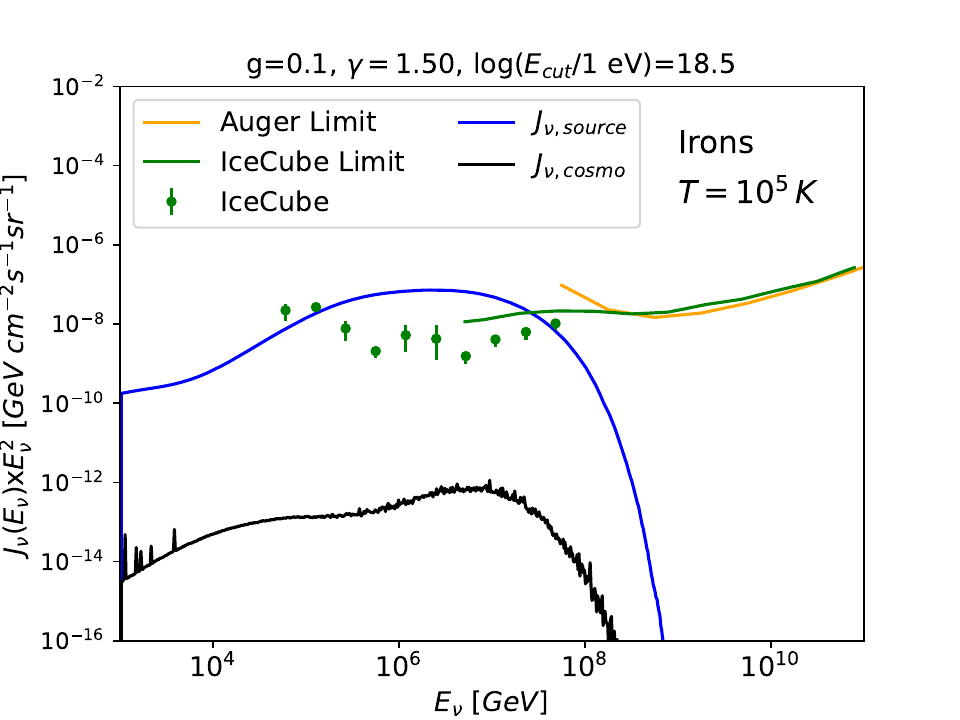}
\end{minipage}
\begin{minipage}{6.5cm}
\centering
\includegraphics[scale=0.42]{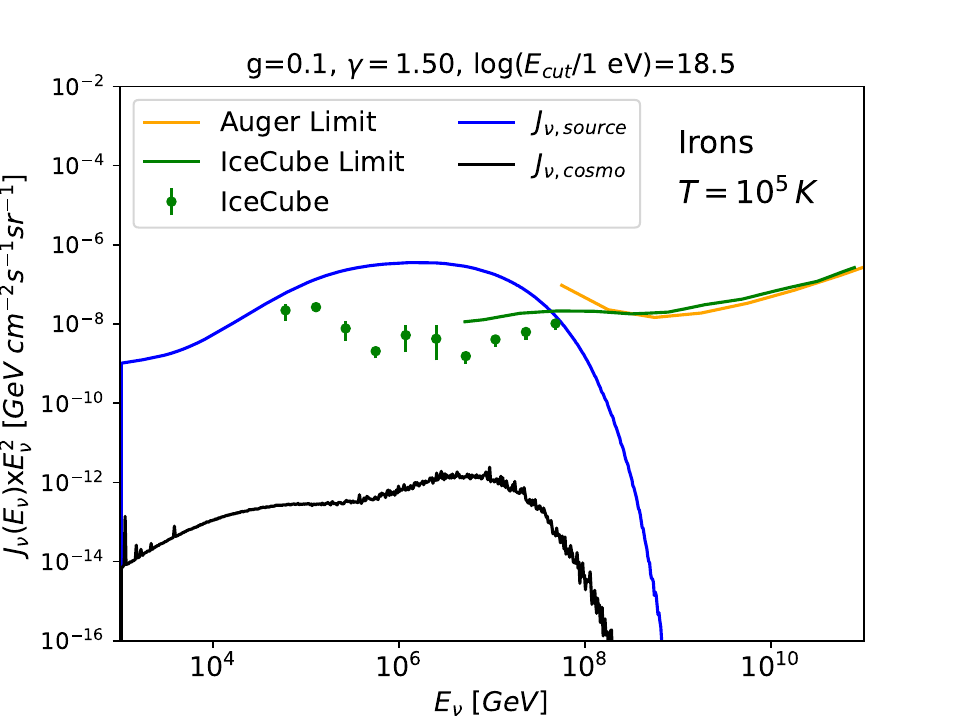}
\end{minipage}
\caption{Same as in Fig.~\ref{neu_flux_p}, for a pure-iron injection. Note the different range of the y-axes with respect to the proton scenarios.}
\label{neu_flux_Fe}
\end{figure}
\par In Fig.~\ref{neu_flux_Fe}, the reference scenario of Fig.~\ref{neu_flux_p} is shown for a pure iron composition at the acceleration. In this case, the scaling factor is fixed to $g=0.1$. The position of the high energy cutoff is shifted to lower energy with respect to the pure-proton case, due to the fact that most of the UHECRs that generate neutrinos are protons produced in the disintegration of iron nuclei. The low-energy region of the spectra is enhanced with respect to the proton scenario, and this can be explained by considering the behavior of the interaction lengths for small values of the Lorentz factors in Figs.~\ref{int_len_p} and~\ref{int_len_Fe}. Moreover, the general effect of a heavier composition injected in the source environment is an increase of the number of produced neutrinos. As discussed in Sec.~\ref{subsec_source-escape}, this is due to the normalization to the observed UHECR spectrum. Photodisintegration processes within the source environment suppress more intensively the UHECRs at the escape, compared to the case of pure protons undergoing photomeson production in the source. Therefore, a larger rate of acceleration at the source is required. The result is an increased neutrino flux at the escape from the source, and then on Earth. The SFR scenario is shown in the right panel of Fig.~\ref{neu_flux_Fe} and it shows the same effects described in the proton case. 

\subsection{Study of source parameters}
\label{subsec_parameter}
In order to quantify the compatibility of the obtained spectra of source and cosmogenic neutrinos with the available data and limits, we introduce two \textit{control quantities}. We define the neutrino spectral ratio at $E_\nu = 10^6 \,\text{GeV}$ as
\begin{equation}
\label{ration_nu}
R_\nu(E_\nu = 10^6 \,\text{GeV})=\dfrac{J_{\nu,\text{source}}(E_\nu = 10^6 \,\text{GeV})}{J_\text{IceCube}(E_\nu = 10^6 \,\text{GeV})}\, ,
\end{equation}
where $J_{\nu,\text{source}}$ is the propagated source-neutrino spectrum and $J_\text{IceCube}$ is the observed IceCube neutrino spectrum previously introduced. Additionally, given the total neutrino exposure $\mathcal{E}(E_\nu)$ of the Pierre Auger Observatory \cite{PierreAuger:2019ens}, we calculate the expected number of neutrinos as
\begin{equation}
\label{N_cosmo}
N_{\nu}= \int dE_\nu \, \mathcal{E}(E_\nu)\left[J_{\nu,\text{source}}(E_\nu)+J_{\nu,\text{cosmo}}(E_\nu) \right]\, .
\end{equation}
and we compare this value to 2.39, being this the Feldman-Cousins factor for non-observation of events in the absence of
expected background \cite{Feldman:1997qc}. We compute both these quantities in Eqs.~\eqref{ration_nu} and~\eqref{N_cosmo} for the reference scenario and for both the source evolution parameterizations. We then vary one parameter, keeping the others unchanged, to investigate their impact on our control quantities. 
\begin{figure}[t]
\centering
\begin{minipage}{6.5cm}
\centering
\includegraphics[scale=0.42]{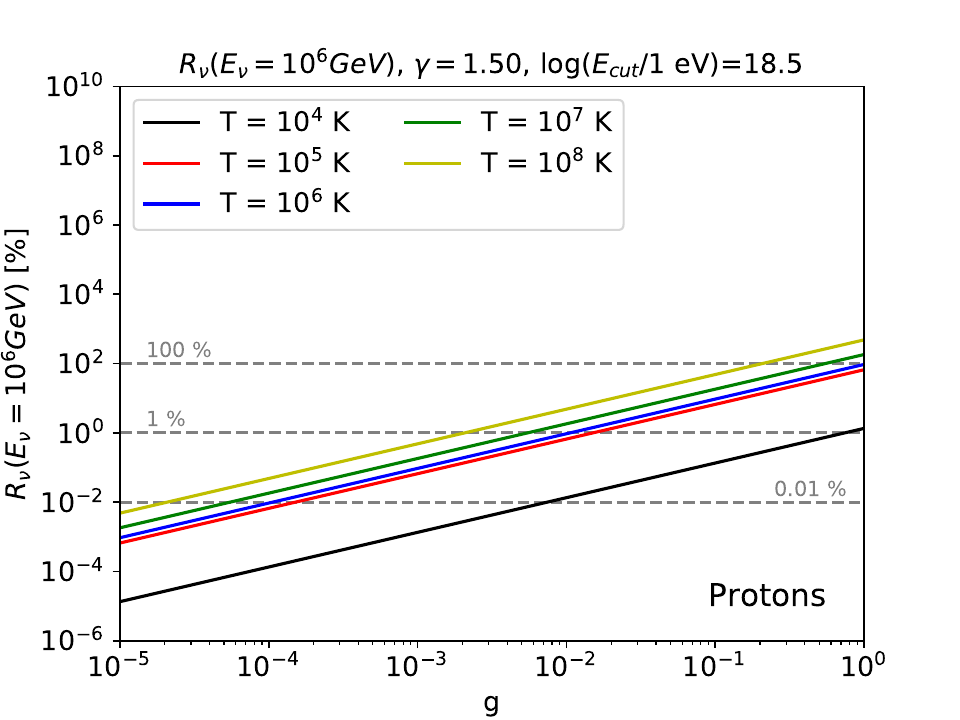}
\end{minipage}
\begin{minipage}{6.5cm}
\centering
\includegraphics[scale=0.42]{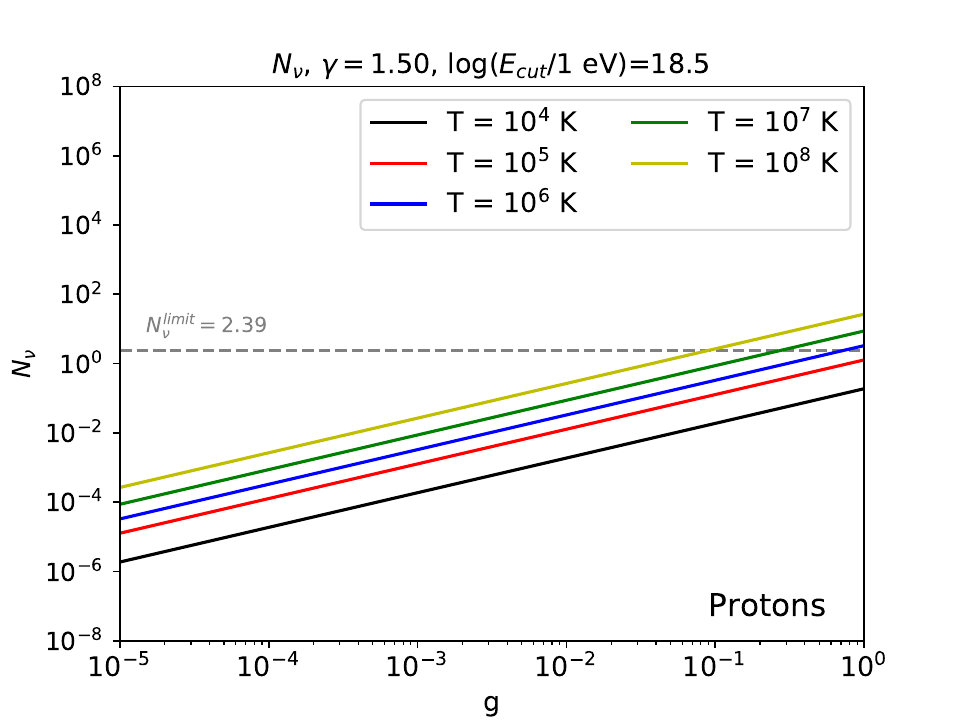}
\end{minipage}
\begin{minipage}{6.5cm}
\centering
\includegraphics[scale=0.42]{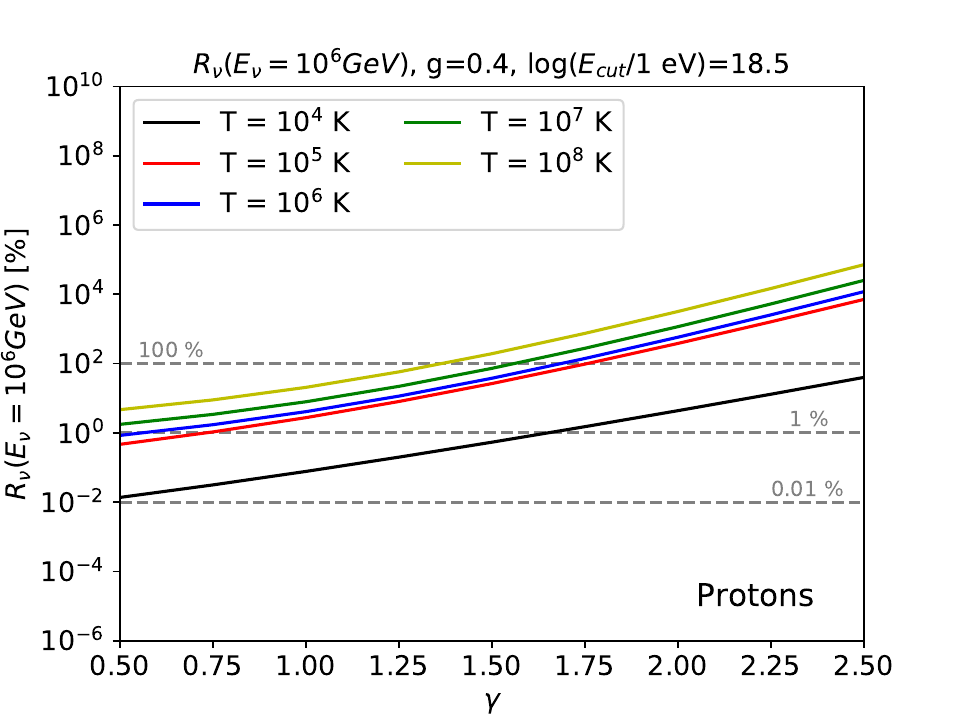}
\end{minipage}
\begin{minipage}{6.5cm}
\centering
\includegraphics[scale=0.42]{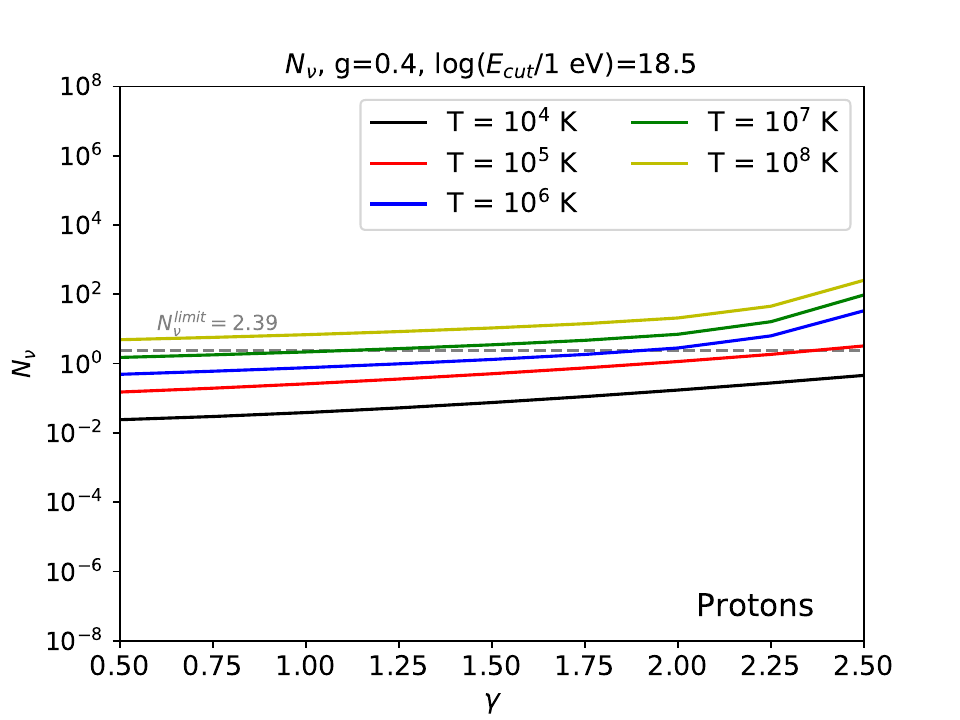}
\end{minipage}
\begin{minipage}{6.5cm}
\centering
\includegraphics[scale=0.42]{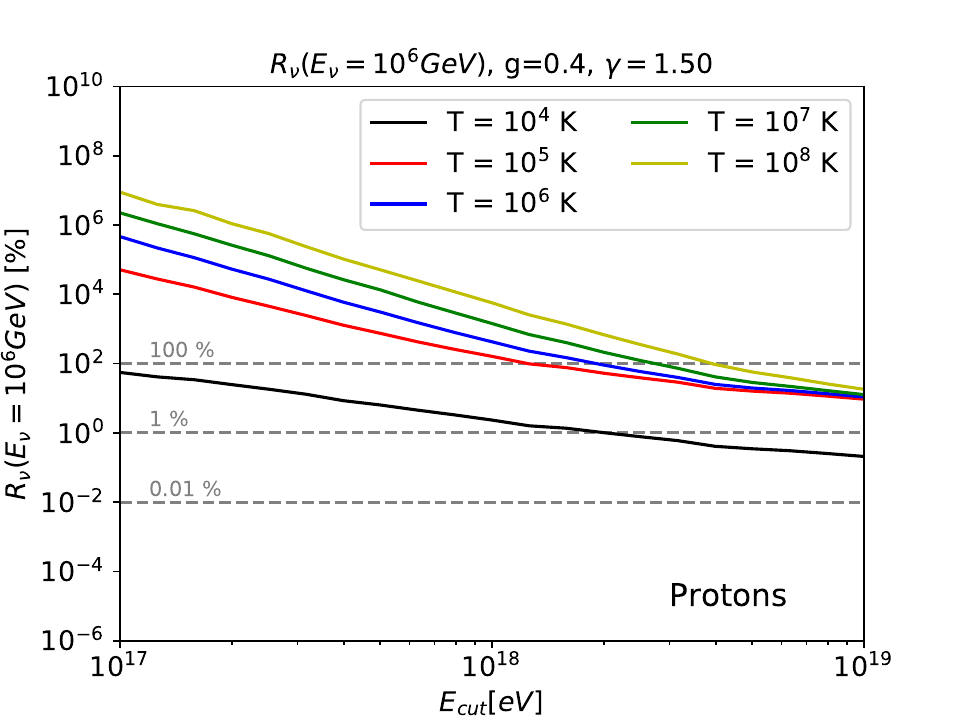}
\end{minipage}
\begin{minipage}{6.5cm}
\centering
\includegraphics[scale=0.42]{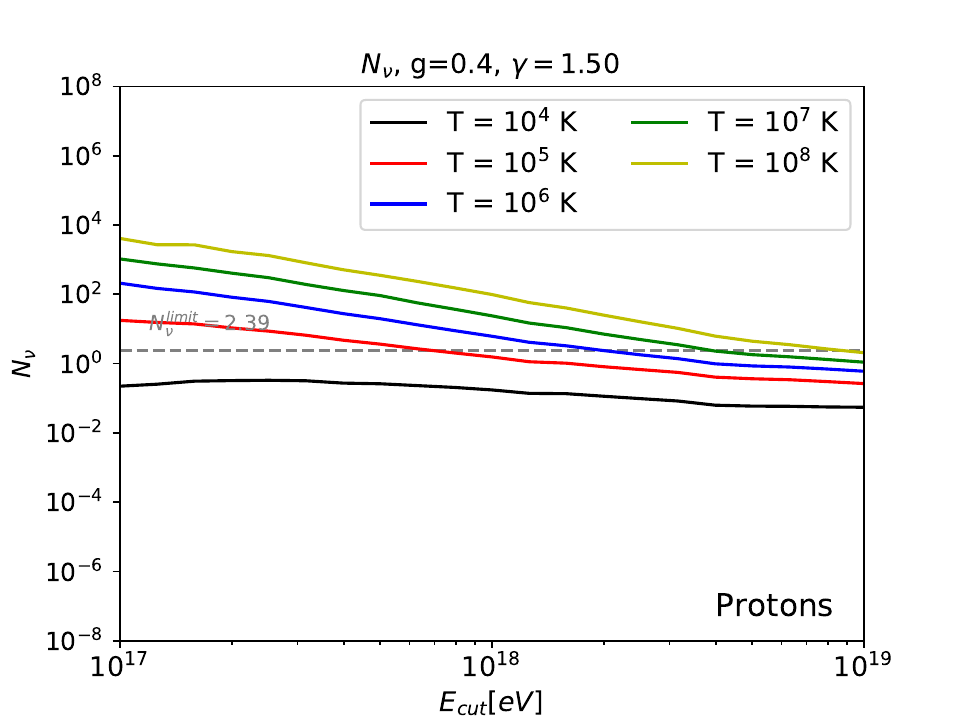}
\end{minipage}
\caption{Left column: neutrino spectral ratio, calculated at $E_\nu=10^{6}\,\text{GeV}$, as defined in Eq.~\eqref{ration_nu}. Right column: number of neutrinos, as defined in Eq.~\eqref{N_cosmo}. These quantities are shown as a function of the normalization factor $g$ (upper), the spectral index $\gamma$ (central) and the high energy cutoff $E_\text{cut}$ (bottom panels). Colors correspond to different temperatures. In the right panels, gray lines indicate some reference ratios. The Feldman-Cousins factor for non-observation of events is indicated with a gray line in the left panels. The injected UHECR composition in the source environment is pure-proton, and the evolution parameter is $m=0$.}
\label{parameter_scan_p}
\end{figure}
\begin{figure}[t]
\centering
\begin{minipage}{6.5cm}
\centering
\includegraphics[scale=0.42]{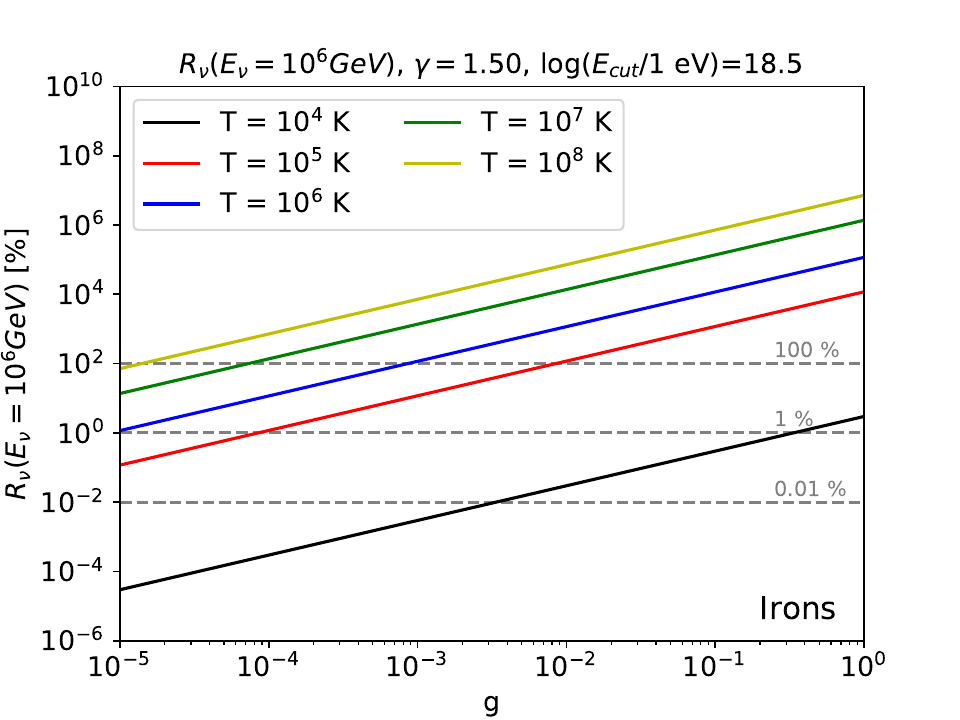}
\end{minipage}
\begin{minipage}{6.5cm}
\centering
\includegraphics[scale=0.42]{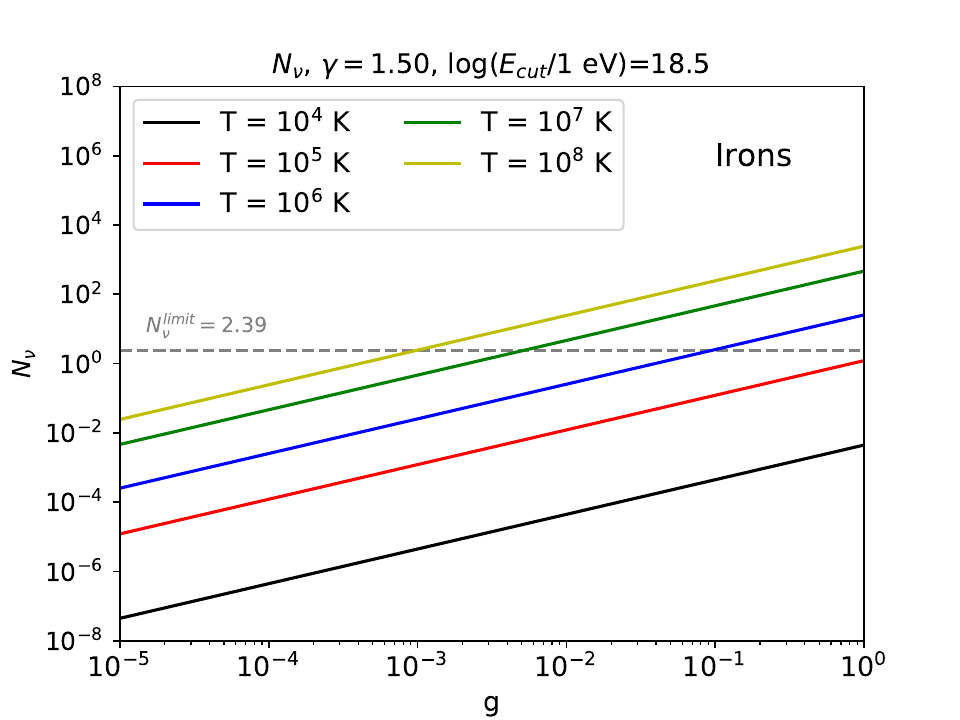}
\end{minipage}
\begin{minipage}{6.5cm}
\centering
\includegraphics[scale=0.42]{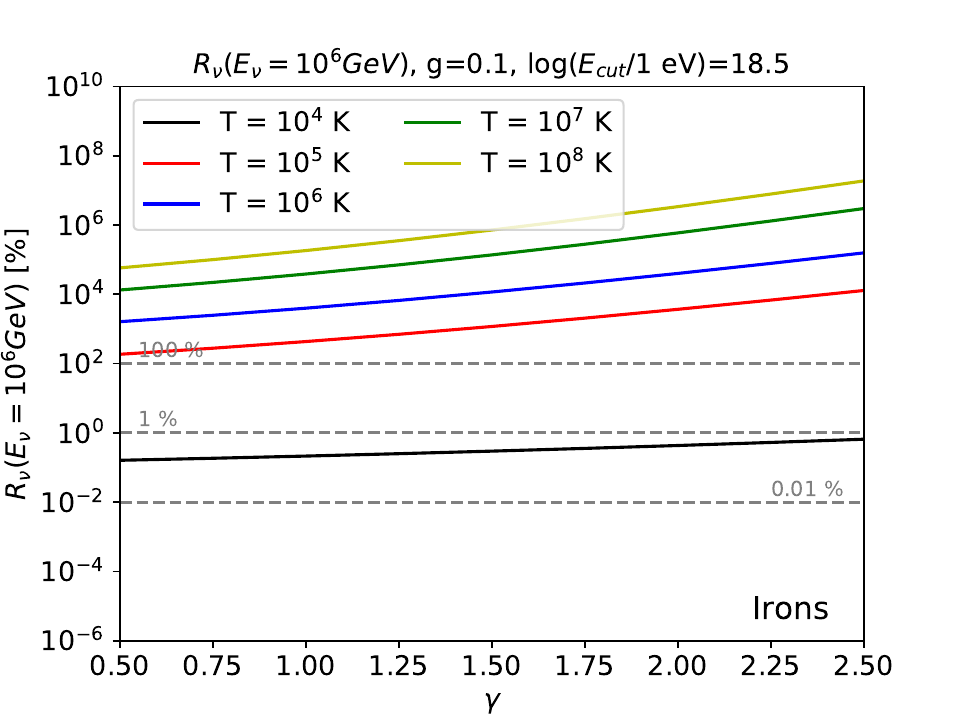}
\end{minipage}
\begin{minipage}{6.5cm}
\centering
\includegraphics[scale=0.42]{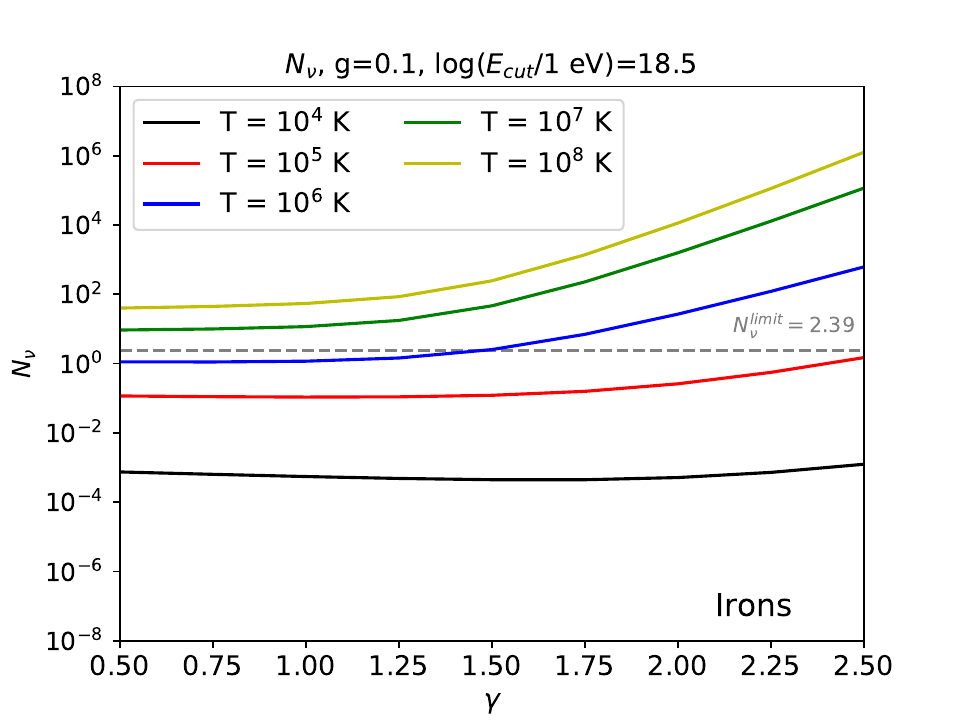}
\end{minipage}
\begin{minipage}{6.5cm}
\centering
\includegraphics[scale=0.42]{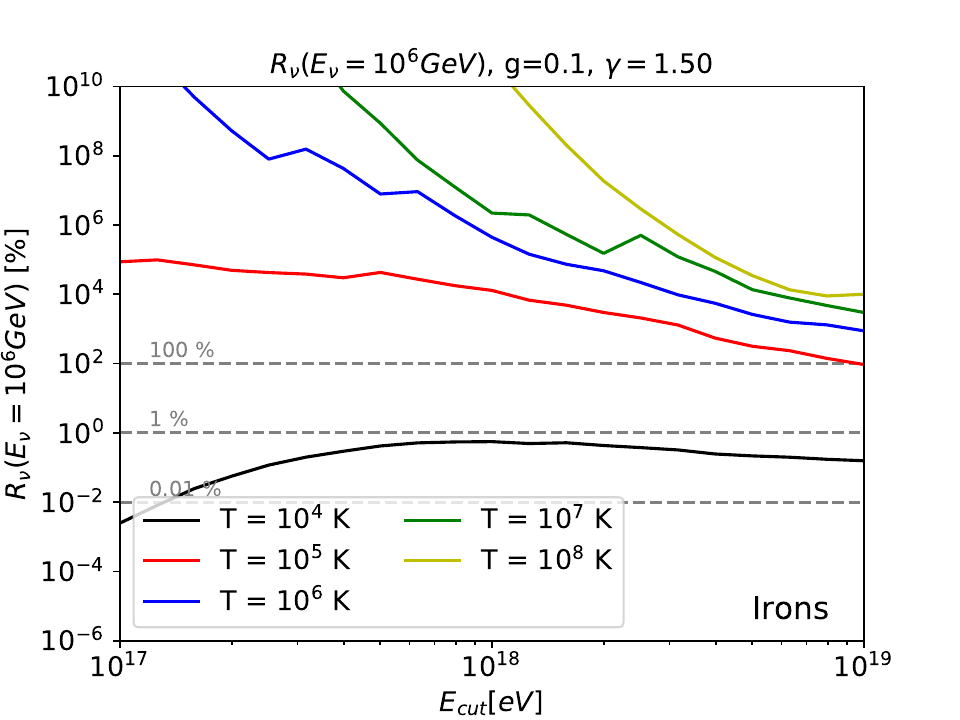}
\end{minipage}
\begin{minipage}{6.5cm}
\centering
\includegraphics[scale=0.42]{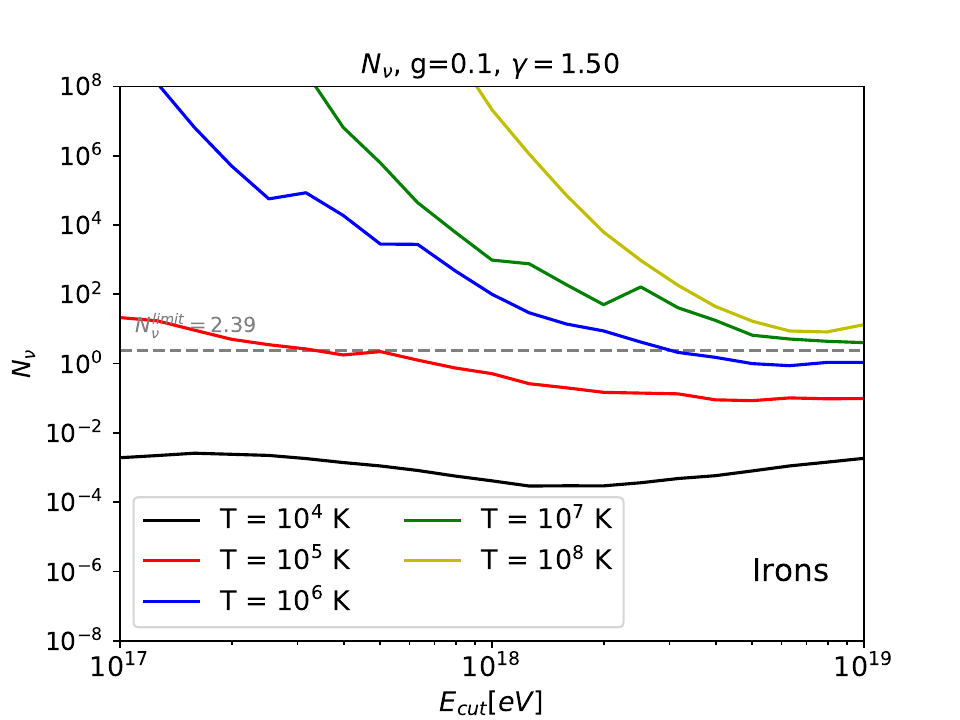}
\end{minipage}
\caption{Same as in Fig.~\ref{parameter_scan_p}, for the case of pure-iron injection and evolution parameter $m=0$.}
\label{parameter_scan_Fe}
\end{figure}
\par In Fig.~\ref{parameter_scan_p}, we show the parametric study of the quantities defined in Eqs.~\eqref{ration_nu} and~\eqref{N_cosmo} for a pure proton composition at the acceleration and evolution parameter $m=0$. In each row of Fig.~\ref{parameter_scan_p}, one parameter among $\gamma$, $E_\text{cut}$ and $g$ is varied for all the source temperatures $T$ (indicated as colored lines). In particular, in the first row the scan over $g$, in the second row the scan over $\gamma$, and in the third row the scan over $E_\text{cut}$ are shown. In the first (second) column of Fig.~\ref{parameter_scan_p} the neutrino spectral ratios at $E_\nu = 10^6 \,\text{GeV}$ (the number of neutrinos) are shown. Gray lines in the first column indicate the $0.01\%$, $1.0\%$ and $100.0\%$ contributions to the measured IceCube neutrino flux, while in the second column the Feldman-Cousins factor for the non observation of neutrinos is shown. We can immediately see that the two control quantities clearly depend on the temperature of the local photon field, as already discussed above and shown in other studies investigating the neutrino production as a function of the density of photons in the source environment \cite{Biehl:2017zlw,Boncioli:2018lrv,Muzio:2019leu,Muzio:2021zud}. The factor $g$ simply scales the UHECR flux at Earth and, as a consequence, the same scaling is found both in the source and cosmogenic neutrinos. The spectral shape of the UHECR spectrum at the acceleration affects the spectral ratio at $E_\nu = 10^6 \,\text{GeV}$ more than the expected number of neutrinos above $E_\nu = 10^8 \,\text{GeV}$. This is mostly due to the fact that a softer UHECR spectrum involves a larger number of low-energy UHECR protons, that is reflected in a larger number of neutrinos at $E_\nu = 10^6 \,\text{GeV}$. This is not contradictory to the results shown in Sec.~\ref{subsec_int-efficiency}: the neutrino production efficiency $f_\nu$ is given by the total neutrino emissivity of the source, while in Fig.~\ref{parameter_scan_p} we calculate the control quantity $R_\nu$ at a specific energy value to compare our simulated fluxes with available experimental results. The same cannot be seen for the number of neutrinos, defined above $10^{8}\,\text{GeV}$ due to the experimental neutrino exposure $\mathcal{E}(E_\nu)$ from \cite{PierreAuger:2019ens}, because only the high-energy part of the expected neutrino flux (due to the highest energy region of the UHECR spectrum) could be possibly measured. Both the spectral ratio at $E_\nu = 10^6 \,\text{GeV}$ and the number of neutrinos above $10^{8}\,\text{GeV}$ increase when $E_\text{cut}$ decreases. This is due to the fact that for low values of $E_\text{cut}$ a higher normalization is required to account for the observed UHECR flux.
\par In Fig.~\ref{parameter_scan_Fe} the parameter scan is shown for the case of pure iron injection at the acceleration in the source environment. In this case, the production of neutrinos is even more dependent on the temperature of the BB than in the pure-proton scenario. We also find that the number of neutrinos increases rapidly as the UHECR spectral index becomes softer. A softer spectral index corresponds to a larger number of low energy iron nuclei in the source environment, compared to the high-energy ones. Therefore, due to photodisintegration, a much larger normalization is required than in the pure proton scenario. The same effect can be seen when $E_\text{cut}$ decreases. The cases of SFR source evolution are shown in Appendix~\ref{app_parameter} in Fig.~\ref{parameter_scan_p_SFR} for proton injection and Fig.~\ref{parameter_scan_Fe_SFR} for iron injection. This source evolution assumption slightly increases the neutrino flux with respect to the $m=0$ case.

\subsection{Source temporal evolution}
\label{subsec_time_int}
In Sec.~\ref{sec_int_source_env} we have discussed the temporal evolution of the quantities that characterize the source environment, after the merger time. In particular, the evolution of SEDs in Eqs.~\eqref{sed_bb} and~\eqref{sed_nt_bis} and of the source radius in Eq.~\eqref{source_radius} is determined by the relation between the time after the merger and the BB temperature in Eq.~\eqref{temperature_time}. To account for the evolution of the interaction region, and thus the evolution of neutrino production, we integrate the energy spectra of the escaped particles over the time after the merger. We consider the time steps shown in Fig.~\ref{img_sed}: the first time step corresponds to $t=10^2\,\text{s}$ after the merger ($T=10^8\,\text{K}$), and the last time step corresponds to $t=10^4\,\text{s}$ after the merger ($T=10^4\,\text{K}$). As discussed in Sec.~\ref{subsec_source-escape}, the maximum acceleration energy depends slightly on the time after the merger. Therefore, as done in the previous sections, we consider $E_\text{cut}$ as a fixed parameter of the acceleration process. We adopt the same normalization procedure described in Sec.~\ref{subsec_extragal-prop}, and for normalizing to the observed UHECR flux at $E_\text{cut}$, we use the all-particle propagated spectrum integrated over the time evolution of the source. The reference scenarios are again characterized by the spectral parameter $\gamma=1.5$ and $E_\text{cut}=10^{18.5}\,\text{eV}$. The same definition of the scale factor $g$ in Eq.~\ref{spectrum_norm} is used: for the reference scenario $g=0.4$ in the proton injection case and $g=0.1$ in the iron case.
\begin{figure}[t]
\centering
\begin{minipage}{6.5cm}
\centering
\includegraphics[scale=0.42]{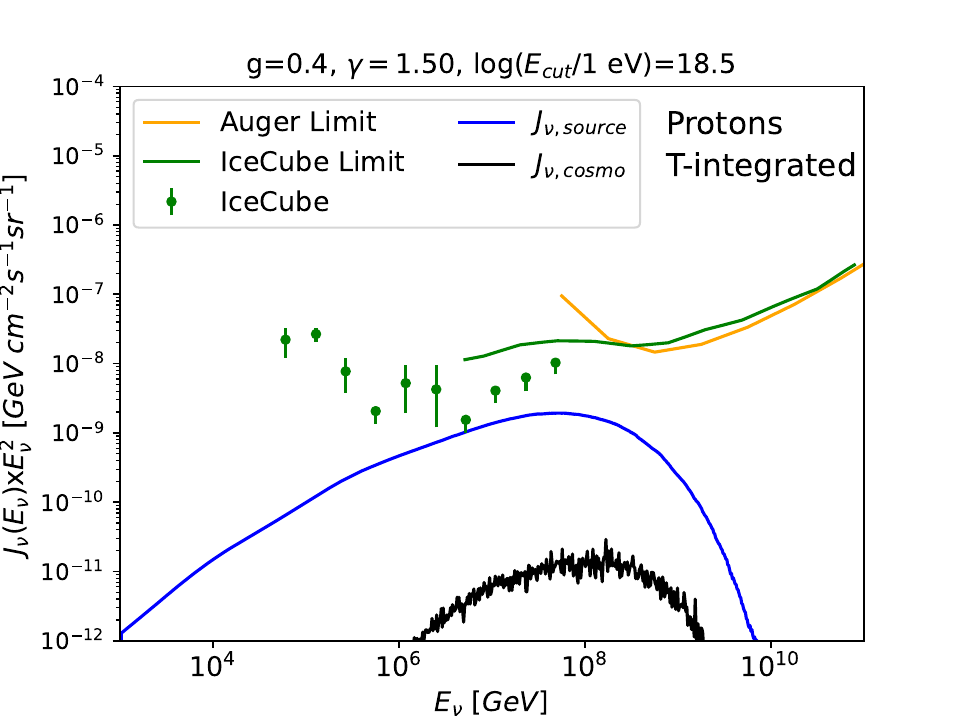}
\end{minipage}
\begin{minipage}{6.5cm}
\centering
\includegraphics[scale=0.42]{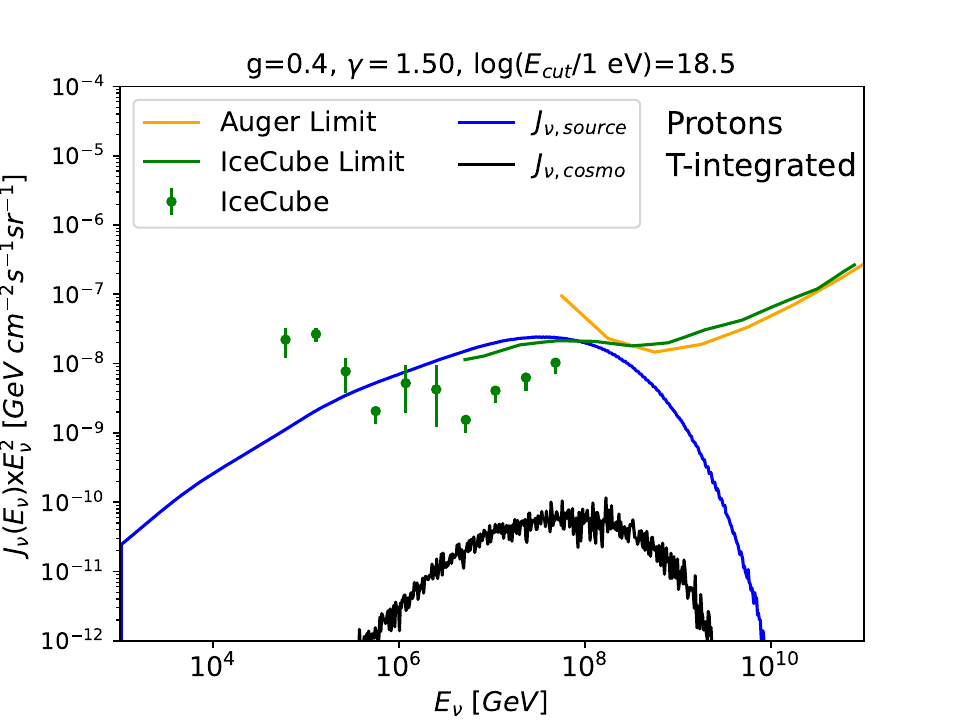}
\end{minipage}
\caption{Single-flavor neutrino energy spectra for a pure-proton injection for a population of BNS mergers where the temporal evolution is taken into account (no-evolution (left) and SFR evolution (right panel) cases): propagated source neutrinos (blue line) and cosmogenic neutrinos (black line). The observed neutrino flux and cosmogenic neutrino limit by IceCube \cite{IceCube:2017zho} and the cosmogenic neutrino limit by the Pierre Auger Observatory \cite{PierreAuger:2019ens} are also shown.}
\label{neu_flux_p_int}
\end{figure}
\begin{figure}[t]
\centering
\begin{minipage}{6.5cm}
\centering
\includegraphics[scale=0.42]{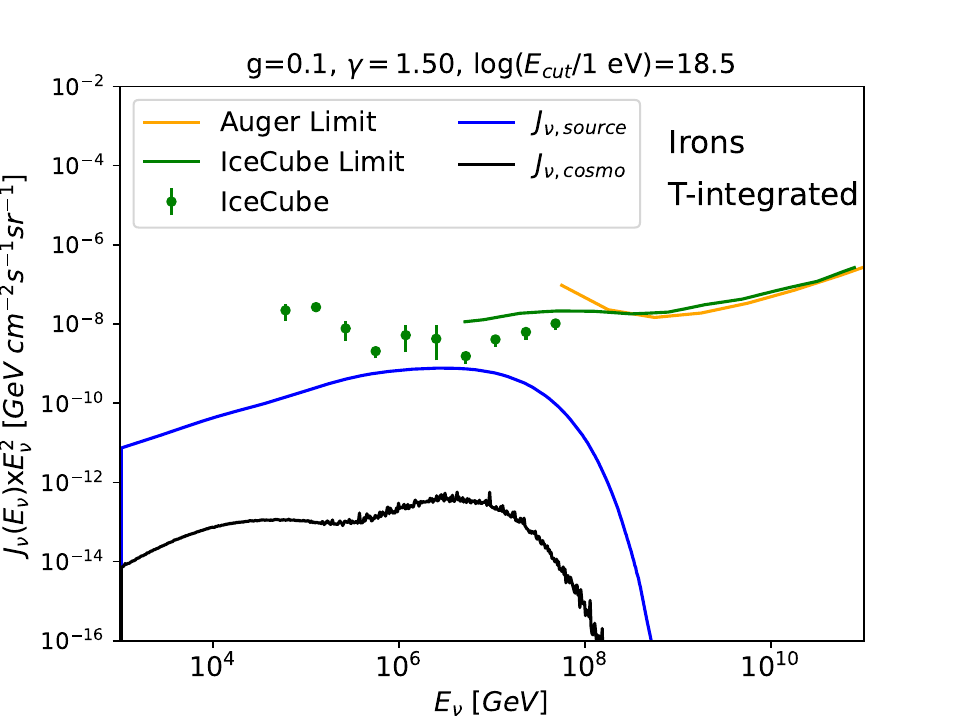}
\end{minipage}
\begin{minipage}{6.5cm}
\centering
\includegraphics[scale=0.42]{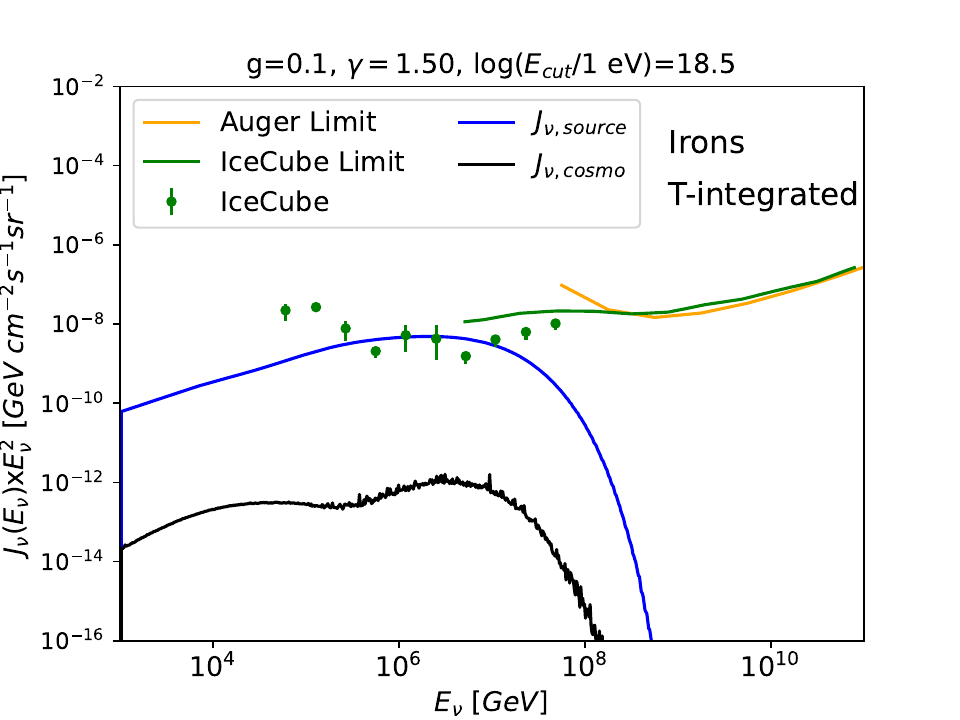}
\end{minipage}
\caption{Same as in Fig.~\ref{neu_flux_p_int}, for a pure-iron injection. Note the different range of the y-axes with respect to the proton scenarios.}
\label{neu_flux_Fe_int}
\end{figure}
\par In Fig.~\ref{neu_flux_p_int} the propagated source neutrino spectra for the reference scenario for a population of BNS mergers with evolution $m=0$ (left) and SFR evolution (right) are shown. Experimental data and limits are the same of Fig.~\ref{neu_flux_p}. In Fig.~\ref{neu_flux_Fe_int} the pure-iron injection scenarios are shown. The general results obtained in Sec.~\ref{subsec_extragal-prop} apply also here.
However, some differences are introduced when the time evolution of the source emission is considered. All the source neutrino spectra in Figs.~\ref{neu_flux_p_int} and~\ref{neu_flux_Fe_int} (in particular the iron scenarios) are suppressed with respect to the corresponding spectra in Figs.~\ref{neu_flux_p} and~\ref{neu_flux_Fe}. This is due to the normalization of the propagated CRs: in particular, if we consider the latest times after the merger ($T\sim10^4\,\text{K}$) most of the injected nuclei escape from the source region without interacting; therefore, a smaller normalization factor is needed to account for the contribution to the observed CR spectrum at $10^{18.5}\,\text{eV}$, with respect to the case in which a fixed time after the merger was considered. In addition, different slopes in the low energy part of source neutrino spectra are obtained as a result of combining different neutrino spectra escaped form the source (see Figs.~\ref{escape_p} and~\ref{escape_Fe} for the energy spectra of the escaped CRs and neutrinos from the source environment).
\begin{figure}[t]
\centering
\begin{minipage}{6.5cm}
\centering
\includegraphics[scale=0.42]{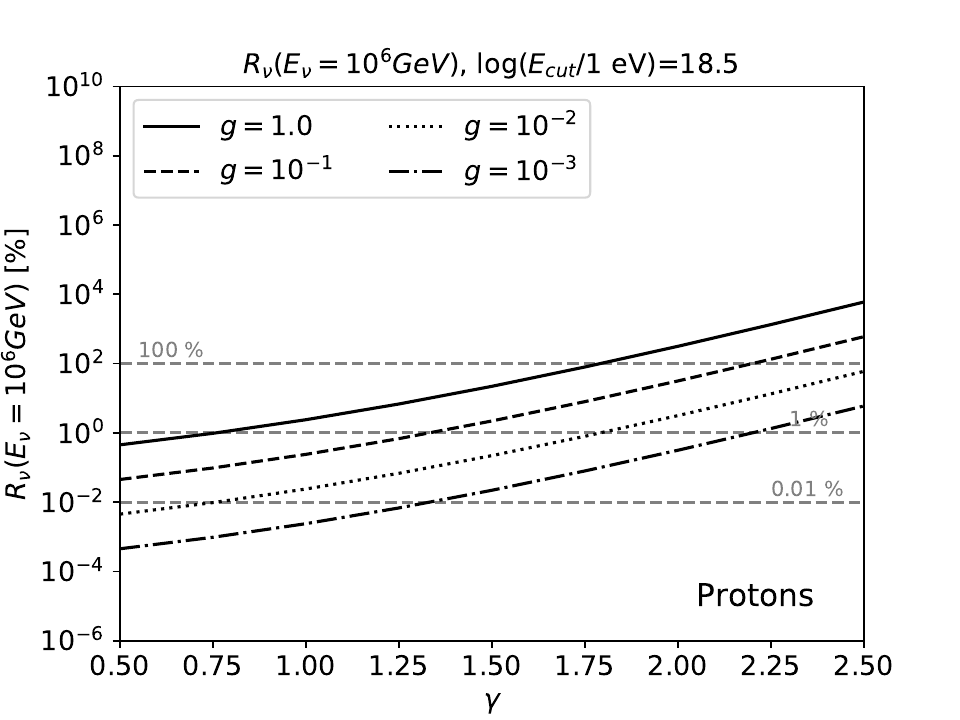}
\end{minipage}
\begin{minipage}{6.5cm}
\centering
\includegraphics[scale=0.42]{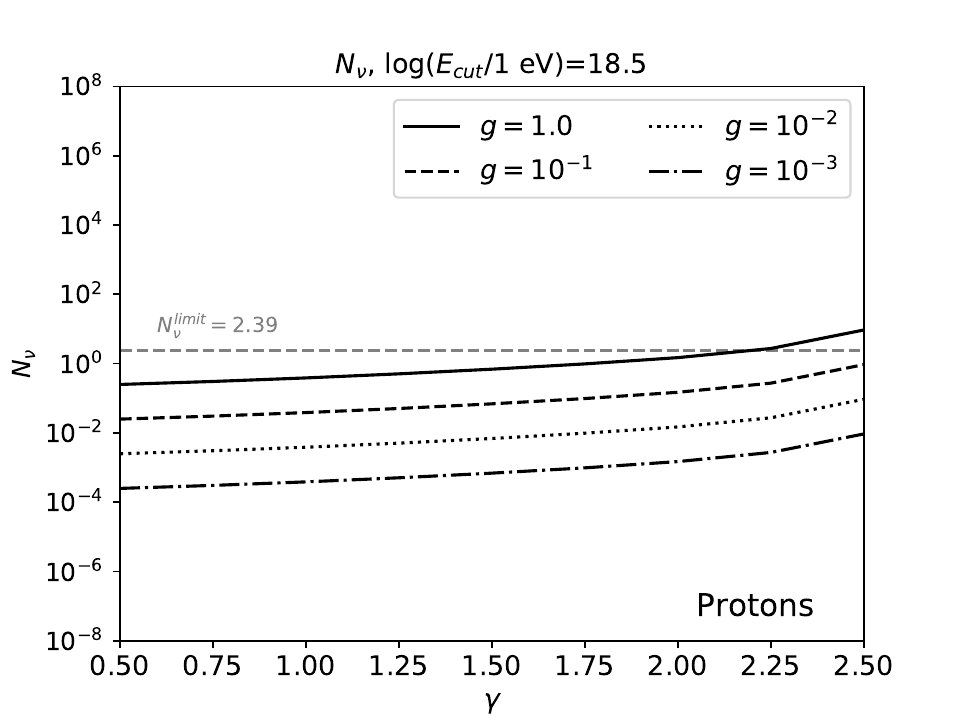}
\end{minipage}
\begin{minipage}{6.5cm}
\centering
\includegraphics[scale=0.42]{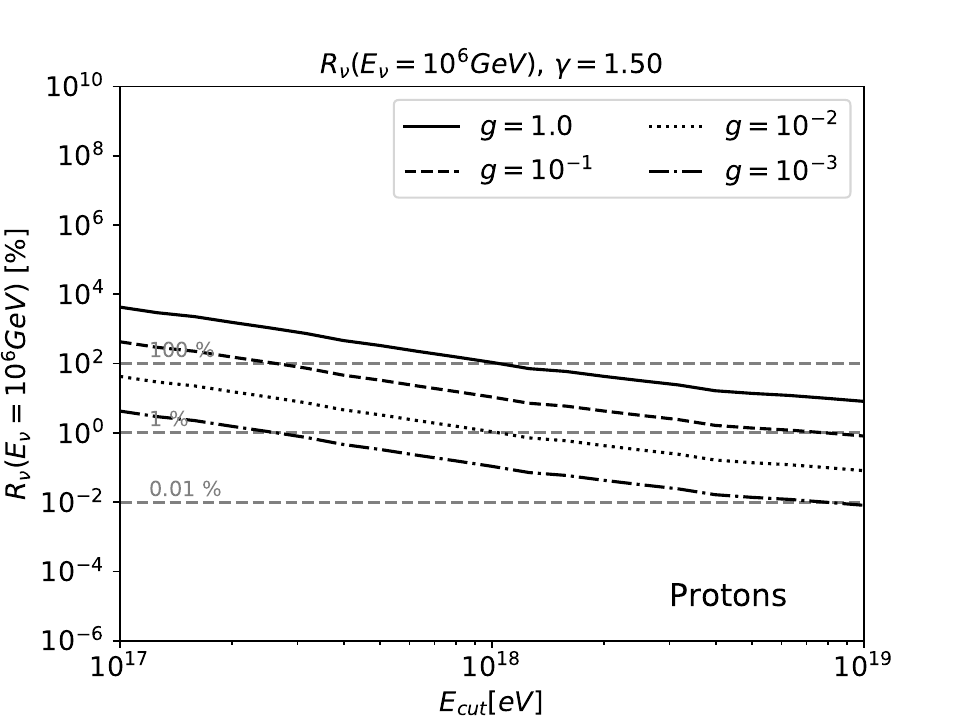}
\end{minipage}
\begin{minipage}{6.5cm}
\centering
\includegraphics[scale=0.42]{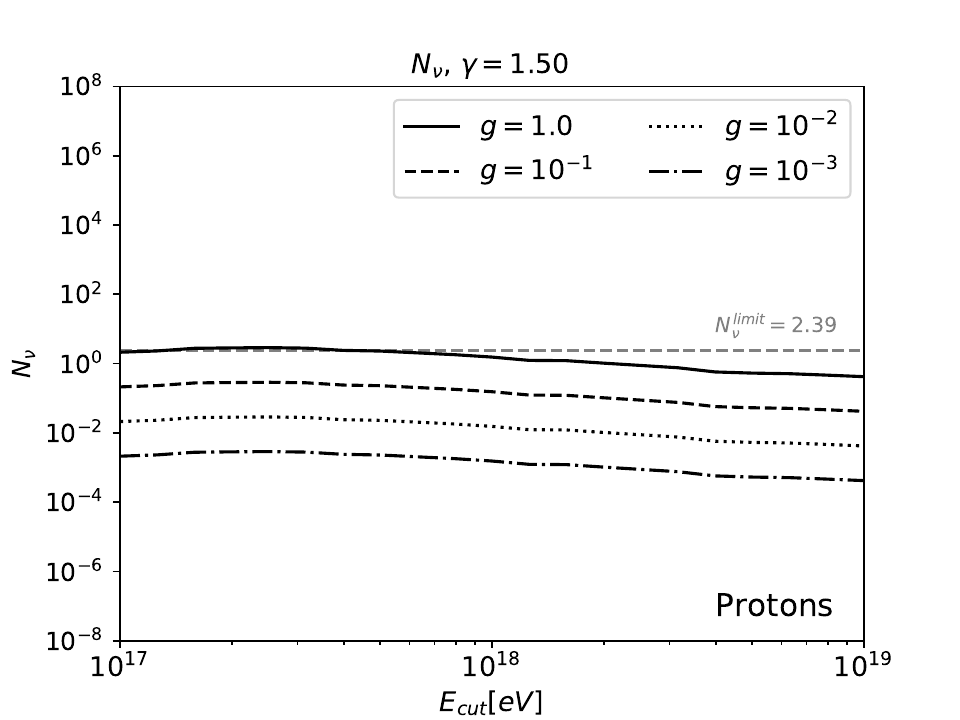}
\end{minipage}
\caption{Left column: neutrino spectral ratio, calculated at $E_\nu=10^{6}\,\text{GeV}$, as defined in Eq.~\eqref{ration_nu}. Right column: number of neutrinos, as defined in Eq.~\eqref{N_cosmo}. These quantities are shown as a function of the spectral index $\gamma$ (upper) and the high energy cutoff $E_\text{cut}$ (bottom panels), both integrating over the temporal evolution of the source. Several normalization factors $g$ are considered in each panel: $g=1.0$ (solid lines), $g=10^{-1}$ (dashed lines), $g=10^{-2}$ (dotted lines) and $g=10^{-3}$ (dashed-dotted lines). In the right panels, gray lines indicate some reference ratios. The Feldman-Cousins factor for non-observation of events is indicated with a gray line in the left panels. The injected CR composition in the source environment is pure-proton, and the evolution parameter is $m=0$.}
\label{parameter_scan_p_int}
\end{figure}
\par
\begin{figure}[t]
\centering
\begin{minipage}{6.5cm}
\centering
\includegraphics[scale=0.42]{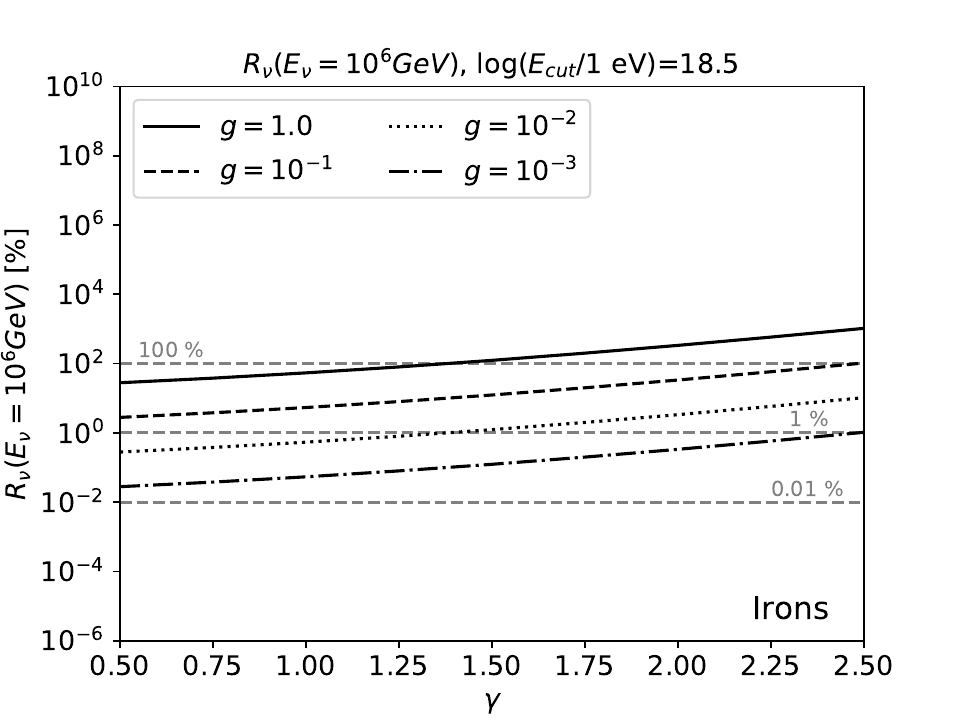}
\end{minipage}
\begin{minipage}{6.5cm}
\centering
\includegraphics[scale=0.42]{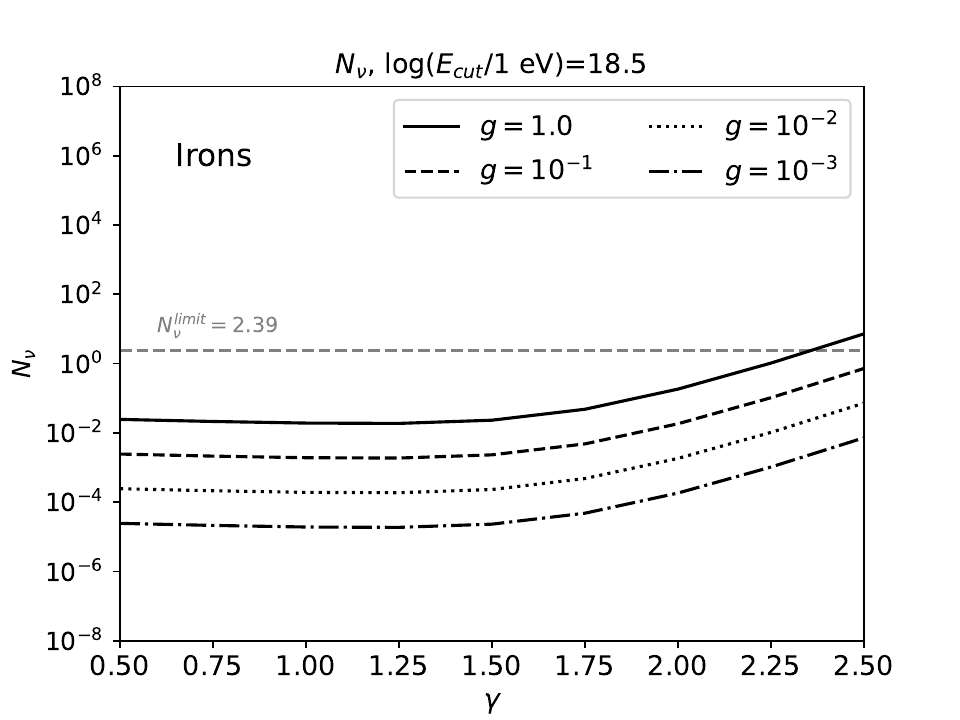}
\end{minipage}
\begin{minipage}{6.5cm}
\centering
\includegraphics[scale=0.42]{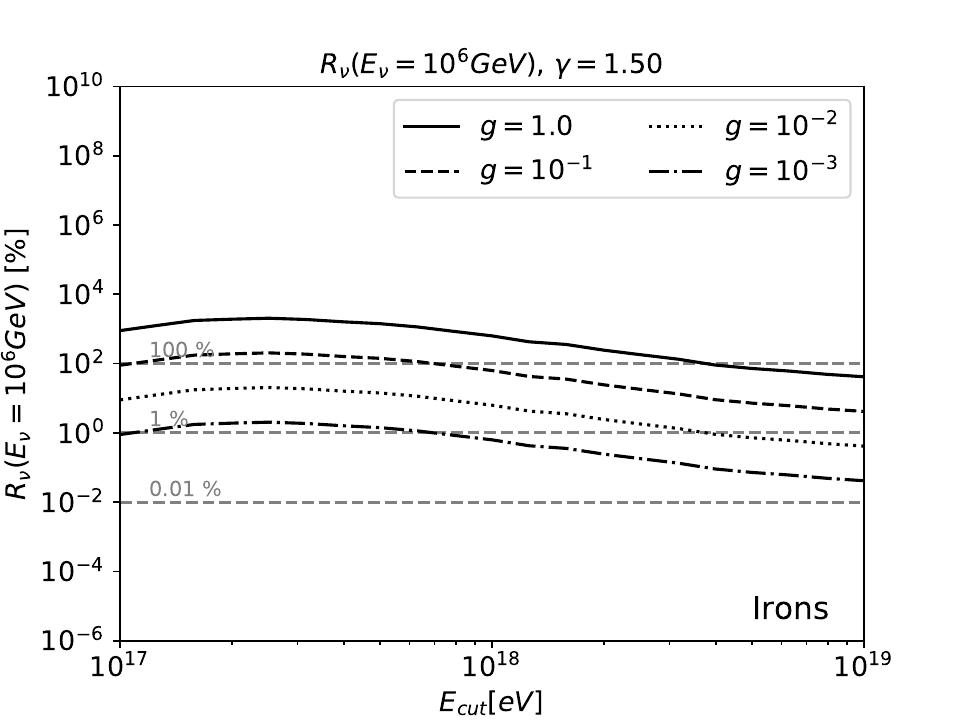}
\end{minipage}
\begin{minipage}{6.5cm}
\centering
\includegraphics[scale=0.42]{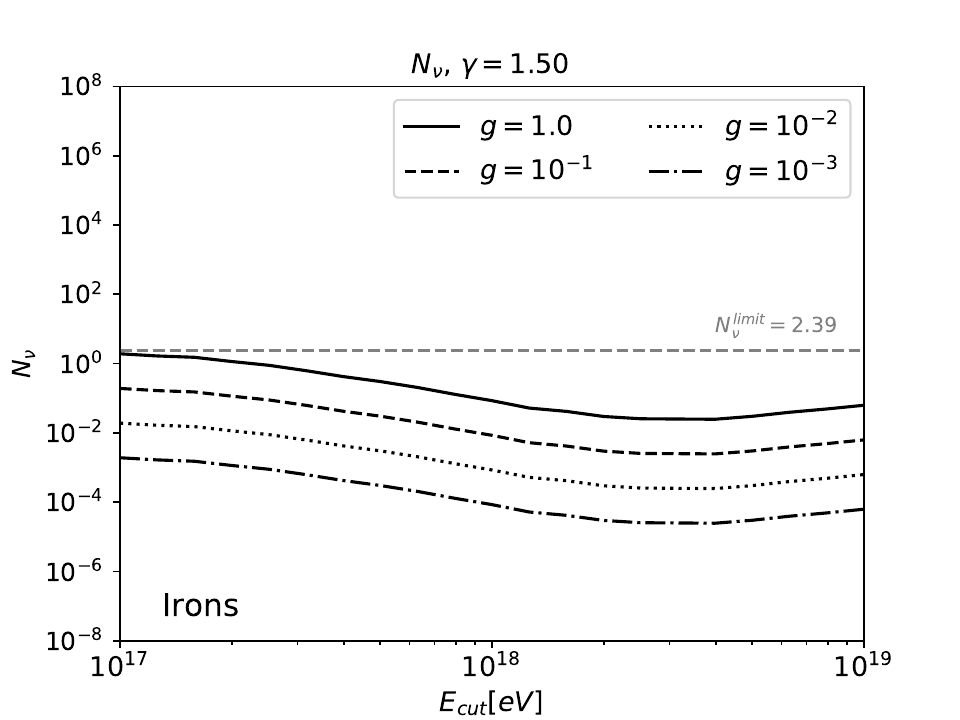}
\end{minipage}
\caption{Same as in Fig.~\ref{parameter_scan_p_int}, for the case of pure-iron injection and evolution parameter $m=0$.}
\label{parameter_scan_Fe_int}
\end{figure}

 We evaluate the two \textit{control quantities} defined in Eqs.~\eqref{ration_nu} and~\eqref{N_cosmo} for the source neutrino spectra discussed above. As done in Sec.~\ref{subsec_parameter}, we vary one parameter, keeping the other ones unchanged. In Fig.~\ref{parameter_scan_p_int} the parametric study of the \textit{control quantities} is shown for a pure-proton composition at the acceleration and source evolution $m=0$. Differently from Fig.~\ref{parameter_scan_p}, in each panel of Fig.~\ref{parameter_scan_p_int} different values of $g$ are shown together. The qualitative behaviors of $R_\nu(E_\nu=10^6\,\text{GeV})$ and $N_{\nu}$ as functions of $\gamma$ and $E_\text{cut}$ are the same discussed for a fixed value of the source temperature. Moreover, $R_\nu(E_\nu=10^6\,\text{GeV})$ and $N_{\nu}$ are reduced when the scaling factor $g$ decreases. It can be seen that the most stringent constraints are given by $R_\nu(E_\nu=10^6\,\text{GeV})$, while $N_\nu$ almost always agrees with the experimental limits and observations, even for the most extreme scenario $g=1.0$. In Fig.~\ref{parameter_scan_Fe_int} the same results are shown for the case of pure-iron acceleration. In general, all the \textit{control quantities} in Figs.~\ref{parameter_scan_p_int} and~\ref{parameter_scan_Fe_int} are in better agreement with the neutrino data and limits than in Figs.~\ref{parameter_scan_p} and~\ref{parameter_scan_Fe}. This is because the integration over the source temperature has the consequence that the normalization is mostly defined by the CR flux at low temperature (i.e. the latest to leave the source). Therefore, a smaller UHECR normalization is required and smaller neutrino fluxes are obtained. The cases of SFR source evolution are shown in Appendix~\ref{app_parameter} in Figs.~\ref{parameter_scan_p_SFR_int} and~\ref{parameter_scan_Fe_SFR_int}.

\section{Discussion and conclusions}
\label{sec_discussion}
In this work, we have realized a source-propagation model where the considered source environment is the end-state of the merger of binary-neutron-star systems. The Monte Carlo code \texttt{SimProp-v2r4} has been adapted to simulate the in-source interactions, while the original version of the code was used for the extragalactic propagation. We have assumed, as supported by \cite{Decoene:2019eux}, that the neutrino production does not take place in relativistic jets, but in the shocks that might be generated behind the ejected material; in addition, we have taken into account cosmic rays that can reach energies possibly contributing to the region just below the ankle in the measured energy spectrum, as shown in \cite{Rodrigues:2018bjg}. We have considered the non-thermal and thermal spectral energy density of the local photons, generated by the synchrotron emission and the nuclear decays of the unstable nuclear species synthesized in the ejected material, respectively. For doing this, we have taken as a reference the event GW170817 \cite{LIGOScientific:2017vwq}, the only event for which a gravitational wave and the electromagnetic counterpart have been detected nowadays.
\par As a result of the study of the propagation of UHECRs in the BNS remnant environment, we have shown that interactions with the non-thermal photons are not efficient enough to produce neutrinos or photodisintegrate the fallback material (see Fig.~\ref{opacities}). For this reason, we have neglected this photon field in our in-source simulations. On the other hand, black-body photons can trigger interactions with both injected protons and iron nuclei. In this case, the opacity of the source is greater than one for $T\gtrsim 10^5 \,\text{K}$ ($t\lesssim 10^{3.5}\,\text{s}$) for protons, and for $T\gtrsim 10^4 \,\text{K}$ ($t\lesssim 10^{4}\,\text{s}$) for iron nuclei. In particular, for Lorentz factors $\Gamma\simeq10^9$, the fallback material interacts with the thermal photon field through photomeson production and photodisintegration. However, very low- and very high-energy nuclei can leave the source environment undisturbed (see total interaction lengths in Figs.~\ref{int_len_p} and ~\ref{int_len_Fe}). As an outcome, we have shown that the efficiency of interactions is higher in the early stages after the merger (i.e. $T\gtrsim10^6\,\text{K}$) than at later stages. In particular, scenarios of proton injection saturate at $10\%$ conversion of cosmic ray energy into neutrinos. However, in the case of iron nuclei we observe that photodisintegration increases the source conversion efficiency in cosmic rays, while the neutrino production is almost unchanged. 
\par In Sections.~\ref{subsec_parameter} and~\ref{subsec_time_int}, we have quantified the diffuse neutrino flux at Earth (the neutrinos produced in the source environment and the ones produced in the extragalactic propagation) as a function of the temperature of the black body, as well as depending on the CR spectral parameters, for a population of identical BNS mergers. We have shown that in general the observed neutrino flux cannot be associated with cosmogenic neutrinos. The neutrino spectral ratio $R_\nu(E_\nu=10^{6}\,\text{GeV})$ and the number of neutrinos $N_\nu$ show a dependence on the photon-field temperature, and in general very high temperatures correspond to a large neutrino flux at Earth. We calculated the propagated neutrino fluxes taking into account the time evolution of the source environment. For both compositions considered, in order to avoid overshooting of the measured neutrino flux, the high energy cutoff must be $\gtrsim 10^{18}\,\text{eV}$, if the reference values of the scaling $g$ of the cosmic-ray expected spectra are considered. The effect of the variation of the spectral index $\gamma$ is almost independent of the composition considered and the scaling factor $g$: for both protons and iron nuclei we find that all the values of gamma are acceptable for $g\lesssim0.4$. These results depend weakly on the source evolution model adopted.
\par The efficiency in producing high energy astrophysical neutrinos can be, as a first approximation, connected to the ratio of the total interaction length of cosmic rays to the typical size of the interaction region. In this study we have assumed that the typical escape length is given by the radius of the ejected material in Eq.~\eqref{source_radius}, and the typical escape time is then given by $\tau_\text{esc}(t) =\beta_\text{ej}t$, where $\beta_\text{ej}=0.3$. This corresponds to a source size ranging from $\lambda_\text{esc}\approx10^{12}\,\text{cm}$ immediately after the merger, to $\lambda_\text{esc}\approx10^{14}\,\text{cm}$ in the last considered stage. Being the typical values of $\beta_\text{ej} = 0.1$-$0.3$ (see \cite{Metzger:2019zeh,Shibata:2019wef}), different values of $\beta_\text{ej}$ with respect to what assumed here will only marginally affect the production of neutrinos. 
\par Another important approximation made regarding the confinement of accelerated cosmic rays is the fact that the escape time does not depend on the rigidity of the particles. The presence of a magnetic field in the post-merger environment (with strength $\mathcal{O}(\text{mG})\lesssim B\lesssim\mathcal{O}(\text{G})$, see \cite{Decoene:2019eux,Rodrigues:2018bjg}) could lead to a longer confinement time for low rigidity nuclei. This effect could be particularly important in the case of heavy nuclei: a longer confinement time would imply more interactions with local fields, and thus higher photodisintegration and photopion efficiency. Additionally, synchrotron energy losses should be evaluated in studying the neutrino production efficiency of this class of sources. Future developments of the present work might include these details. 
\par The comparison of the propagated diffuse UHECR spectrum to the measured flux offers the possibility of studying the \textit{source-parameters}, such as the baryonic loading $\eta$, as well as the number density of BNS mergers. The CR emissivity at acceleration is related to the number density of mergers as 
\begin{equation}
\label{emi}
\mathcal{E}_{\text{acc}}=
E_{\text{acc}} \, \dot{n}\, ,
\end{equation}
where $E_{\text{acc}}$ is the total accelerated CR energy and $\dot{n}$ the event rate per volume (i.e. the number of mergers per unit of volume and unit of time) of BNS mergers. The energy in cosmic rays is related to the fallback luminosity $\mathcal{L}_{\text{fb}}$, being this the luminosity of the outflow powered by accretion. We parameterize the fall-back luminosity as in the \textit{optimistic} scenario of \cite{Decoene:2019eux} (i.e. the ejected mass by the merger is $10^{-4}\,M_{\odot}$ and $\beta_\text{ej}=0.3$) obtaining 
\begin{equation}
\label{fallback_lum}
\mathcal{L}_\text{fb}(t) = 1.3 \cdot 10^{43} \cdot\left(\dfrac{t}{10^3\,\text{s}}\right)^{-5/3}\,\text{erg} \,\text{s}^{-1} \, , 
\end{equation}
where $t$ is the time after the merger; the time dependence is taken as in \cite{Decoene:2019eux}, corresponding to the fall-back mass dynamics (see also \cite{Metzger:2019zeh,Rosswog:2006rh}). We then define the baryonic loading $\eta$ as the conversion coefficient of fall-back material into accelerated UHECRs, i.e. $\mathcal{L}_\text{acc}=\eta\cdot\mathcal{L}_\text{fb}$. Thus, the total accelerated CR energy is given by
\begin{equation}
E_\text{acc} = \int dt\,\mathcal{L}_\text{acc}(t) = \eta \int dt\,\mathcal{L}_\text{fb}(t) \, ,
\end{equation}
and then
\begin{equation}
\label{baryonic_loading}
\dfrac{\mathcal{E}_\text{acc}}{\dot{n}\,\eta} = \int dt\,\mathcal{L}_\text{fb}(t) \, .
\end{equation}
\begin{figure}[t]
\centering
\begin{minipage}{6.5cm}
\centering
\includegraphics[scale=0.47]{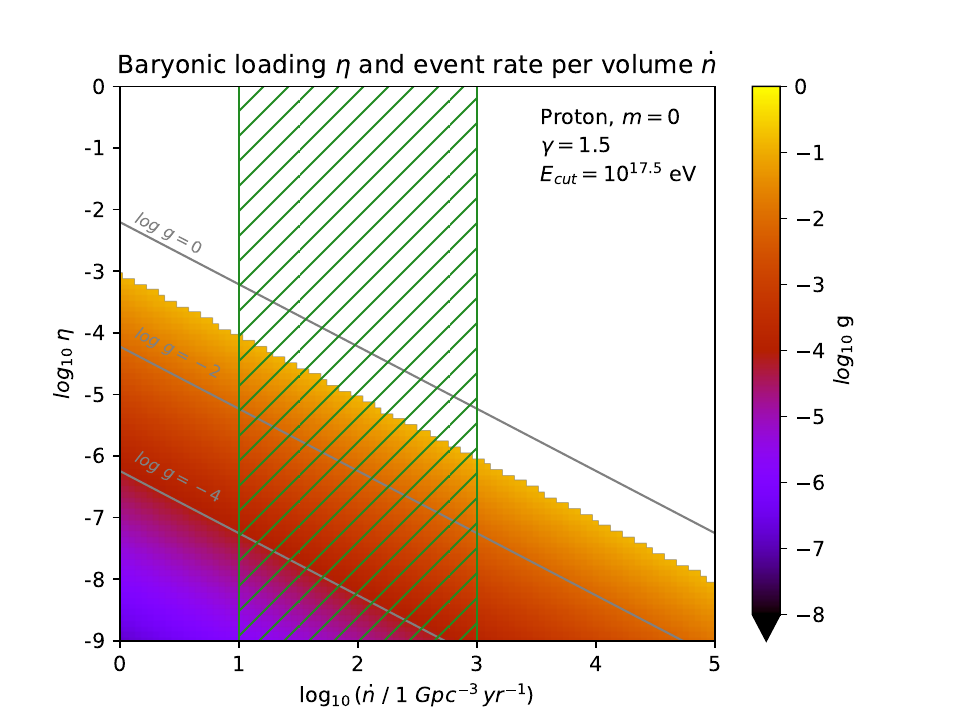}
\end{minipage}
\begin{minipage}{6.5cm}
\centering
\includegraphics[scale=0.47]{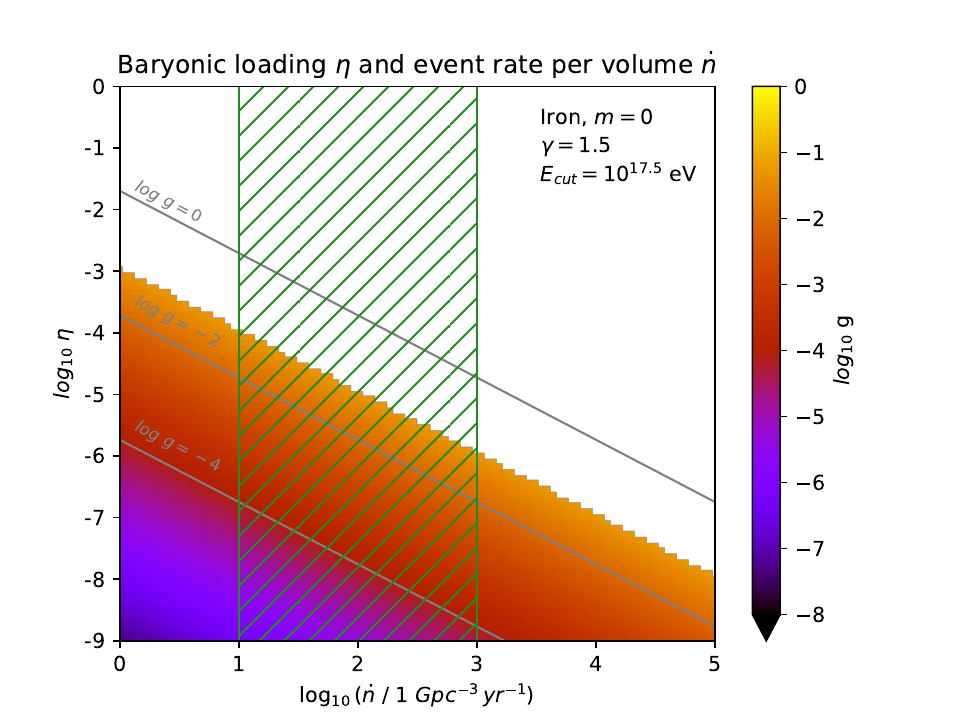}
\end{minipage}
\begin{minipage}{6.5cm}
\centering
\includegraphics[scale=0.47]{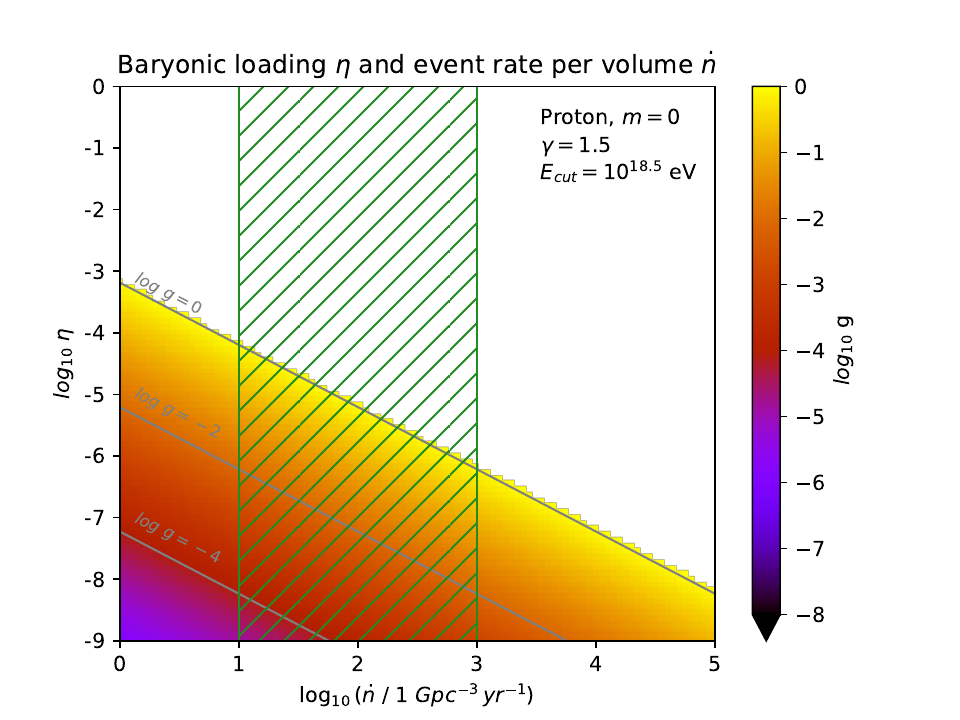}
\end{minipage}
\begin{minipage}{6.5cm}
\centering
\includegraphics[scale=0.47]{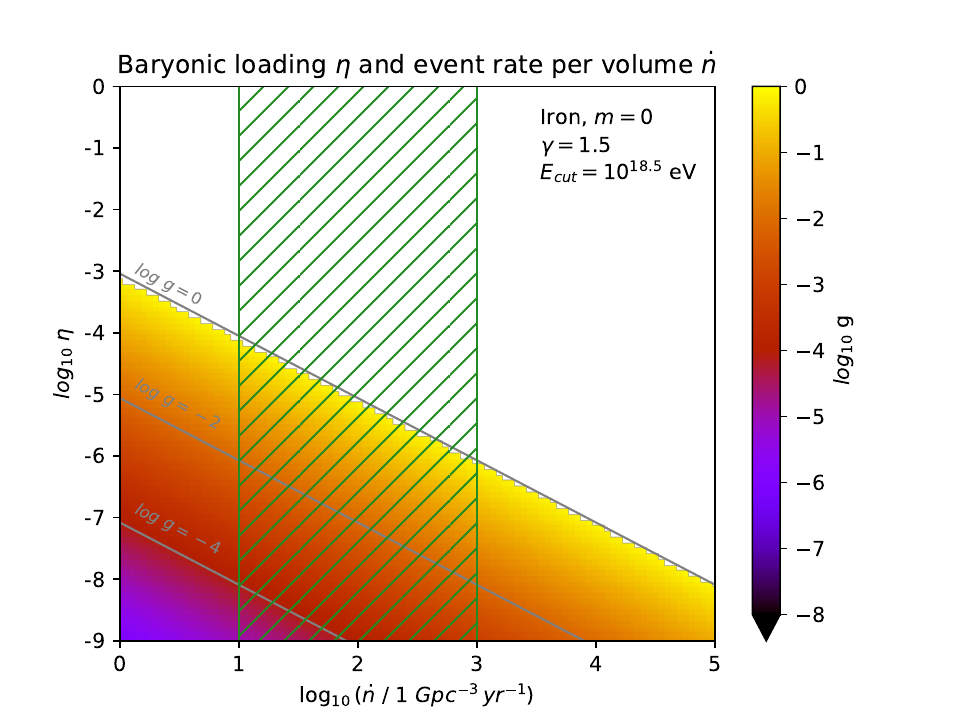}
\end{minipage}
\caption{Scaling parameter $g$ as a function of the baryonic loading $\eta$ and the BNS merging event rate per volume $\dot{n}$, as defined in Eq.~\eqref{baryonic_loading}. The value of $\log_{10}{g}$ is indicated by the color bar.
The gray lines correspond to the reference values of $\log_{10}{g}$, as shown in the panels. The acceleration parameters of the sources are $\gamma=1.5$ for the spectral index and $E_\text{cut}=10^{17.5}\,\text{eV}$ (upper panels) and $E_\text{cut}=10^{18.5}\,\text{eV}$ (lower panels). The left panels correspond to pure-proton injection and the right panels to pure-iron injection. The BNS merger event rate per volume estimated by \cite{LIGOScientific:2017vwq,KAGRA:2021vkt} is shown as an green region. The cosmological source evolution model is $m=0$.}
\label{baryonic_load_density_rate}
\end{figure}
The CR luminosity at acceleration is converted into the propagated CR spectrum by \textit{in-source} and extragalactic interactions. As shown in Sec.~\ref{subsec_extragal-prop}, we scale the propagated CR spectrum to the observed spectrum by the coefficient $g$, defined in Eq.~\eqref{spectrum_norm}. Therefore, for a given value of $g$ corresponding to a propagated CR flux that does not overshoot the CR data, different combinations of the parameters $\eta$ and $\dot{n}$ are acceptable, i.e. $g=g(\dot{n}\cdot\eta)$. By construction, the possible values of $g$ are limited by the condition $g\leq1$ and, in addition, thanks to the model developed in this work, by the fact that the corresponding neutrino flux must be such that $R_\nu(E_\nu=10^6\,\text{GeV})\leq1$ (see Eq.~\eqref{ration_nu}) and $N_\nu\leq 2.39$ (see Eq.~\eqref{N_cosmo}).
\par In Fig.~\ref{baryonic_load_density_rate} the possible values of $g(\dot{n}\cdot \eta)$ are shown for different injection scenarios at the acceleration; in other words, the scenarios corresponding to each value of $g$ are degenerate in terms of the product of $\eta$ and $\dot{n}$, as shown in Eq.~\eqref{baryonic_loading}. The acceleration scenarios shown in Fig.~\ref{baryonic_load_density_rate} are $\gamma=1.5$ for the spectral index and $E_\text{cut}=10^{17.5}\,\text{eV}$ (upper panels) and $E_\text{cut}=10^{18.5}\,\text{eV}$ (lower panels). The left panels correspond to pure-proton composition at the acceleration, and the right panels to pure-iron composition. The BNS merger event rate per volume estimated by \cite{LIGOScientific:2017vwq,KAGRA:2021vkt} is shown as an green region. We show the case of source evolution model $m=0$; the corresponding scenarios for the SFR evolution are shown in Fig.~\ref{baryonic_load_density_rate_SFR}. We also indicate some reference values of $\log_{10}{g}$ with gray lines in each panel of Figs.~\ref{baryonic_load_density_rate} and~\ref{baryonic_load_density_rate_SFR}. In particular, the gray lines for $\log_{10}{g}=0$ correspond to the value of the product $\eta \,\dot{n}$, for which the CR flux is maximum at Earth in the region below the ankle (i.e. it saturates the measured flux at the fixed energy). We can immediately notice that, due to the more intense energy loss experienced by nuclei with respect to protons (in the source and in the extragalactic propagation) the required $\eta \, \dot{n}$ is in general larger. For $E_\text{cut}=10^{17.5}\,\text{eV}$, a scaling factor $g\lesssim10^{-1}$ is required by the constraint on the number of neutrinos produced. This because a larger normalization factor is needed, corresponding to a larger neutrino flux. On the other hand, for $E_\text{cut}=10^{18.5}\,\text{eV}$ the CR saturation scenario $g=1.0$ is possible and it corresponds to a baryonic loading $\eta\simeq10^{-4}-10^{-6}$ for the values of $\dot{n}$ in the allowed region. In the SFR scenario, due to the large number of high-redshift sources, for the same values of $\dot{n}$ we obtain a slightly higher required baryonic.
\par Differently from what done in this work, in \cite{Decoene:2019eux} a baryonic loading $\eta\simeq0.1$ is fixed a-priori; for an event rate per volume of $\dot{n}\simeq1000\,\text{Gpc}^{-3}\,\text{yr}^{-1}$, a contribution to the observed IceCube flux of $\simeq10\%$ is therefore attributed to BNS mergers. With our parametric study, we are instead able to consistently constrain the neutrino production efficiency from BNS mergers by using CR data between the \textit{knee} and the \textit{ankle}. In particular, we derive that corresponding to $\dot{n}=1000\,\text{Gpc}^{-3}\,\text{yr}^{-1}$, a baryonic loading of $\eta\lesssim10^{-5}$ is required in most of the considered scenarios, and in addition we can account for the contribution of the BNS to the cosmic rays below the ankle, if a scaling factor $g\lesssim 10^{-1}$  for $E_\text{cut}\simeq10^{17.5}\,\text{eV}$, and  $g\lesssim 1$ for $E_\text{cut}\simeq10^{18.5}\,\text{eV}$ are considered respectively. 
\par In conclusion, thanks to the source-propagation model proposed in this work, we have shown that a region of the parameter space of the baryonic loading and rate of merger events can be excluded, depending on what fraction of the cosmic ray flux below the ankle is ascribed to BNS mergers and to the constraints from neutrino measurements and upper limits. Further measurements by LIGO/Virgo, such as next-generation gravitational wave and neutrino detectors \cite{Mukhopadhyay:2023niv,Mukhopadhyay:2024lwq,Mukhopadhyay:2024ehs}, might improve the constraining power of such a model, and future possible multimessenger observations might provide further tests of source-propagation models as the one here developed.

\acknowledgments
S.R. and G.S. acknowledge support by the Bundesministerium für Bildung und Forschung, under grants 05A20GU2 and 05A23GU3. The authors acknowledge their participation to the Pierre Auger Collaboration. 

\appendix

\section{Ballistic approximation}
\label{sec_ballistic}
The validity of the ballistic approximation can be expressed considering the Larmor radius 
\begin{equation}
r_g=3.1\cdot10^{12}\left(\dfrac{E/Z}{10^{15}\,\text{eV}}\right)\left(\dfrac{B}{1\,\text{G}}\right)^{-1}\,\text{cm} \, , 
\end{equation}
where $B$ is the magnetic field within the interaction region and $E$ and $Z$ are the energy and the atomic number of the CR, respectively. Given the radius of interaction region in Eq.~\eqref{source_radius}, the ballistic approximation is defined by $\lambda_\text{esc}(t)\lesssim r_g$. Therefore, the condition on the CR Lorentz factor $\Gamma$ is
\begin{equation}
\Gamma\gtrsim3\cdot10^{5} \left(\dfrac{Z}{A}\right) \left(\dfrac{t}{10^{2}\,\text{s}}\right)\left(\dfrac{B}{1\,\text{G}}\right) \, ,
\end{equation}
where $A$ is the CR atomic mass. If we consider that $Z/A\sim1/2$ and that the neutrino production is maximal for $t\sim10^2\,\text{s}$, considering a limiting magnetic field strength value equal to $B\sim1\,\text{G}$ (see \cite{Decoene:2019eux,Rodrigues:2018bjg}), we obtain that the ballistic approximation can be used for $\Gamma\gtrsim10^5$ values of the CR Lorentz factor ($E\gtrsim10^{14}\,\text{eV}$ for protons and $E\gtrsim6\cdot10^{15}\,\text{eV}$ for iron nuclei).

\section{UHECR interaction rate}
\label{sec_uhecr-int}
In this appendix we report some details of the computation of UHECR interactions.
Due to the very high relativistic boost of UHECRs, cosmic photons appear as high-energy gamma ray in the rest frame of the particle. Therefore, photohadronic interactions between UHECRs and cosmic photons become possible, and the corresponding interaction rate $\tau_{ij}^{-1}$ for the process $i$ between the cosmic ray nucleus and the background photon field $j$ is given by 
\begin{equation}
\label{int_rate}
\tau_{ij}^{-1}=\dfrac{c}{2\Gamma^2}\int_{\epsilon_\text{th}}^\infty d\epsilon\,\sigma_i \left(\epsilon \right)\epsilon \int_{\epsilon/2\Gamma}^\infty d\bar{\epsilon} \, n_j \left( \bar{\epsilon} \right) \bar{\epsilon}^{-2} \, ,
\end{equation}
where $\bar{\epsilon}$ is the photon energy in the laboratory rest frame, $n_j (\bar{\epsilon})$ is the photon spectral energy density (SED, i.e. the number of photons per unit of volume and energy) in the laboratory rest frame, $\sigma_i (\epsilon)$ is the total cross-section expressed as a function of the photon energy in the nucleus rest frame, $\epsilon_\text{th}$ is the threshold photon energy in the nucleus rest frame and $\Gamma$ is the nucleus Lorentz factor (a complete derivation of Eq.~\eqref{int_rate} can be found in \cite{Boncioli:2022ojf,Boncioli:2023gbl}). 
\par The interaction rate in Eq.~\eqref{int_rate} is a linear function in both the cross-section and photon SED. We can then compute the total interaction rate of different interaction processes as  
\begin{equation}
\label{tot_rate}
\tau_\text{tot}^{-1} = \sum_{i,j} \tau_{ij}^{-1} \, ,
\end{equation}
where the sum is over the considered photon fields and the possible interaction processes. The corresponding probability associated with the combination $i$ and $j$ is 
\begin{equation}
\label{ij_prob}
p_{ij}=\dfrac{\tau_{ij}^{-1}}{\tau_{\text{tot}}^{-1}}\, .
\end{equation}
The latter relation shows the fact that several interaction processes can be evaluated separately. The escape condition of a particle from the source environment 
can be associated with an escape rate $\tau_\text{esc}^{-1}$ and compared with the interaction rates to determine when the particle is free to leave the interaction region.  In this work, we adopt this strategy. 
\begin{figure}[t]
\centering
\begin{minipage}{6.5cm}
\centering
\includegraphics[scale=0.42]{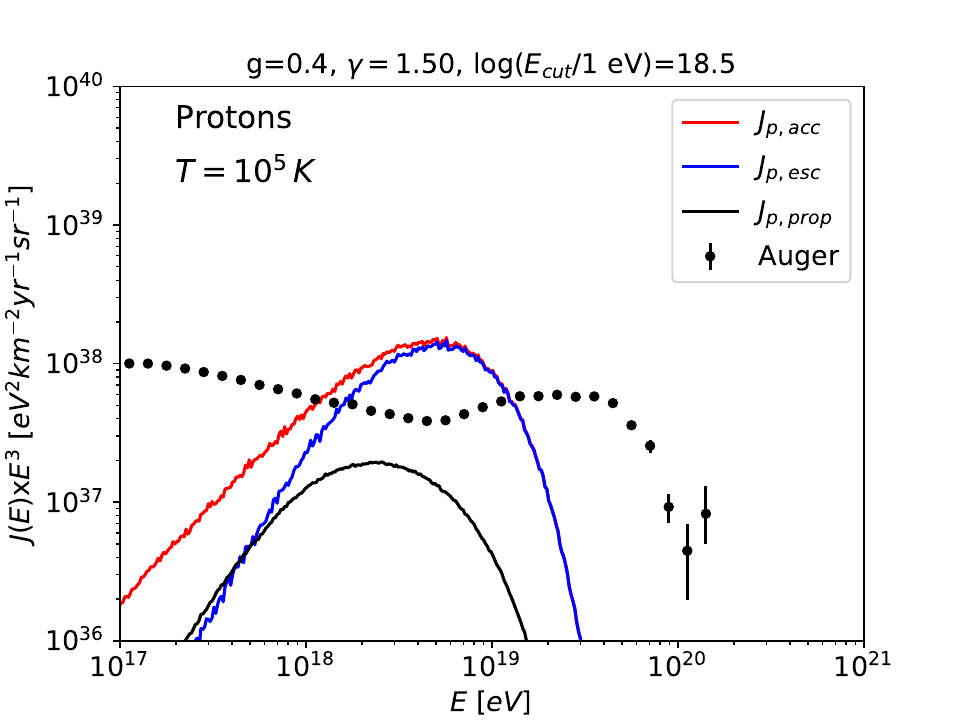}
\end{minipage}
\begin{minipage}{6.5cm}
\centering
\includegraphics[scale=0.42]{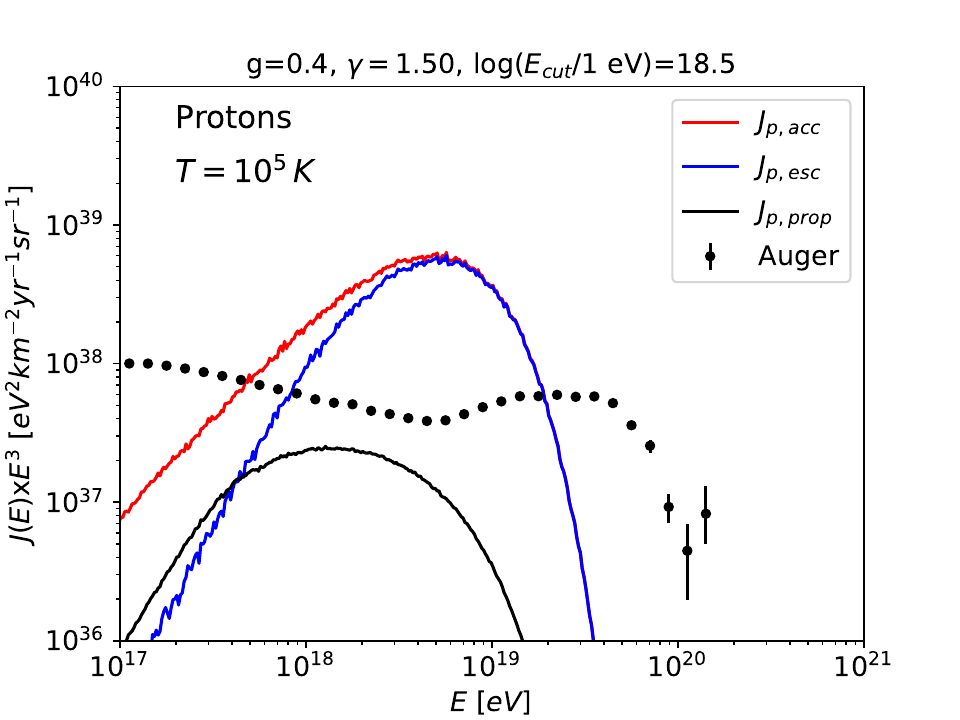}
\end{minipage}
\caption{Energy spectra for a pure-proton injection: injected in the source (red line), escaped from the source (blue line) and propagated at Earth (black line). Observed cosmic ray flux by the Pierre Auger Observatory \cite{PierreAuger:2019phh} is also shown (black dots). Left panel refers to no cosmological source evolution, right panel refers to SFR source evolution in Eq.~\eqref{SFR_function}.}
\label{uhecr_flux_p}
\end{figure}
\begin{figure}[t]
\centering
\begin{minipage}{6.5cm}
\centering
\includegraphics[scale=0.42]{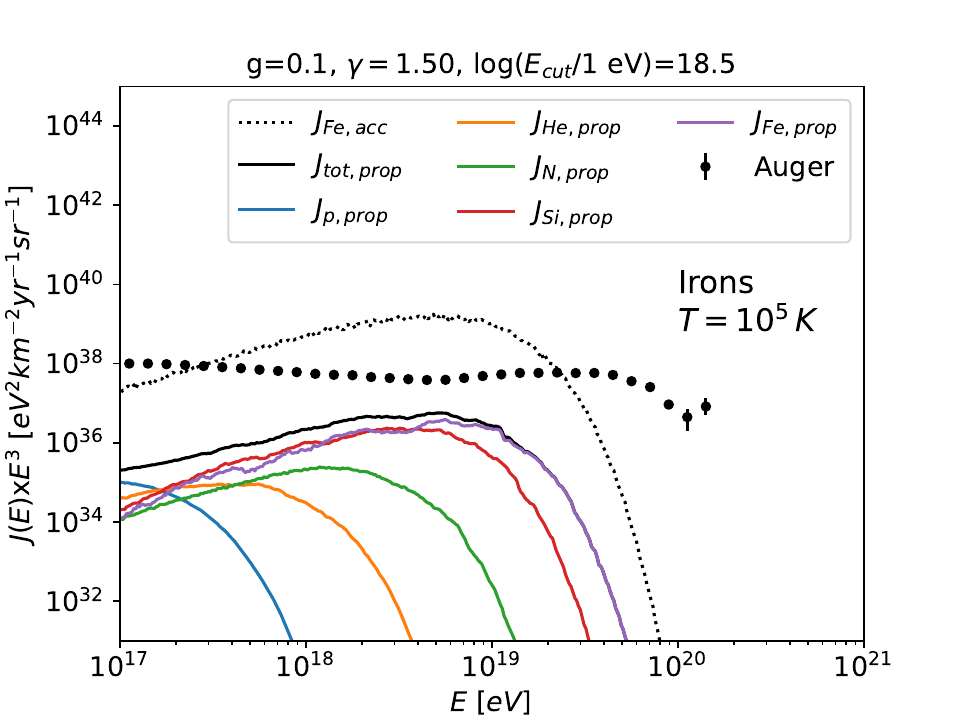}
\end{minipage}
\begin{minipage}{6.5cm}
\centering
\includegraphics[scale=0.42]{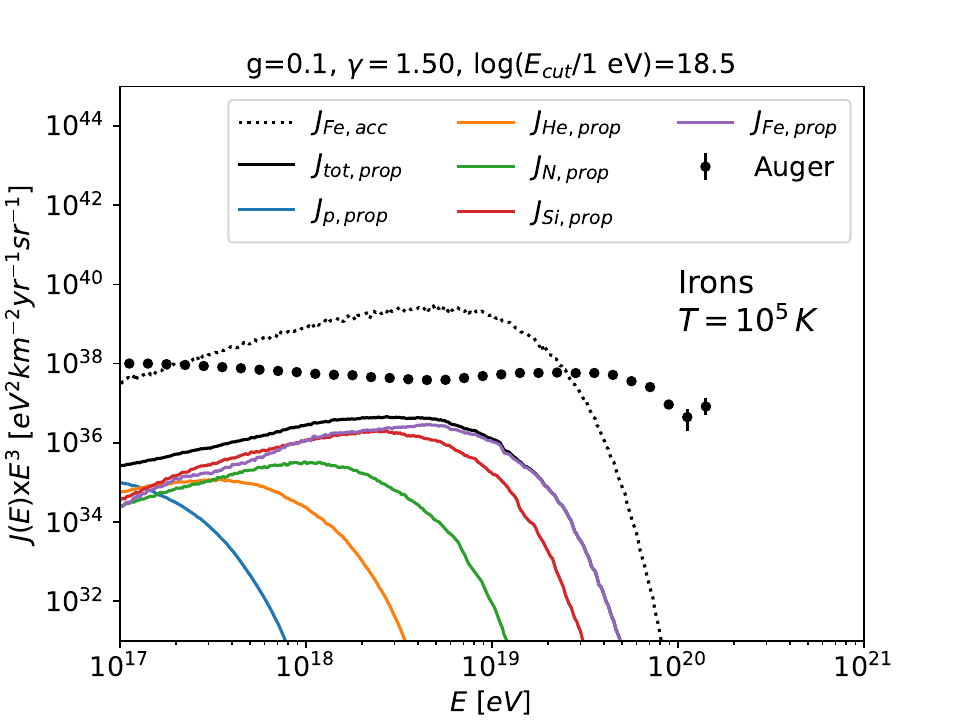}
\end{minipage}
\caption{Same of Fig.~\ref{uhecr_flux_p}, but for pure-iron injection. Injected iron nuclei are shown with black dotted line. Propagated mass groups (colored lines) and the total flux (black line) are shown. Note the different range of the y-axes with respect to the proton scenarios.}
\label{uhecr_flux_Fe}
\end{figure}
\begin{figure}[t]
\centering
\begin{minipage}{6.5cm}
\centering
\includegraphics[scale=0.42]{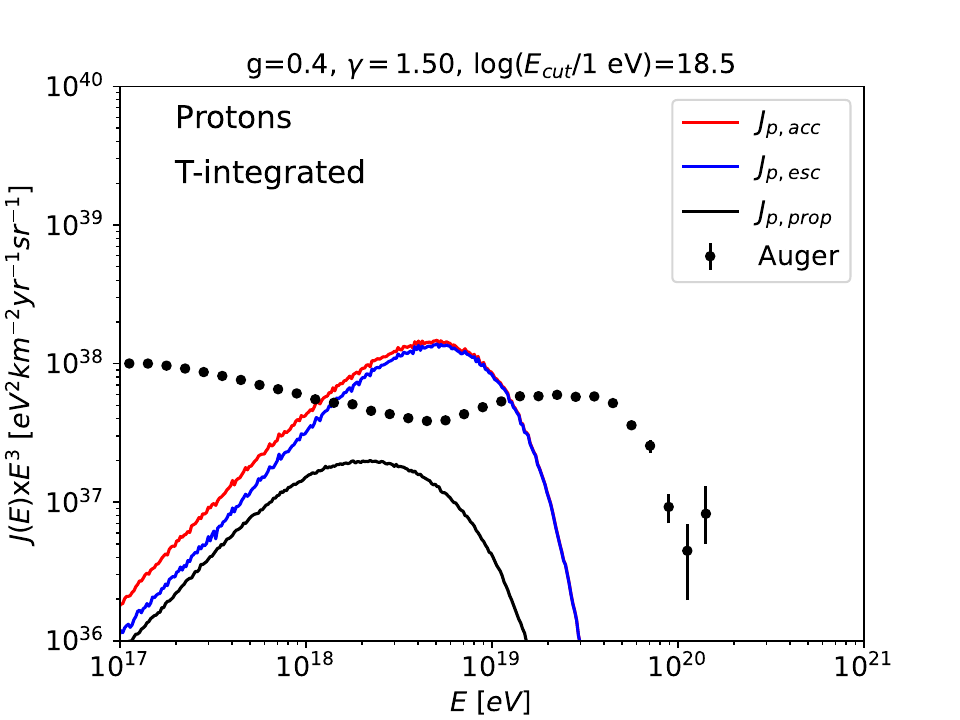}
\end{minipage}
\begin{minipage}{6.5cm}
\centering
\includegraphics[scale=0.42]{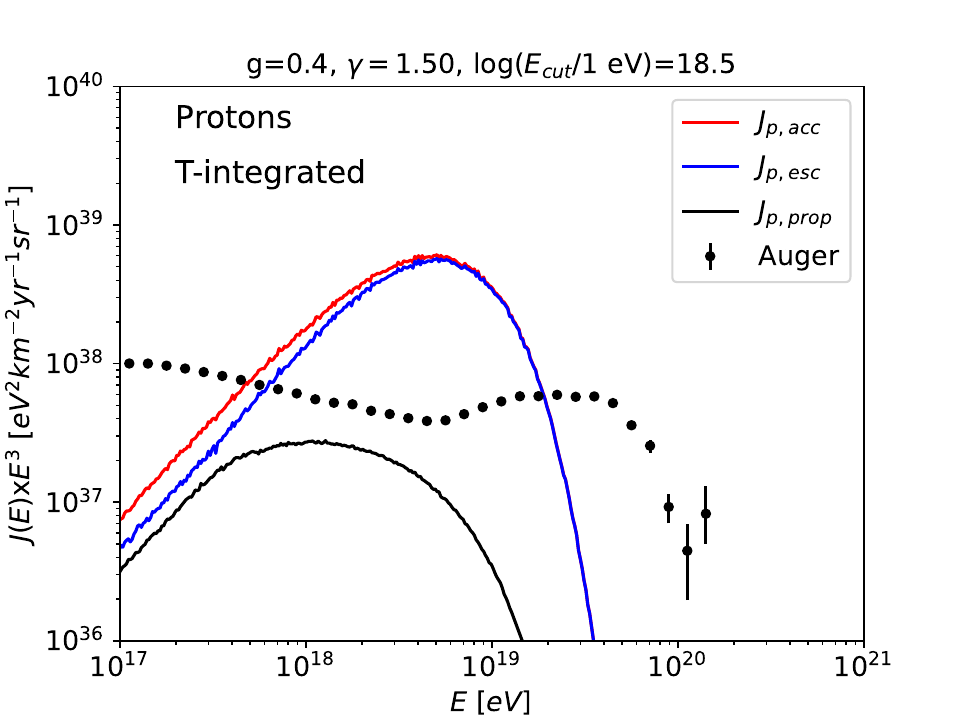}
\end{minipage}
\caption{Same of Fig.~\ref{uhecr_flux_p}, but integrating on the temporal evolution of the source.}
\label{uhecr_flux_p_int}
\end{figure}
\begin{figure}[t]
\centering
\begin{minipage}{6.5cm}
\centering
\includegraphics[scale=0.42]{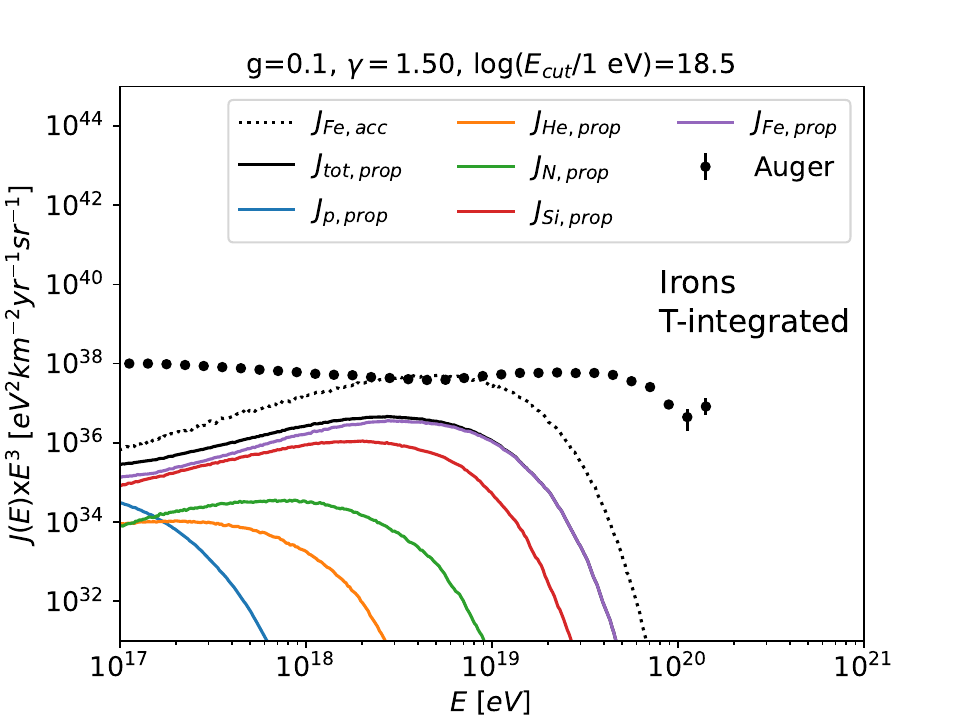}
\end{minipage}
\begin{minipage}{6.5cm}
\centering
\includegraphics[scale=0.42]{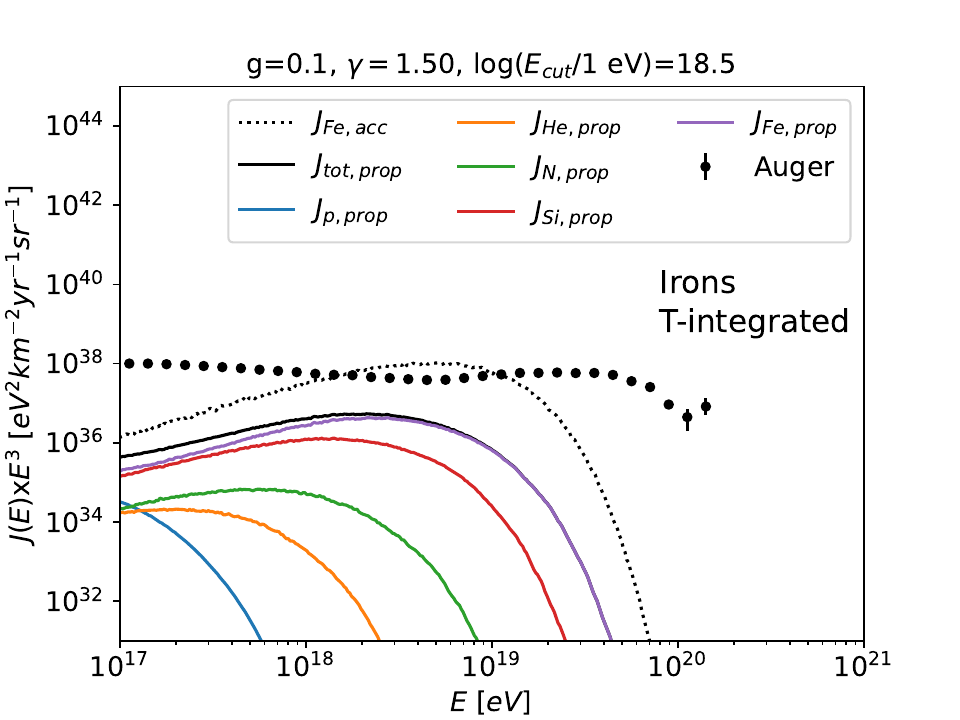}
\end{minipage}
\caption{Same as in Fig.~\ref{uhecr_flux_p_int}, for a pure-iron injection. Note the different range of the y-axes with respect to the proton scenarios.}
\label{uhecr_flux_Fe_int}
\end{figure}
\section{Cosmic-ray spectra}
\label{sec_uhecr-spectra}
In this appendix, the CR spectra corresponding to the predicted neutrino fluxes shown in Figs.~\ref{neu_flux_p} and~\ref{neu_flux_Fe} are shown. In particular, the pure proton scenario is shown in Fig.~\ref{uhecr_flux_p}, and the pure iron scenario is shown in Fig.~\ref{uhecr_flux_Fe}. In the case of protons, the accelerated (red), the escaped (blue) and the propagated (black) spectra are shown together with observed data\footnote{In general, accelerated and escaped spectra are given in different units than the observed spectrum. Accelerated and escaped spectra here refer to the spectra predicted in the absence of interactions within the source or during propagation, respectively.}. For iron nuclei, the accelerated spectrum (black dotted) is shown with the total propagated spectrum (black) and the different mass components (colored lines). Observed cosmic ray flux by the Pierre Auger Observatory \cite{PierreAuger:2019phh} is also shown (black dots). In Figs.~\ref{uhecr_flux_p_int} and~\ref{uhecr_flux_Fe_int} the CR spectra corresponding to the predicted neutrino fluxes shown in Figs.~\ref{neu_flux_p_int} and~\ref{neu_flux_Fe_int} are shown.
\section{Parameter study with the SFR}
\label{app_parameter}
\begin{figure}[t]
\centering
\begin{minipage}{6.5cm}
\centering
\includegraphics[scale=0.42]{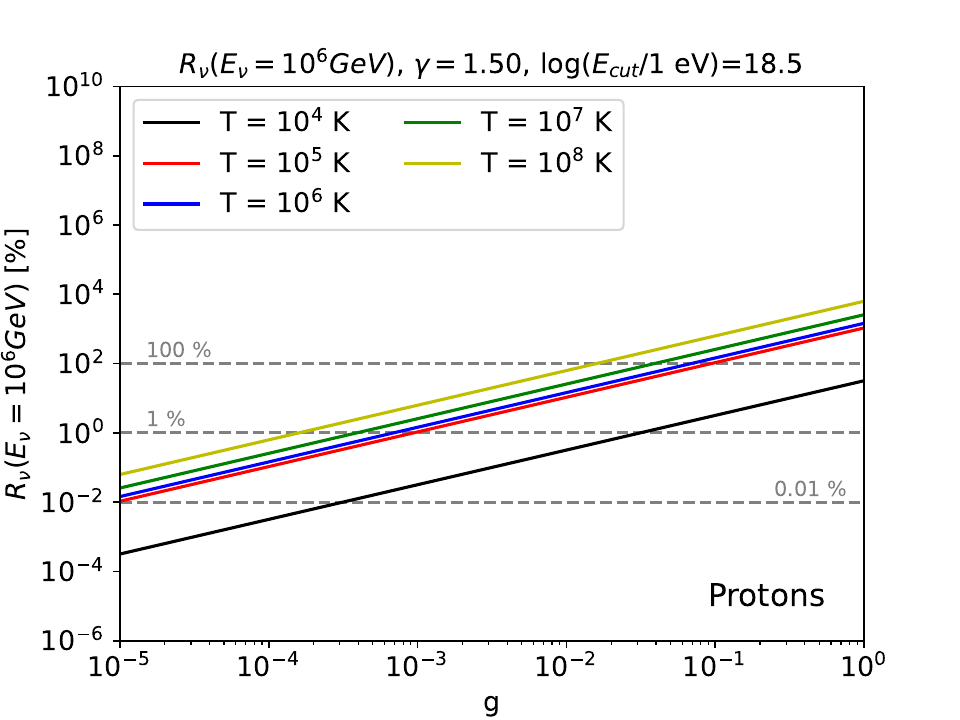}
\end{minipage}
\begin{minipage}{6.5cm}
\centering
\includegraphics[scale=0.42]{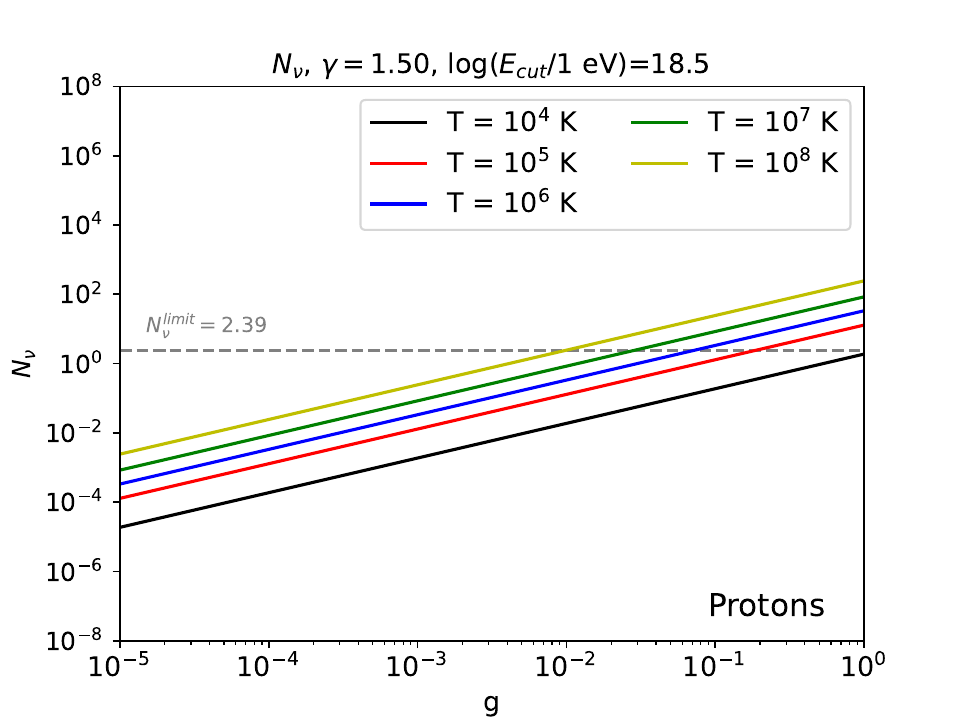}
\end{minipage}
\begin{minipage}{6.5cm}
\centering
\includegraphics[scale=0.42]{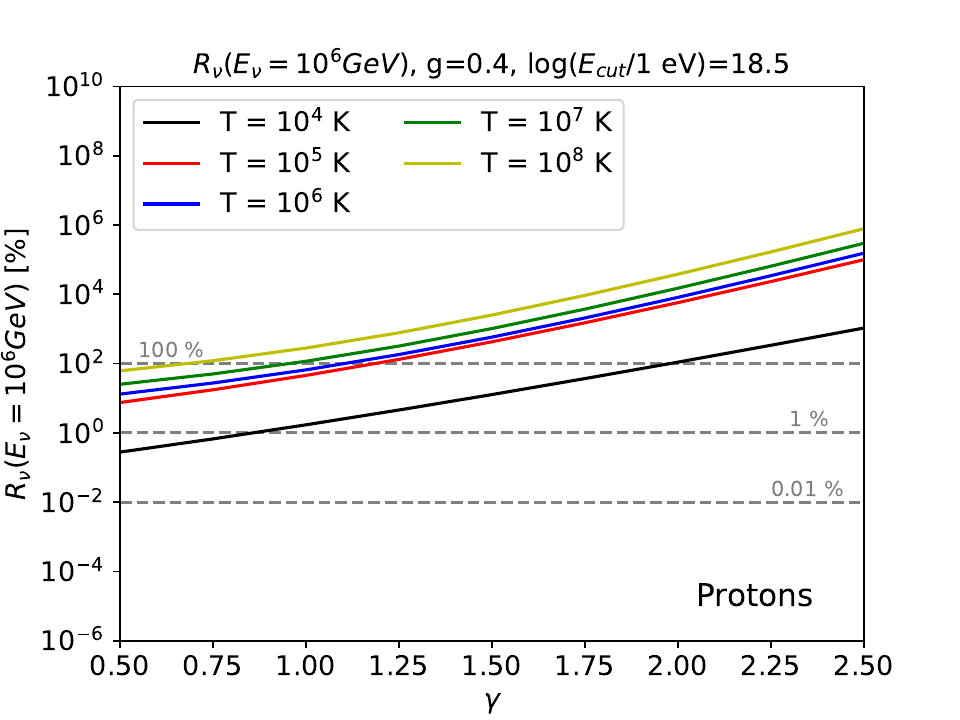}
\end{minipage}
\begin{minipage}{6.5cm}
\centering
\includegraphics[scale=0.42]{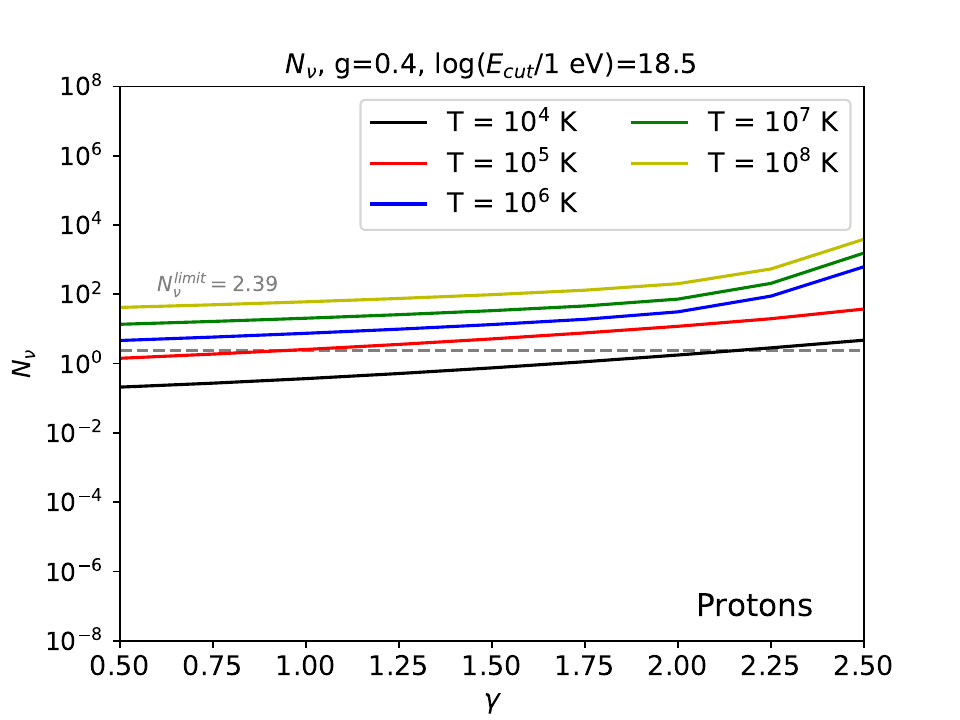}
\end{minipage}
\begin{minipage}{6.5cm}
\centering
\includegraphics[scale=0.42]{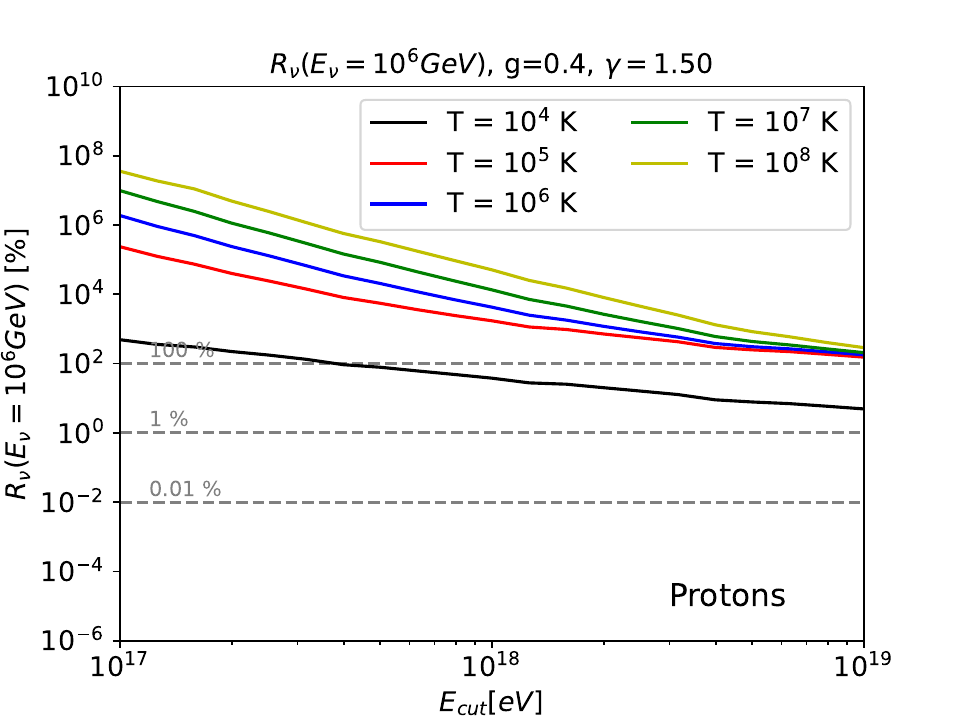}
\end{minipage}
\begin{minipage}{6.5cm}
\centering
\includegraphics[scale=0.42]{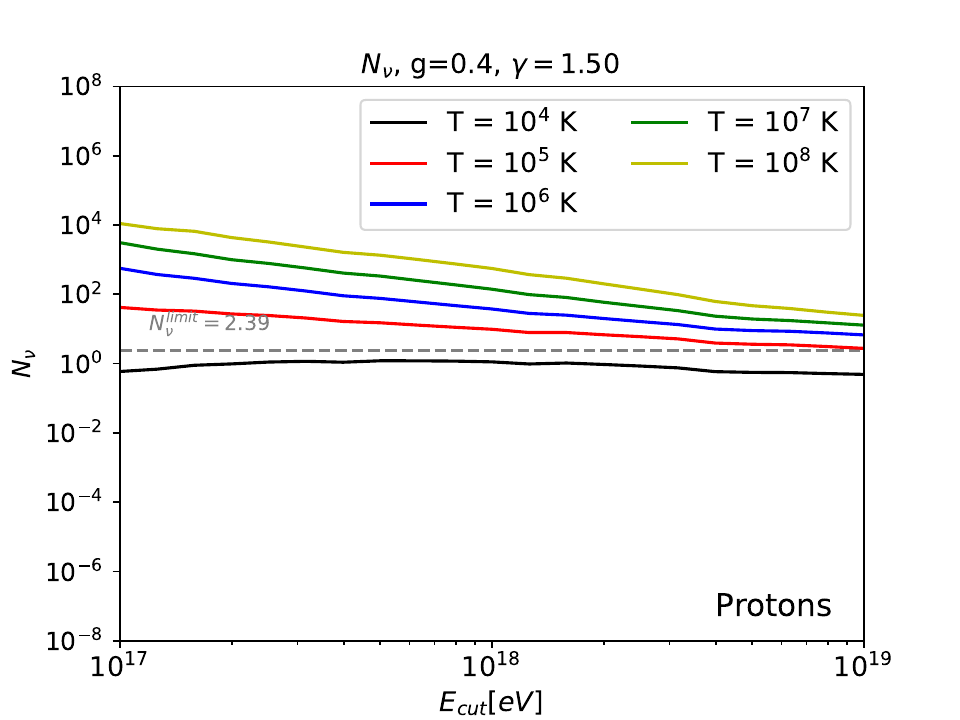}
\end{minipage}
\caption{Same of Fig.~\ref{parameter_scan_p}, but for the SFR source evolution in Eq.~\eqref{SFR_function}.}
\label{parameter_scan_p_SFR}
\end{figure}

\begin{figure}[H]
\centering
\begin{minipage}{6.75cm}
\centering
\includegraphics[scale=0.42]{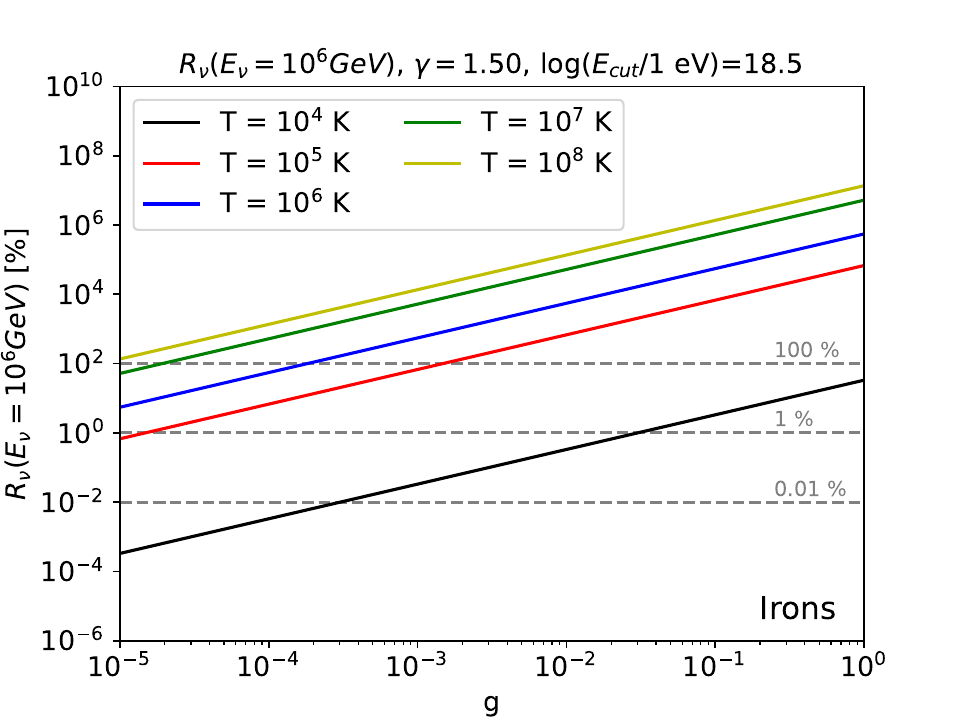}
\end{minipage}
\begin{minipage}{6.5cm}
\centering
\includegraphics[scale=0.42]{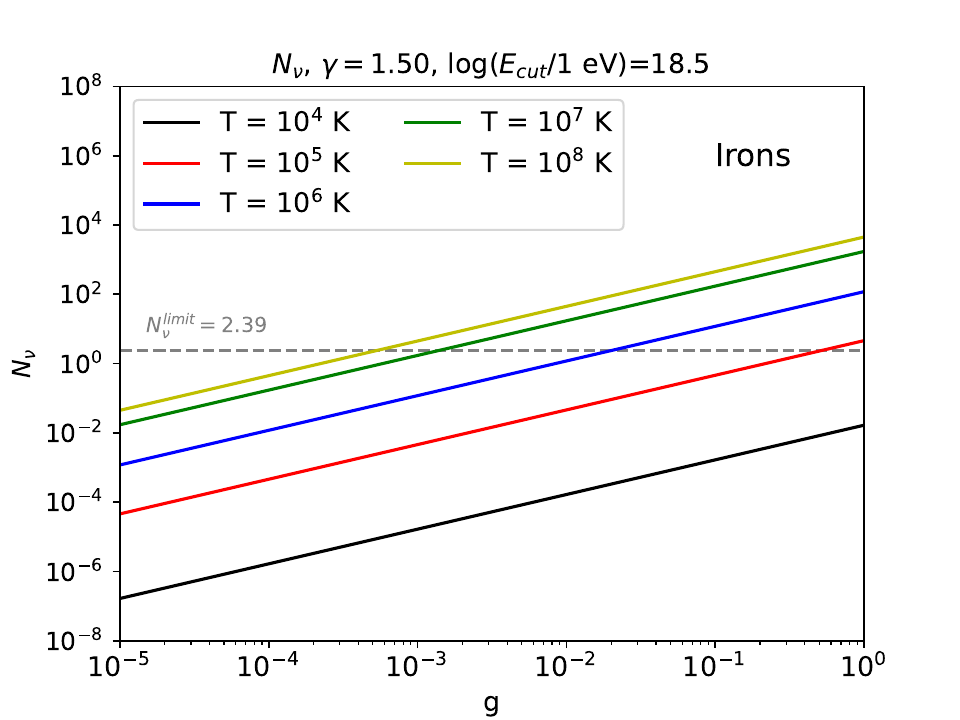}
\end{minipage}
\begin{minipage}{6.5cm}
\centering
\includegraphics[scale=0.42]{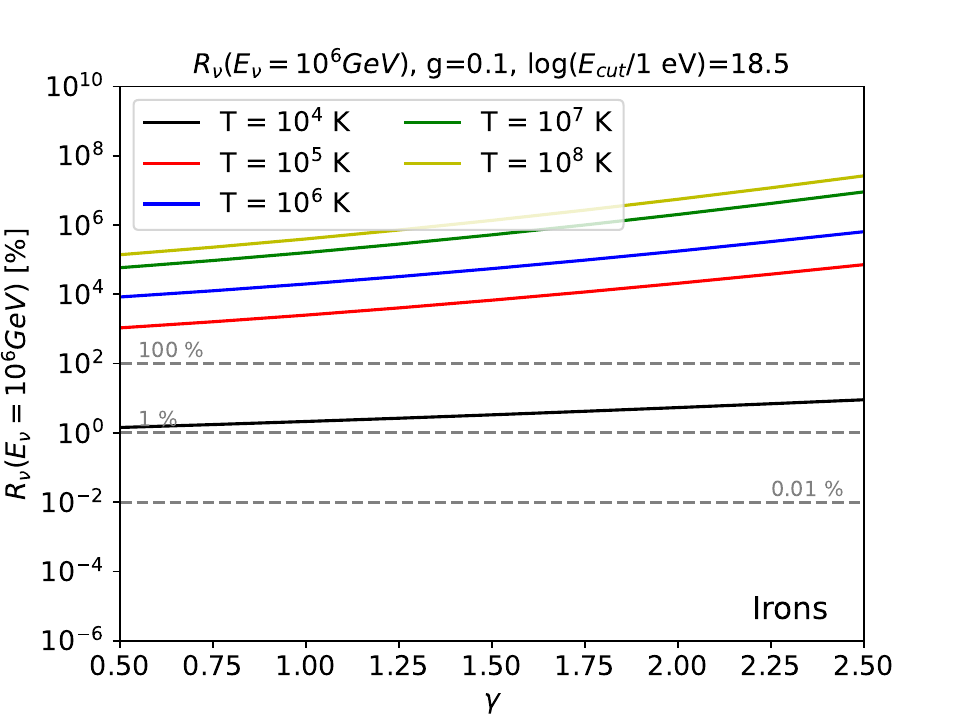}
\end{minipage}
\begin{minipage}{6.5cm}
\centering
\includegraphics[scale=0.42]{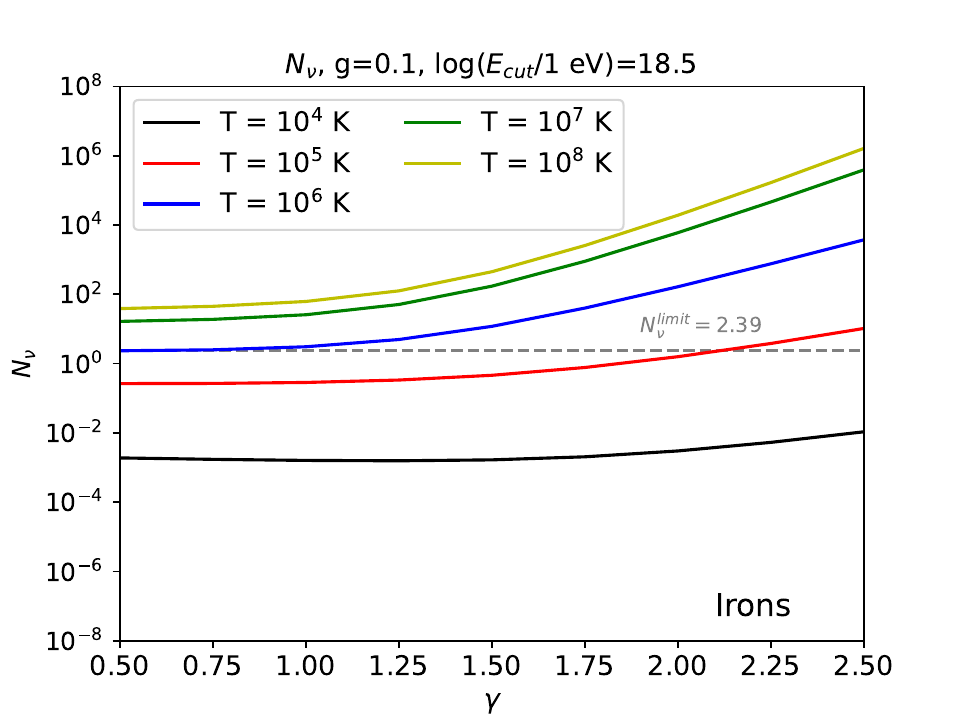}
\end{minipage}
\begin{minipage}{6.5cm}
\centering
\includegraphics[scale=0.42]{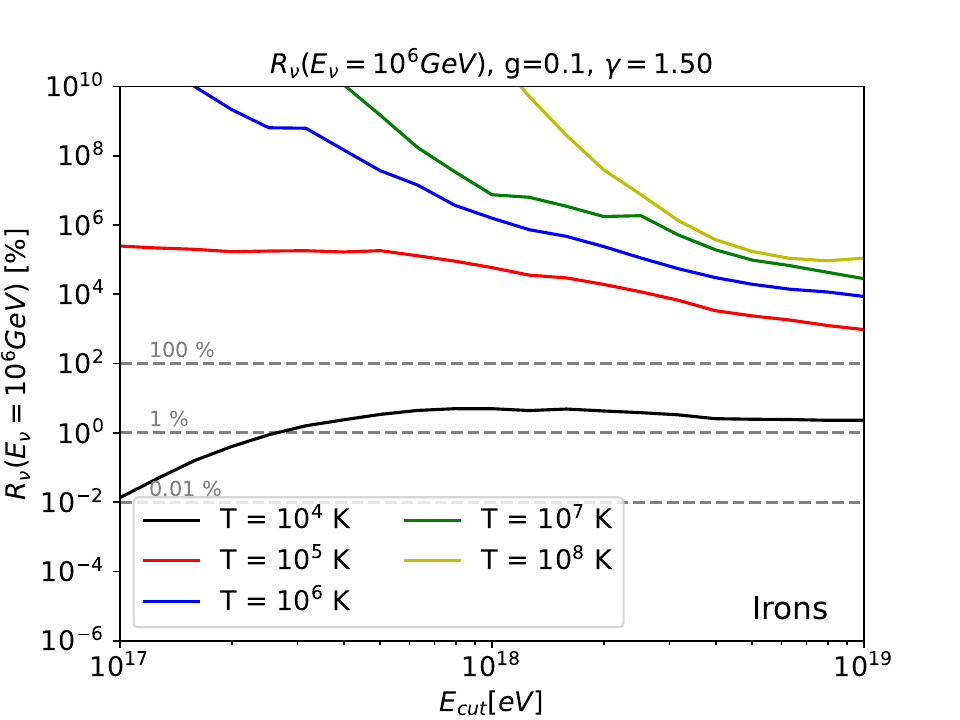}
\end{minipage}
\begin{minipage}{6.5cm}
\centering
\includegraphics[scale=0.42]{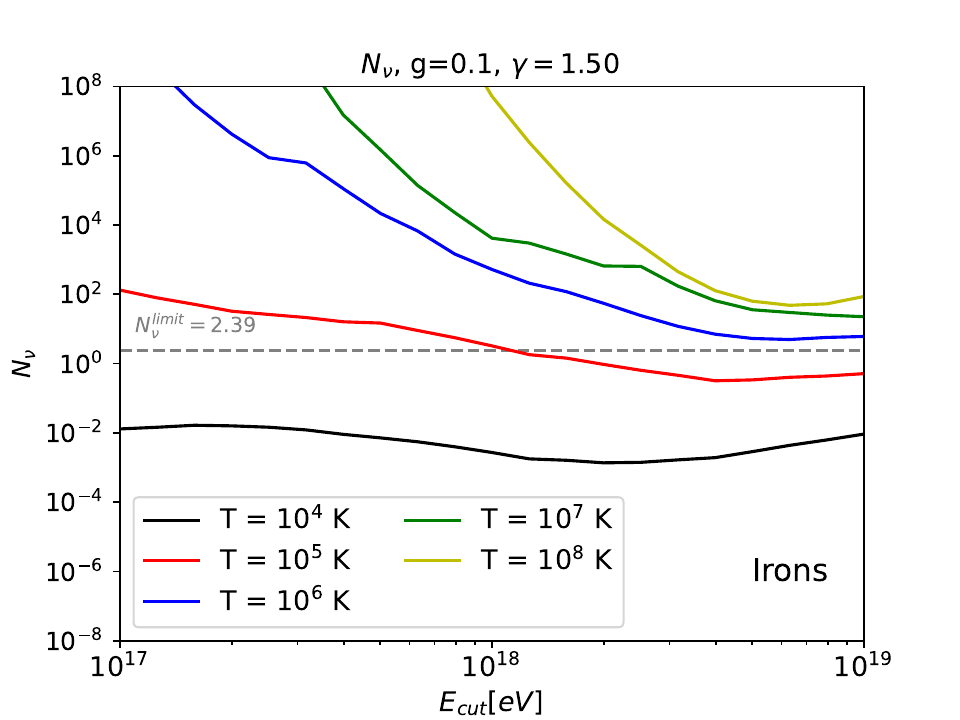}
\end{minipage}
\caption{Same of Fig.~\ref{parameter_scan_Fe}, but for the SFR source evolution in Eq.~\eqref{SFR_function}.}
\label{parameter_scan_Fe_SFR}
\end{figure}

\begin{figure}[t]
\centering
\begin{minipage}{6.5cm}
\centering
\includegraphics[scale=0.42]{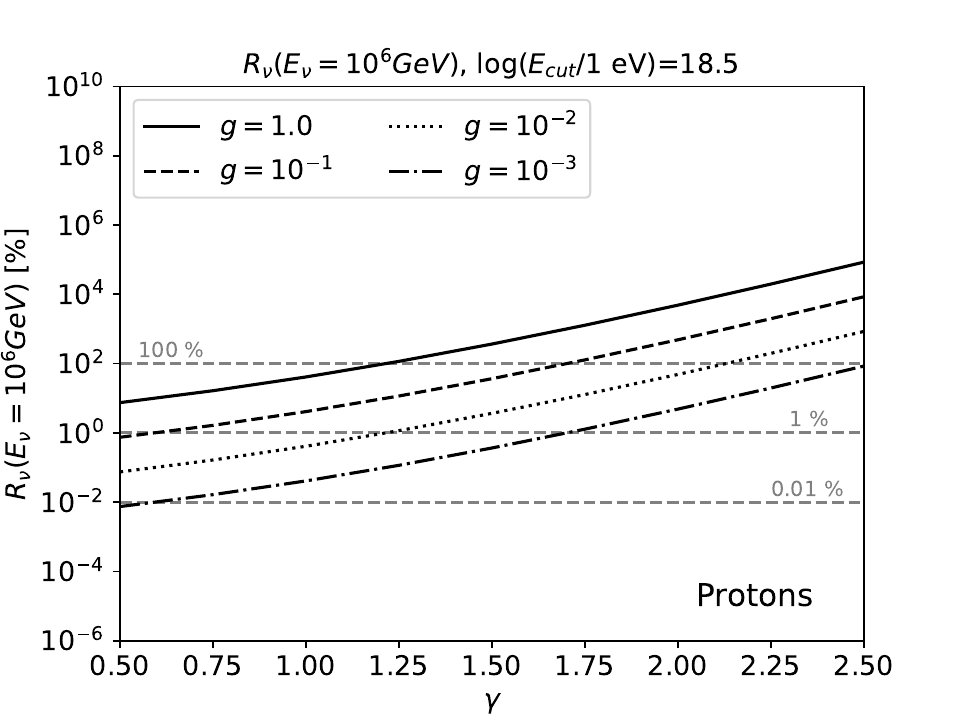}
\end{minipage}
\begin{minipage}{6.5cm}
\centering
\includegraphics[scale=0.42]{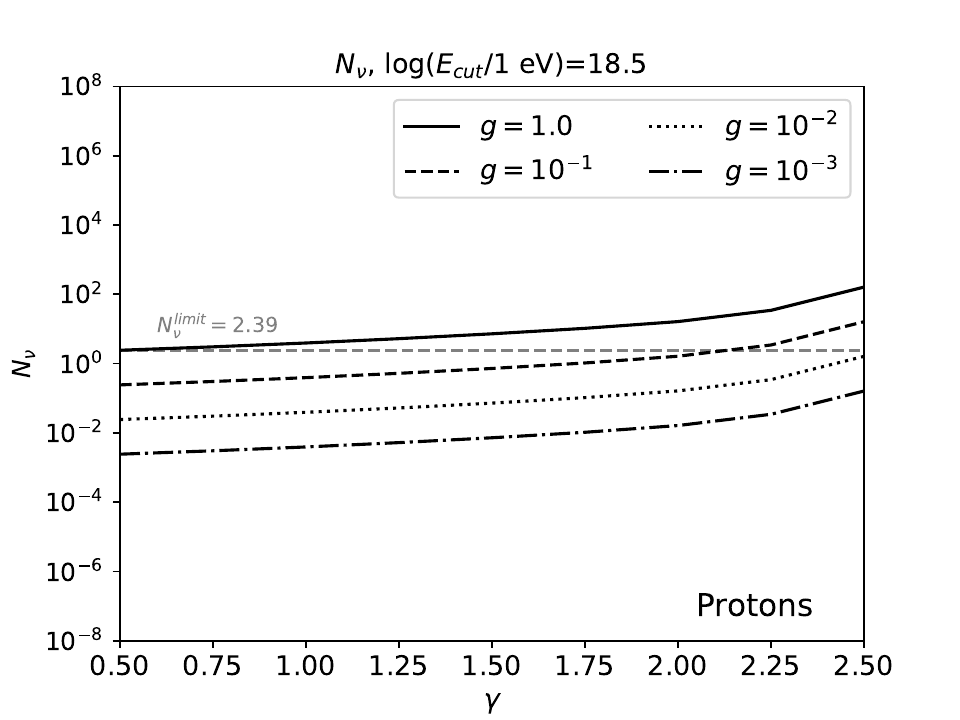}
\end{minipage}
\begin{minipage}{6.5cm}
\centering
\includegraphics[scale=0.42]{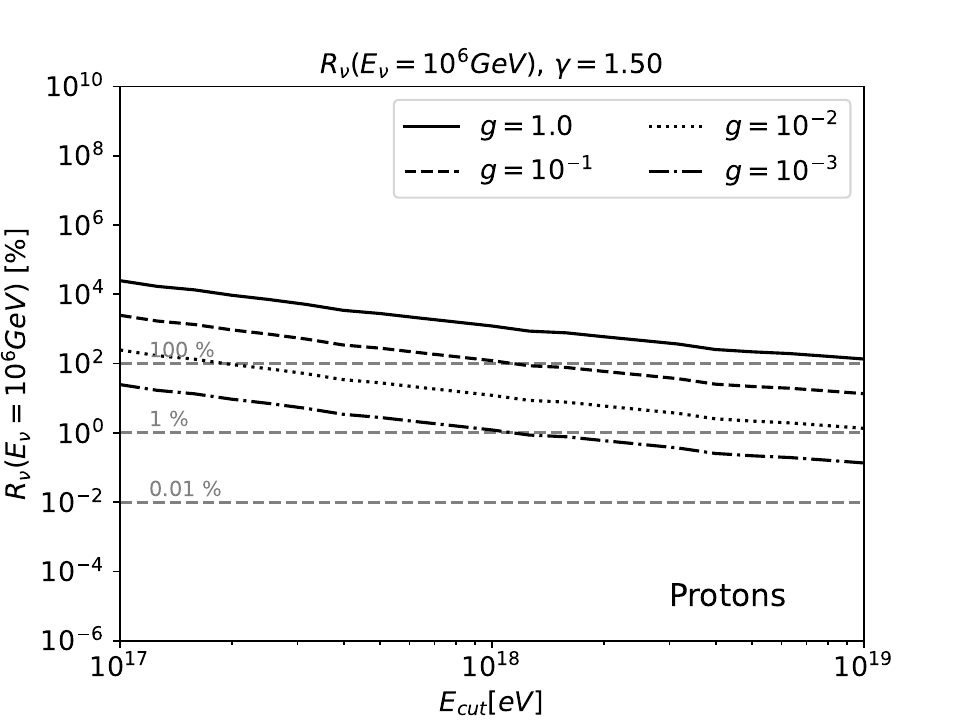}
\end{minipage}
\begin{minipage}{6.5cm}
\centering
\includegraphics[scale=0.42]{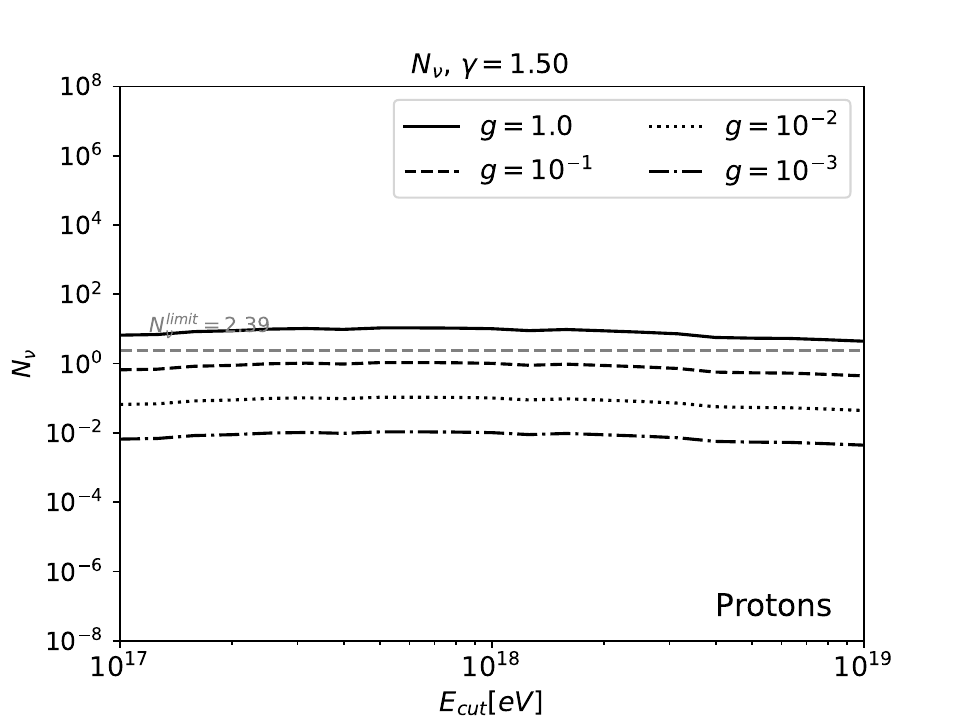}
\end{minipage}
\caption{Same of Fig.~\ref{parameter_scan_p_int}, but for the SFR source evolution in Eq.~\eqref{SFR_function}.}
\label{parameter_scan_p_SFR_int}
\end{figure}

\begin{figure}[H]
\centering
\begin{minipage}{6.5cm}
\centering
\includegraphics[scale=0.42]{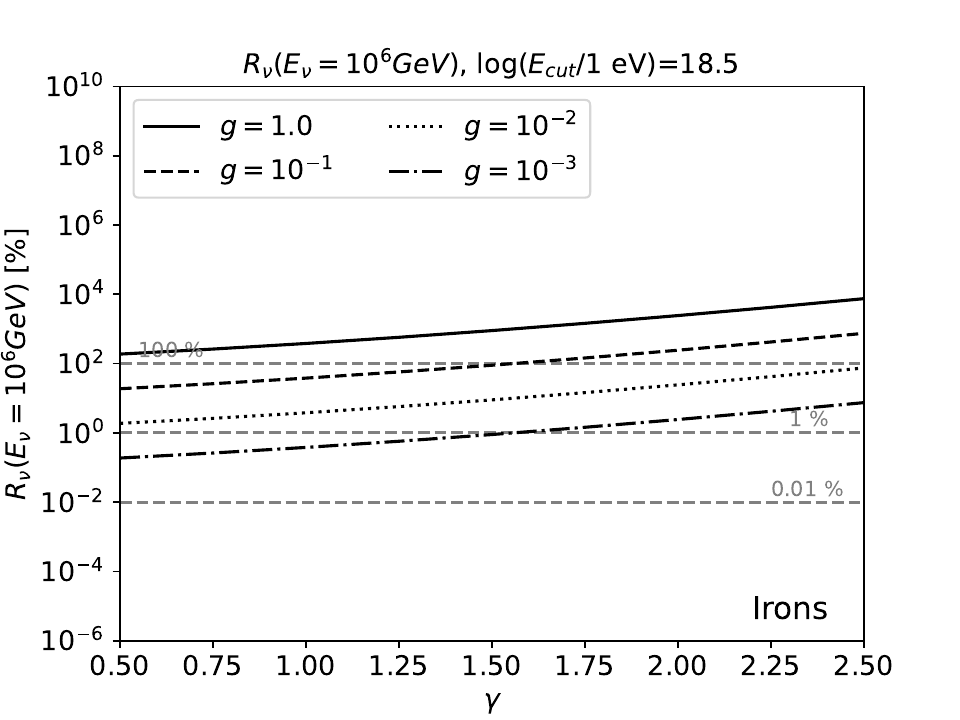}
\end{minipage}
\begin{minipage}{6.5cm}
\centering
\includegraphics[scale=0.42]{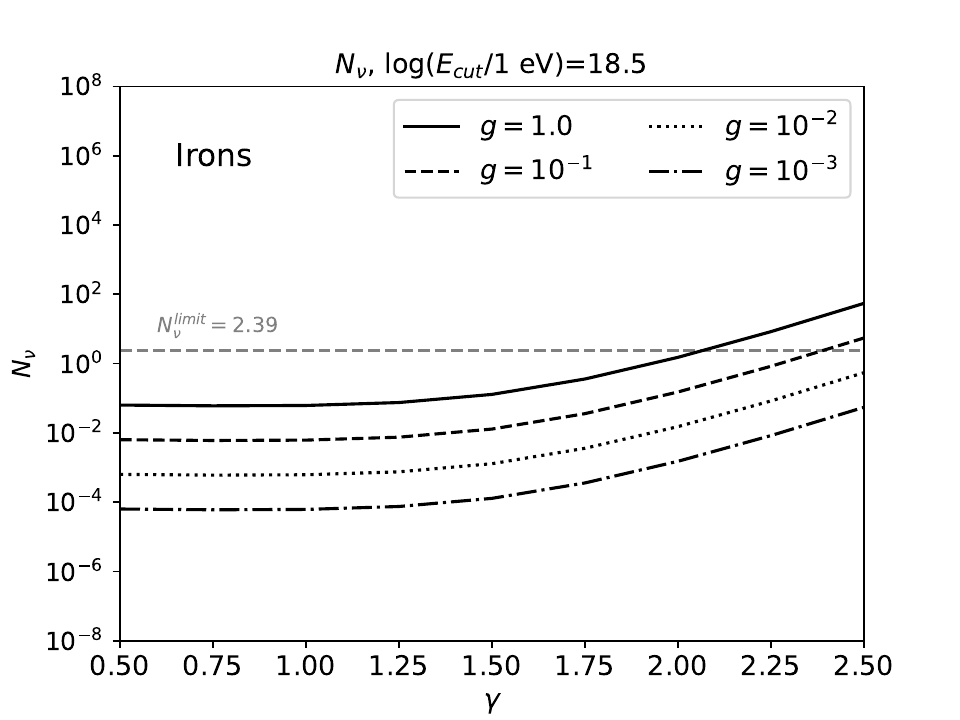}
\end{minipage}
\begin{minipage}{6.5cm}
\centering
\includegraphics[scale=0.42]{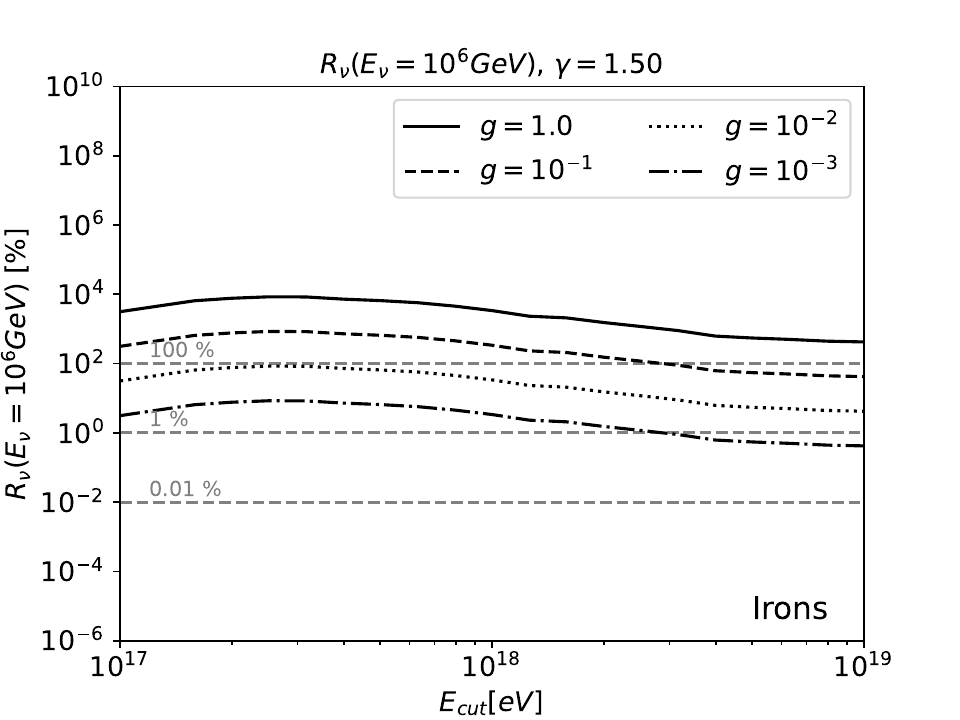}
\end{minipage}
\begin{minipage}{6.5cm}
\centering
\includegraphics[scale=0.42]{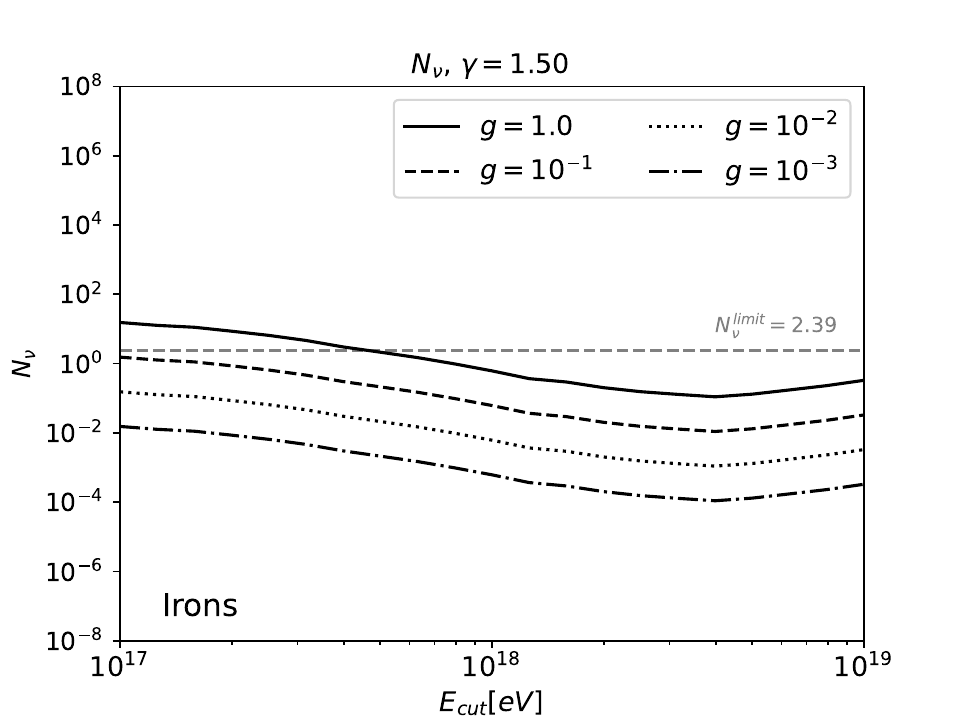}
\end{minipage}
\caption{Same of Fig.~\ref{parameter_scan_Fe_int}, but for the SFR source evolution in Eq.~\eqref{SFR_function}.}
\label{parameter_scan_Fe_SFR_int}
\end{figure}

\begin{figure}[H]
\centering
\begin{minipage}{6.5cm}
\centering
\includegraphics[scale=0.47]{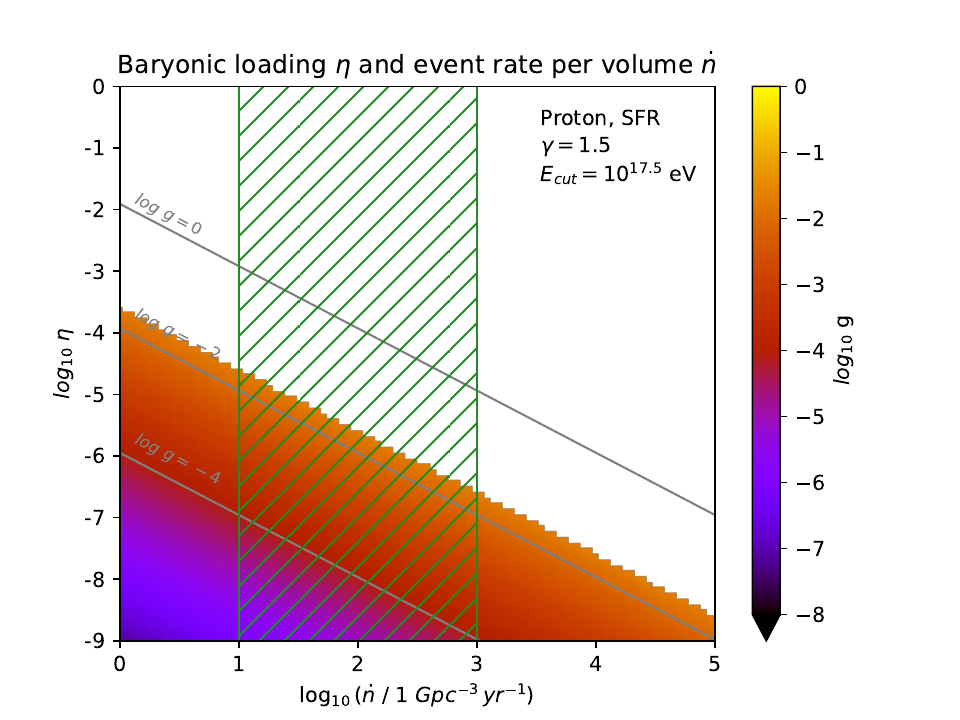}
\end{minipage}
\begin{minipage}{6.5cm}
\centering
\includegraphics[scale=0.47]{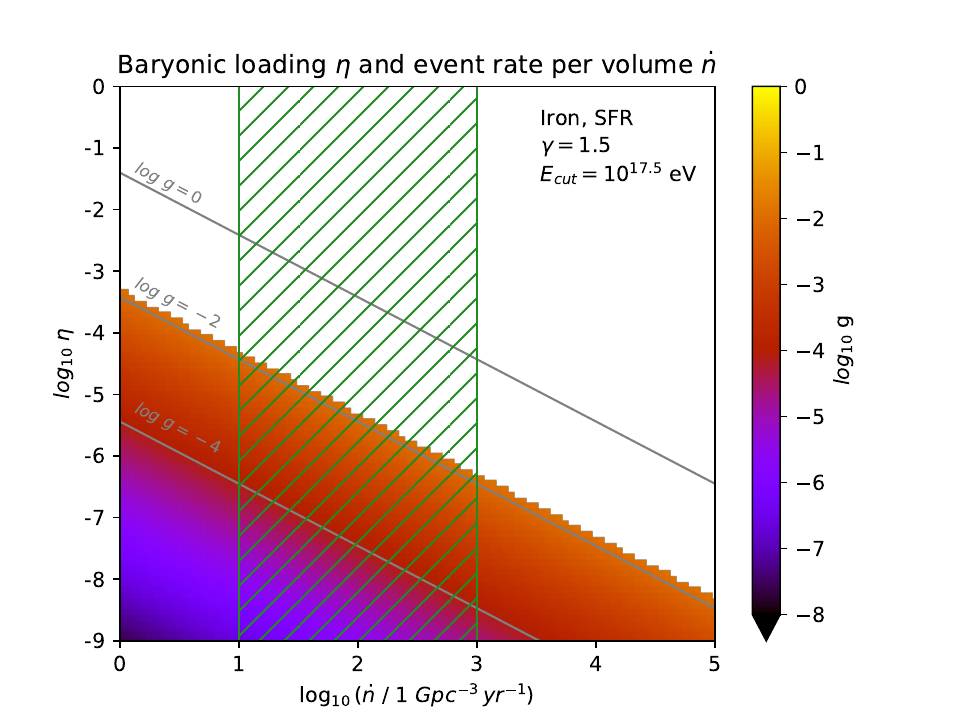}
\end{minipage}
\begin{minipage}{6.5cm}
\centering
\includegraphics[scale=0.47]{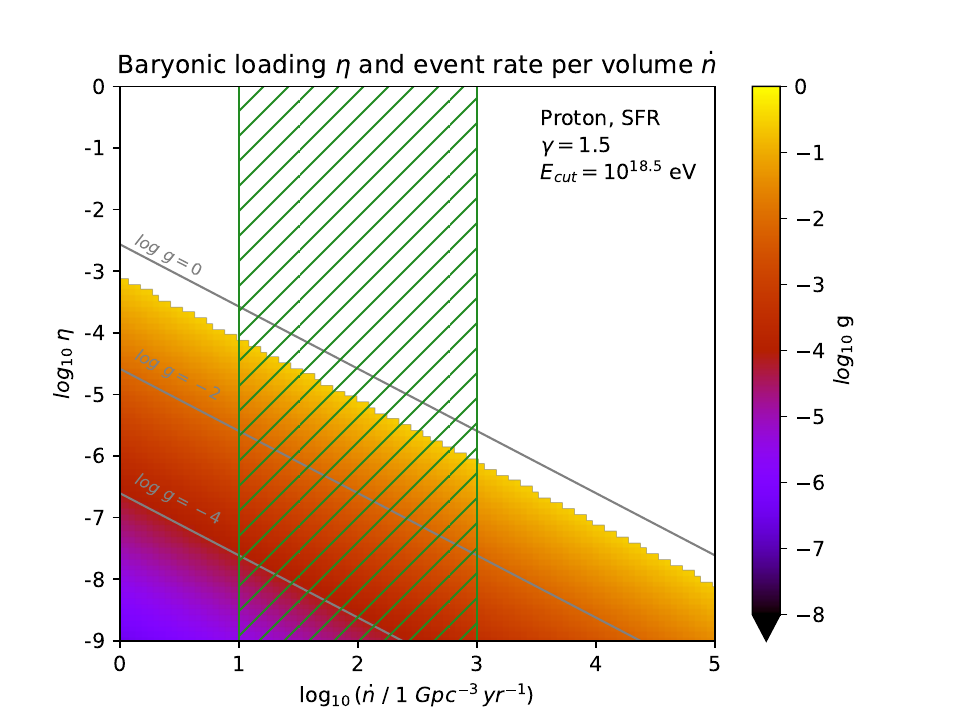}
\end{minipage}
\begin{minipage}{6.5cm}
\centering
\includegraphics[scale=0.47]{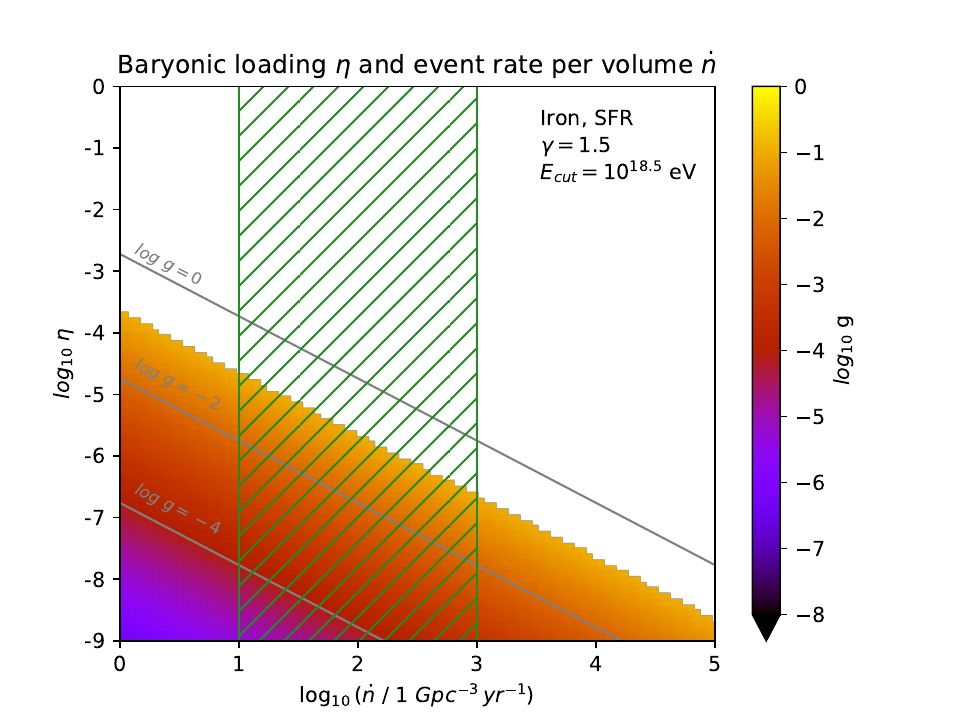}
\end{minipage}
\caption{Same of Fig.~\ref{baryonic_load_density_rate}, but for the SFR source evolution in Eq.~\eqref{SFR_function}.}
\label{baryonic_load_density_rate_SFR}
\end{figure}

In this appendix, the same analysis presented in Sec.~\ref{subsec_parameter} is shown for the scenarios in which the source evolution is that of the SFR , given in Eq.~\eqref{SFR_function}. The results for proton and iron acceleration are shown in Figs.~\ref{parameter_scan_p_SFR} and ~\ref{parameter_scan_Fe_SFR}, respectively. No relevant differences are observed with respect to the case $m=0$. In Figs.~\ref{parameter_scan_p_SFR_int} and~\ref{parameter_scan_Fe_SFR_int} we show the same analysis of Figs.~\ref{parameter_scan_p_int} and~\ref{parameter_scan_Fe_int} for the SFR source evolution model. In Fig.~\ref{baryonic_load_density_rate_SFR} the same constraints of Fig.~\ref{baryonic_load_density_rate} are shown for the SFR source evolution model.

\end{document}